\documentclass{article}
\usepackage{epsfig}
\begin{document}

\sloppy
\newcommand{\rp}{\right)}
\newcommand{\lp}{\left(}
\newcommand \be  {\begin{equation}}
\newcommand \bea {\begin{eqnarray}}
\newcommand \ee  {\end{equation}}
\newcommand \eea {\end{eqnarray}}

\title{Critical Ruptures}

\author{Anders Johansen$^1$ and Didier Sornette$^{1,2,3}$
\footnote{Corresponding author: D. Sornette,
University of California, Los Angeles, 3845 Slichter Hall, Box 951567 Los
Angeles, CA 90095-1567. Tel.: +1 (310) 825 28 63 Fax.: +1 (310) 206 3051-
e-mail: sornette@moho.ess.ucla.edu.}\\
$^1$ Institute of Geophysics and
Planetary Physics\\ University of California, Los Angeles, California 90095\\
e-mail: anders@moho.ess.ucla.edu. \\
$^2$ Department of Earth and Space Science\\
University of California, Los Angeles, California 90095\\
$^3$ Laboratoire de Physique de la Mati\`{e}re Condens\'{e}e\\ CNRS UMR6622 and
Universit\'{e} de Nice-Sophia Antipolis\\ B.P. 71, Parc
Valrose, 06108 Nice Cedex 2, France}

\maketitle

\begin{abstract}
The fracture of materials is a catastrophic phenomenon of considerable 
technological and scientific importance. Here, we analysed experiments 
designed for industrial applications in order to test the concept that, 
in heterogeneous materials such as fiber composites,  rocks, concrete 
under compression and materials with large distributed residual stresses, 
rupture is a genuine critical point, {\it i.e.}, the culmination of a 
self-organization of damage and cracking characterized by power law 
signatures. Specifically, we analyse the acoustic emissions recorded 
during the pressurisation of spherical tanks of kevlar or carbon fibers 
pre-impregnated in a resin matrix wrapped up around a thin metallic liner 
(steel or titanium) fabricated and instrumented by A\'erospatiale-Matra Inc.
These experiments are performed as part of a routine industrial procedure 
which tests the quality of the tanks prior to shipment and varies in nature. 
We find that the seven acoustic emission recordings of seven pressure tanks 
which was brought to rupture exhibit clear acceleration in agreement with a 
power law ``divergence'' expected from the critical point theory. In addition, 
we find strong evidence of log-periodic corrections that quantify the 
intermittent succession of accelerating bursts and quiescent phases of the 
acoustic emissions on the approach to rupture. An improved model accounting 
for the cross-over from the non-critical to the critical region close to the 
rupture point exhibits interesting predictive potential.
\end{abstract}

\thispagestyle{empty}
\vspace{1cm}

\newpage

\pagenumbering{arabic}

\section{Plan of the study}

In this paper, we first present in section 2 a brief review of the
``critical rupture''
concept with an emphasis on the role of heterogeneity. Section 3 describes the
experimental systems and the properties of the acoustic emission time series
that we analyse with three theoretical formulas derived from the critical
rupture
concept. We present a brief justification for these three power laws.
Section 4 gives the results obtained on the acoustic emission energy
release rate on
seven systems. Section 5 analyses the cumulative energy releases of these seven
systems. Section 6 describes
the relative merits of the three power law formulas for the prediction of the
critical pressure of rupture and section 7 concludes.

\section{Review of the ``critical rupture'' concept}

\subsection{Background}

The damage and fracture of materials is of enormous technological
interest due to their economic and human cost. They cover a wide range of
phenomena like, {\it e.g.}, cracking of glass, aging of concrete, the failure 
of fiber networks in the formation of paper and the breaking of a metal bar 
subject to
an external load. Failure of composite systems are of utmost importance in
naval, aeronautics and space industry \cite{rocket}. By the term composite, we
refer to materials with heterogeneous microscopic structures and also to
assemblages of macroscopic elements forming a super-structure. Chemical and 
nuclear plants suffer from cracking due to corrosion either of chemical or 
radioactive origin, aided by thermal and/or mechanical stress.

Despite the large amount of
experimental data and the considerable effort that has been undertaken by
material scientists \cite{inge}, many questions about fracture have
not been answered yet. There is no comprehensive understanding of rupture
phenomena but only a partial classification in
restricted and relatively simple situations. This lack of fundamental
understanding is indeed reflected in the absence of reliable prediction
methods for rupture based on a suitable monitoring of the stressed 
system. Not only is there a lack of theoretical understanding of the 
reliability of a system, but the empirical laws themselves have often 
limited value. What we need are models that
incorporate the underlying physics to identify and use
relevant precursory patterns. Here, we propose innovative steps in this
direction that are based on two key concepts\,: the role of heterogeneity
and the
possible existence of a hierarchy of characteristic scales.

Many material ruptures occur by a ``one crack'' mechanism and a lot of
effort is
being devoted to the understanding, detection and prevention of the
nucleation of cracks  \cite{instab,roughness}.
Exceptions to the ``one crack'' rupture mechanism are heterogeneous
materials such as fiber composites,
rocks, concrete under compression and
materials with large distributed residual stresses. The common property
shared by these systems is the existence of large inhomogeneities,
that often limit the use of effective medium theories for the
elastic and more generally the mechanical properties.
In these systems, failure may occur as the culmination of a progressive
damage involving complex interactions between multiple defects and the growing
of micro-cracks. In addition, other relaxation, creep, ductile, or plastic
behaviors, possibly coupled with corrosion effects, may come into play.
Many important practical applications involve the coupling between
mechanic and chemical effects with the competition between several
characteristic
time scales. Application of stress may act as a catalyst of
chemical reactions \cite{mechanochemi} or, reciprocally, chemical
reactions may lead to bond weakening \cite{chemimech} and thus promote
failure. A dramatic example is the aging of present aircrafts due to repeating
loading in a corrosive environment \cite{Airforce}.
The interaction between multiple defects and the existence of
several characteristic scales present a considerable challenge to the modeling
and prediction of rupture. Those are the systems and problems that will guide
our modeling efforts.

\subsection{Previous Statistical Physics models}

\subsubsection{Scaling and critical point}

Motivated by the multi-scale nature of ruptures in heterogeneous systems and by
analogies with the percolation model \cite{Stauffer}, de Arcangelis et al.
first suggested \cite{ARH} in the mid-eighties that rupture of sufficiently 
heterogeneous media would exhibit some universal properties similar to critical
phase transitions. The idea was to build on the knowledge accumulated in 
statistical physics on the so-called $N-$body problem and cooperative effects 
in order to describe multiple interactions between defects. Applying these
concepts, scaling laws were found to describe size effects and damage 
properties [10-14]. However, being essentially quasi-static in nature they 
neglected important aspects of the rupture process and had very limited 
potential in terms of time-to-failure analysis. In 1992 a dynamical model 
of rupture introduced in 1992 was introduced which achieved a higher 
degree of realism. It was initially formulated in
the framework of electric breakdown under the name of the ``thermal fuse 
model'' \cite{Vanneste}\,: when subjected to a given current, a fuse heats up 
due to a generalized Joule effect  and eventually breaks down when its 
temperature reaches the melting threshold. Later, it was reformulated in 
\cite{dendrite} by showing that it is exactly equivalent to a (scalar) 
anti-plane mechanical model of rupture with elastic interaction in which 
the temperature becomes a local damage variable. This model accounts for 
space-dependent elastic and rupture properties, has a realistic loading and 
produces many growing interacting micro-cracks with an organization which is 
a function of the damage-stress law.
It was found that, under a step-function stress loading,
the total rate of damage, as measured for instance by
the elastic energy released per unit time, on average increases as a power
law of the
time-to-failure. In this model, rupture was indeed found
to occur as the culmination of the progressive nucleation, growth and fusion
between micro-cracks, leading to a fractal network, but the exponents
were found to be non-universal and a function of the damage law. This
model has since then been found to describe correctly the experiments on
the electric breakdown of insulator-conducting composites \cite{Lamai}.
Another application of the thermal fuse model is damage
by electro-migration of polycrystalline metal films \cite{Bradley}.
See also \cite{dendrite} for relations with dendrites and fronts propagation.

In 1991-1995, it was proposed and tested on a real engineering composite
structure the
concept that failure in fiber composites may be described similarly, namely
that the rate of damage would exhibit a ``critical'' behavior
\cite{Anifrani}. This critical behavior corresponds to an acceleration
of the
rate of energy release or to a deceleration, depending on the nature and
range of the
stress transfer mechanism and on the loading procedure. Based on
general consideration on the nature of the experimental signatures of critical
rupture, it was proposed that the power law behavior of the
time-to-failure analysis should be corrected for the presence of log-periodic 
modulations \cite{Anifrani} as signatures of a hierarchy of characteristic 
scales in the rupture process. This method is now been used by
the French Aerospace company A\'erospatiale-Matra on pressure tanks made of
kevlar-matrix
and carbon-matrix composites embarked on the European Ariane 4 and 5 rockets.
In a nutshell, the method consists in this application in
recording acoustic emissions under constant stress rate
and the acoustic emission energy as a function of
stress is fitted by the above log-periodic critical theory \cite{Anifrani}.
One of the parameter is the
time of failure and the fit thus provides a ``prediction'' when the sample
is not brought to failure in the first test \cite{Anifrani2}. Good predictive
performances have been reported (Anifrani, private communication).
Since we now have a better understanding of the mechanisms at the origin
hierarchical self-organization in rupture
\cite{logperiodic,Huang,logperiodicreport} which seems to apply to
some other systems as well \cite{otherlogperiodic,DLA}, here we
re-examine the critical rupture concept and the evidence for the existence
of log-periodic corrections to scaling \cite{Anifrani,canonical}.
This study is
based on the analysis of 7 acoustic emission recordings of 7 pressure tank
structures brought to rupture made available to us.
We also present preliminary tests of the predictive
skills, in particular using an extension of the theory which allows us to
incorporate the cross-over regime from the non-critical to the critical regime
\cite{Andersen2}.

\subsubsection{The role of heterogeneities}

A key parameter is the degree and nature of disorder.
This was considered early by Mogi \cite{Mogi}, who showed experimentally on
a variety
of materials that, the larger
the disorder, the stronger and more useful are the precursors to rupture.
For a long time,
the Japanese research effort for earthquake prediction and risk assessment
was based on this very idea \cite{Mogirecent}.

The role of heterogeneities on the nature of rupture has been quantified
using a spring-block model
with stress transfer over limited range \cite{Andersen1}. This model
does not claim realism but attempts rather to capture the role of limited
stress
transfer and heterogeneity. The heterogeneity was found to play the role 
of a relevant field\,: systems with limited
stress amplification exhibit a tri-critical transition \cite{tricri},
from a Griffith-type abrupt rupture (first-order) regime to a
progressive damage (critical) regime as the disorder
increases. This effect has also been demonstrated  on a
simple mean-field model of rupture, known as the
democratic fiber bundle model \cite{DFBM}. In a two-dimensional
spring-block model of surface fracture, the stress can be released
by breaking of springs {\it and\/} block slips \cite{Andersen1}.
This spring-block model may represent schematically the experimental
situation where a balloon covered with paint or dry resin is progressively
inflated. An industrial application may be for instance a metallic tank
with carbon
or kevlar fibers impregnated in a resin matrix wrapped up around it  which is
slowly pressurized \cite{Anifrani}, as we report in this paper. As a 
consequence, it elastically deforms,
transferring tensile stress to the overlayer. Slipping (called fiber-matrix
delamination) and cracking can thus occur in the overlayer. In
\cite{Andersen1},
this process is modeled by an array of blocks which represents the
overlayer on a coarse
grained scale in contact with a surface with solid friction contact. The solid
friction will limit stress amplification.
The stress-strain curves for different values of the disorder $\Delta$, here
quantified by the width of the distribution of initial positions of the
blocks which
captures the effect of residual stresses in the material (but
does not explore the other dimensions of disorder), show a larger softening
and rounding
as disorder increases. The phase diagram of the
fracturing in the  $(\Delta; F_c / F_s)$  plane, where $F_c$ (resp. $F_s$) is
the rupture (resp. sliding) threshold shows that, for fixed $F_c / F_s < 2.9$,
increasing the disorder $\Delta$ allows the system to go from a first-order
behaviour to a
critical regime. The fact that the disorder is so relevant as to create the
analog of a tri-critical behavior can be traced back to the existence of solid
friction on the blocks which ensures that the elastic forces in the springs are
carried over a bounded distance (equal to the size of a slipping ``avalanche'')
during the stress transfer induced by block motions. In this context, we
note that
the importance of heterogeneity in the
context of fiber composites has also been stressed in \cite{Zhou}.

In the presence of long-range elasticity, disorder is found to be always
relevant leading to a critical rupture. However,
the disorder controls the width of the critical region \cite{Andersen2}. The
smaller it is, the smaller will be the critical region, which
may become too small to play any role in practice. This has been confirmed by
simulations of the thermal fuse model mentioned above \cite{Vanneste}. 
The damage rate on approach to failure for different disorder can be 
rescaled onto a universal master curve \cite{Andersen2}.

Numerical simulations of Sahimi and Arbati \cite{Arbati}
have recently confirmed that, near the global failure point,
the cumulative elastic energy released during fracturing of heterogeneous
solids with long-range elastic interactions
follows a power law with log-periodic corrections to the
leading term  consistent with previous results [22-30].
The presence of log-periodic correction to scaling in the elastic energy
released
has also been demonstrated numerically for the thermal fuse model
\cite{canonical,logperiodicreport} using a novel
averaging procedure, called the ``canonical ensemble averaging''.
A recent experimental study of rupture of fiber-glass
composites has also confirmed the critical scenario \cite{Ciliberto}.

These results indicate that the ``critical'' behavior is not restricted to
limited stress amplification but may well pertain to a much broader class
of systems. This needs to be investigated more. In quasi-static models of 
rupture \cite{Herrmann,Roux2}, numerical simulations and
perturbation expansions have shown the existence of
three main regimes, depending on the distribution $p(x)$ of rupture
thresholds $x$.
If $p(x) \sim x^{\phi_0 -1}$ for $x \rightarrow 0$ and
$p(x) \sim x^{-(1+\phi_{\infty})}$ for $x \rightarrow +\infty$, then the
three regimes
depend on the relative value of $\phi_0$ and $\phi_{\infty}$ compared to
two critical values $\phi_0^c$ and $\phi_{\infty}^c$. The ``weak disorder''
regime
occurs for $\phi_0 >\phi_0^c$ (few weak elements) and
$\phi_{\infty}>\phi_{\infty}^c$
(few strong elements) and boils down
essentially to the nucleation of a ``one-crack'' run-away. For
$\phi_0 \leq \phi_0^c$ (many weak elements) and $\phi_{\infty}>\phi_{\infty}^c$
(few strong elements), the rupture is
controlled by the weak elements, with important size effects.
The damage is diffuse but presents a structuration at large scales.
For $\phi_0 > \phi_0^c$ (few weak elements) and $\phi_{\infty} \leq
\phi_{\infty}^c$
(many strong elements), the
rupture is controlled by the strong elements\,: the final
damage is diffuse and the density of broken elements goes to a non-vanishing
constant. This third case is very similar to the percolation models of rupture
and it has been shown that percolation is retrieved
in the limit of very large disorder \cite{Roux}.

\subsubsection{Qualitative physical scenario}

A qualitative physical picture for the progressive damage of an heterogeneous
system leading  to global failure emerges from all these results.
First, single isolated defects and
micro-cracks nucleate which then, with the increase of load or time of
loading, both grow
and multiply leading to an increase of the density of defects per unit volume.
As a consequence, defects begin to merge until a ``critical density'' is
reached. Uncorrelated percolation \cite{Stauffer} provides a starting modeling
point valid in the limit of very large disorder \cite{Roux,Gilabert}. For
realistic systems, long-range correlations transported by the stress field
around defects and cracks make the problem much more subtle. Time dependence is
expected to be a crucial aspect in the process of correlation building in these
processes. As the damage increases, a new ``phase'' appears, where
micro-cracks
begin to merge leading to screening and other cooperative effects.
Finally, the main fracture is formed causing global failure. The nature of
this global failure may be abrupt (``first-order'') or ``critical''
depending of the type of heterogeneities influencing  load transfer and
stress relaxation mechanisms. In the ``critical'' case,
the failure of composite systems
may often be viewed, in simple intuitive terms, as the result of a correlated
percolation process. However, the challenge is to describe the transition from
damage and corrosion processes at a microscopic level to macroscopic
failure.

\section{Data and methodology}

\subsection{The experimental systems}

The systems used in our study are spherical tanks of radius
or 0.2 to 0.42 m, made of kevlar or carbon fibers pre-impregnated
in a resin matrix wrapped up around a thin metallic liner (steel or titanium).
They are fabricated and instrumented by A\'erospatiale-Matra Inc.
In a typical experiment, each tank is pressurized by increasing the
internal water content
at a constant pressure rate of 3 to 6 bars per second. Acoustic emission
signals
are recorded from three to six acoustic transducers with resonant frequency
of 150 KHz,
placed at equal distances on the equator. Acoustic emissions characterize
rather
faithfully the irreversible motions and damages occuring within the
composites under
increasing load. The recording of acoustic emissions is performed by
a Locan-At from Euro Physical Acoustics Inc, with tunings (thresholds, gains,
Peak Definition Time, Hit Definition Time, Hit Lock-out Time) adjusted for
each experiment.
The output of the Locan-At is a list of acoustic events with their time,
the pressure
at which they occured, their duration, their amplitude and a measure of
their energy. In the sequel,
we analyse the files giving the energy of all recorded acoustic emission
events as a function of pressure.

Acoustic emissions are mechanical waves produced by sudden movements in
stressed
materials. They occur in a wide range of materials, structures and
processes, from
the largest scale (earthquakes) to the smallest on (dislocation motions).
Acoustic emission has been
found to be a delicate technique to use since each loading is unique and
tests the whole
structure. Contamination by noise is a real problem. Notwithstanding the
development
of numerous acoustic emission structural testing procedures
\cite{AEbooks1,AEbooks2}, their practical
implementations for prediction purpose have not been found reliable.
The acoustic emission technique differs from most other non-destructive methods
in that acoustic emissions originate from within the material and results
from high-frequency
motions, while most methods detect existing geometrical heterogeneities. A
large body of
research in the mechanical literature has thus focused on the
identification of the
types of motions that generate acoustic emissions and how their signatures
can be associated
with their sources. Such sudden material motions can be due to crack
nucleation and growth,
fiber-matrix delamination, fiber rupture, etc. In the present work, we
focus rather on the
global view that emerges by analyzing the acoustic emission times series
over the
whole lifetime of the pressure ramp up to rupture.

We analyse 7 acoustic emission data sets recorded during the pressure ramp
up to rupture
of 7 distinct composite pressure tanks listed  below in table 1 with some
of the characteristics
of the experiments.
The analysis was not performed on the raw data sets due to a number of
experimental factors such as unreliable measurements, limited resolution
as well as physical considerations. First, the data was truncated in the upper
and lower ends. The reason for the latter is that the recordings made
for very low pressures are irrelevant to the rupture process
 and are unreliable since the low level of acoustic emissions can easily be
confused
 with exterior noise. This truncation was made at $100$ bars so that all
recordings
 start at pressure larger or equal to this value.
 Some data sets did not have recordings
for such low pressures and was hence not truncated. The upper endpoint was
identified as the first point where the maximum pressure was recorded.
The data files were changed into files of binned data using a binning equal
to the resolution of the pressure measurement, specifically $1$ bar.
The maximum value of the acoustic emission for each data set has been
normalized to $1000$
in order to make the numerical treatment similar for all 7 experiments.


\subsection{Theory}

We have used three increasingly sophisticated mathematical formulas to model
the acoustic emission time series. The first one reads
\be
f(p)=A + B(p_c-p)^z     \label{eq1sa}
\ee
and has a priori 3 adjustable parameters. The important parameters are 
the critical value $p_c$ of the pressure at rupture
and the exponent $z$ (in general negative), which quantifies the
acceleration of
the acoustic emission rate. This is the fundamental representation of
rupture seen as
a critical point in the time-to-failure analysis.

The second formula
\be
f(p)=A + B(p_c-p)^z+C(p_c-p)^z\cos(\omega \ln(p_c-p)-\phi)   \label{logpera}
\ee
contains an additional term with relative weight $C/B$, describing a so-called
log-periodic correction to scaling \cite{Anifrani,logperiodicreport}.
Basically, this formula  means that the power law acceleration
is modulated by downs and ups organized as a geometrical series converging
to $p_c$.
In other words, the intermittent accelerations and quiescents of the
acoustic emissions
around the average power law acceleration are more and more closely spaced
as rupture is
approached. Mathematically, this log-periodic structures can
be represented as the real part of a correction to scaling of the form
$C (p_c-p)^{z+i\omega}$, {\it i.e.}, by a complex exponent. The imaginary part
$\omega$
has the meaning of a logarithmic angular frequency and defines the scaling
factor
$\lambda = \exp [\pi/\omega]$
of the geometrical series of alternating peaks and troughs. The phase
$\phi$ is
of no consequence as it accommodates the specific choice of the pressure unit:
$\omega \ln(p_c-p)-\phi = \omega \ln[(p_c-p)/p_0]$ by the definition
$\phi=-\omega \ln p_0$.
From a general view point, log-periodic oscillations are the hallmark of
a discrete hierarchical structure obeying a discrete scale invariance symmetry
\cite{logperiodicreport}.
Expression (\ref{logpera}) has been proposed previously on the basis of a
discrete
renormalization group approach to rupture \cite{Anifrani,logperiodicreport}.
Detailed theoretical and numerical analysis of ensemble of interacting cracks
have shown that such discrete hierarchy can self-organize from a cascade of
Mullins-Sekerka instabilities \cite{Huang}.

The third formula
\bea
f(p)=A &+& B(\mbox{tanh}((p_c - p)/\tau)))^z  \nonumber  \\
&+&C(\mbox{tanh}((p_c - p)/\tau)))^z
\cos(\omega\ln(\mbox{tanh}((p_c - p)/\tau)) - \phi)   \label{fmkakka}
\eea
adds a new ingredient and is obtained from (\ref{logpera}) by replacing
$p_c - p$ by $\mbox{tanh}((p_c - p)/\tau)$.
It is based on a parametric representation of the
numerical study of Sornette and Andersen \cite{Andersen2}, who found
clear evidence of scaling of the macroscopic elastic modulus and of the
elastic energy release rate as a function of
time-to-rupture in the thermal fuse model \cite{Vanneste} beyond the pure
critical power law regime: this allowed them to collapse
neatly the numerical simulations over more than five decades in time and
more than one decade
in disorder amplitude onto a single master curve that has the following
properties. It
is a pure power law like (\ref{eq1sa}) close to rupture (critical region);
far from rupture where only
few damage events occur (non-critical region), it relaxes exponentially to
a constant value. The
simplest functional form that captures these two regimes and interpolates
smoothly
between them is $\left[{\rm tanh}((p_c - p)/\tau))\right]^z$, which reduces to
(\ref{eq1sa}) for $p_c - p \ll \tau$ and goes exponentially to the constant
$1$ for
$p_c - p \gg \tau$. The characteristic pressure $\tau$ sets the cross-over scale
between the critical and non-critical regime. The analysis \cite{Andersen2} was
based on averages of several tens of independent samples. As shown in
\cite{canonical},
ensemble averaging destroy log-periodic oscillations due to the random
phase $\phi$ which can
vary from sample to sample. When studying specific realisations as
performed below,
these log-periodic structures have to be considered as potentially
important. This is
why we enrich the hyperbolic tangent formula with the log-periodic
corrections associated
with equation (\ref{logpera}), so that expression (\ref{fmkakka}) reduces to
(\ref{logpera}) for $p_c - p \ll \tau$.

For each acoustic emission time series, we have analysed both the energy
rate as well as
its cumulative. The rational for studying the cumulative acoustic
emission
as a function of applied pressure is that taking the cumulative is a
low-pass filter
that smoothen very significantly the noise and usually provides better
signals with higher
signal-over-noise ratio. For our purpose however, it has been shown to 
reduce significantly genuine log-periodic oscillations present in the 
original data \cite{huanglee}.
We thus find useful to perform the analysis of both the binned (energy rate)
and its cumulative, which present
complementary values. Since the cumulative data is a low-pass filtered
version of the
binned data, we first use equation (\ref{eq1sa}) to fit it. The cumulative data
will also be
presented in a non-parametric fashion in double logarithmic plots to show
direct visual
confirming evidence of the power law regime (\ref{eq1sa}) close to rupture.
The two other formulas (\ref{logpera}) and (\ref{fmkakka}) are applied to
the cumulative data where a better performence by eq. (3) would indicate
the presence of a transition from exponential to a power law increase in
the energy release rate as discussed above.

\section{Analysis of the energy release rate} \label{rate}

For the analysis of the energy release rate,
a second truncation was made for both the lower and upper ends of the
pressure interval. The
lower endpoint was defined as the point where the acceleration in the
cumulative energy release takes place. The size of this truncation varies
considerably from data set to data set. For the files containing data all
the way up to $p_c$ (data sets 1,2,3,4,6), the upper endpoint was simply
chosen as the point where the energy release rate had its maximum, thus
removing only a few points. For the two data sets (5,7), where the last point
is far from $p_c$, no truncation was performed.

\begin{figure}
\parbox[l]{0.45\textwidth}{
\epsfig{file=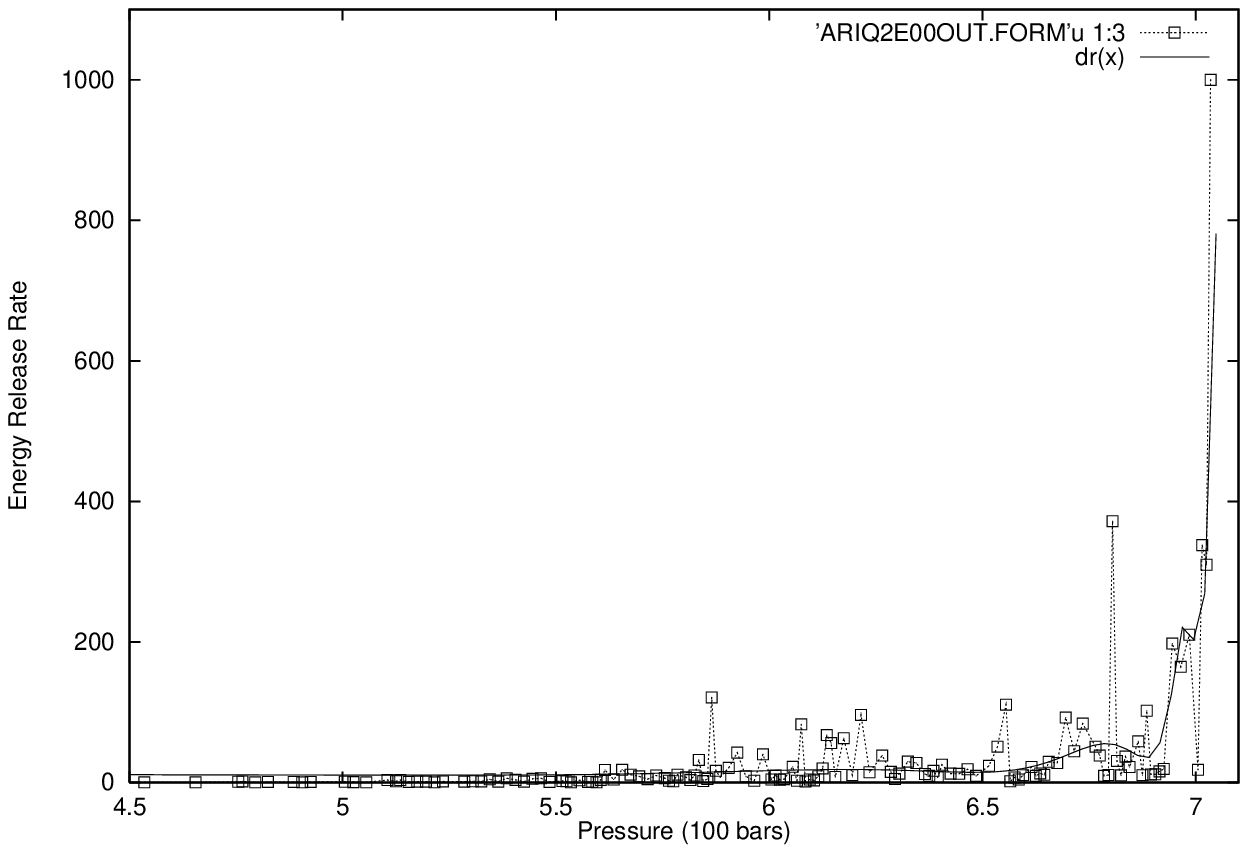,width=0.45\textwidth}}
\parbox[r]{0.45\textwidth}{
\epsfig{file=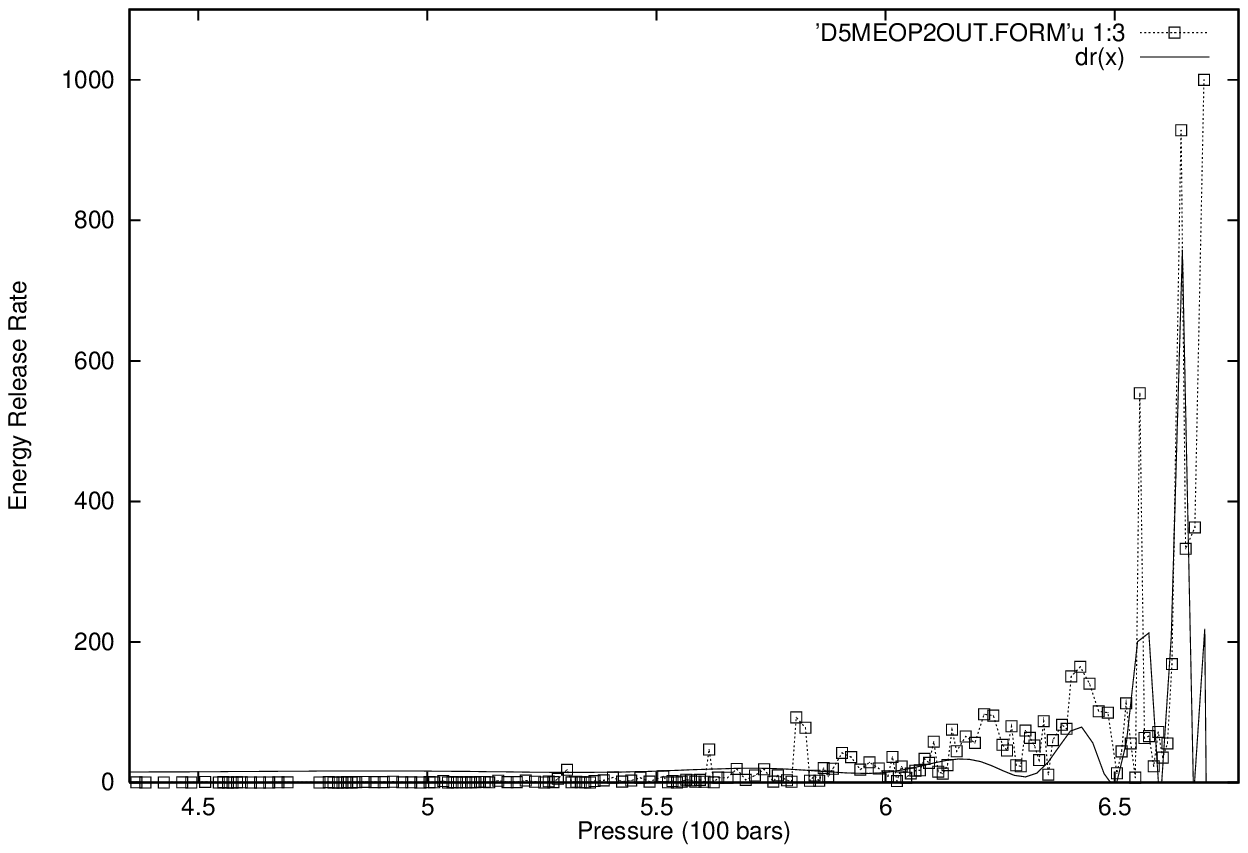,width=0.45\textwidth}}
\parbox[l]{0.45\textwidth}{
\epsfig{file=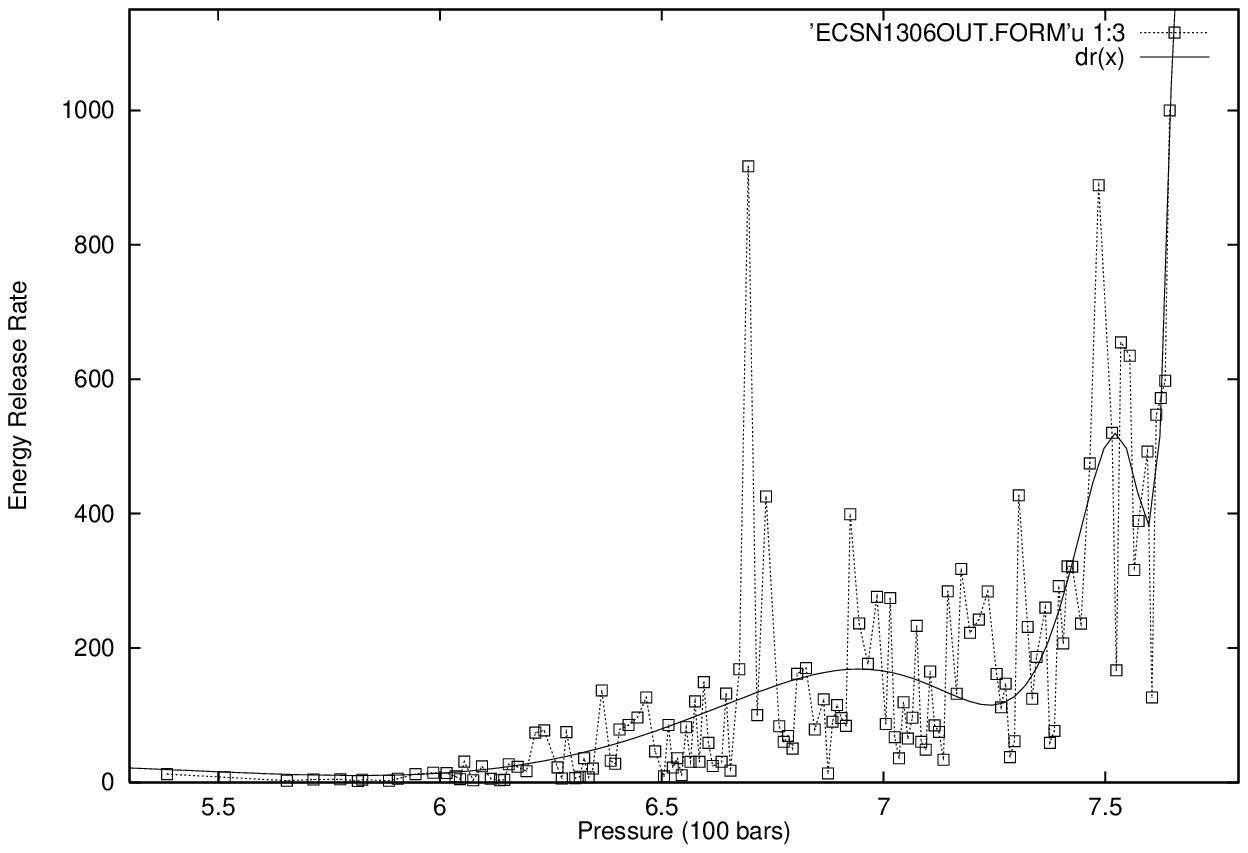,width=0.45\textwidth}}
\parbox[r]{0.45\textwidth}{
\epsfig{file=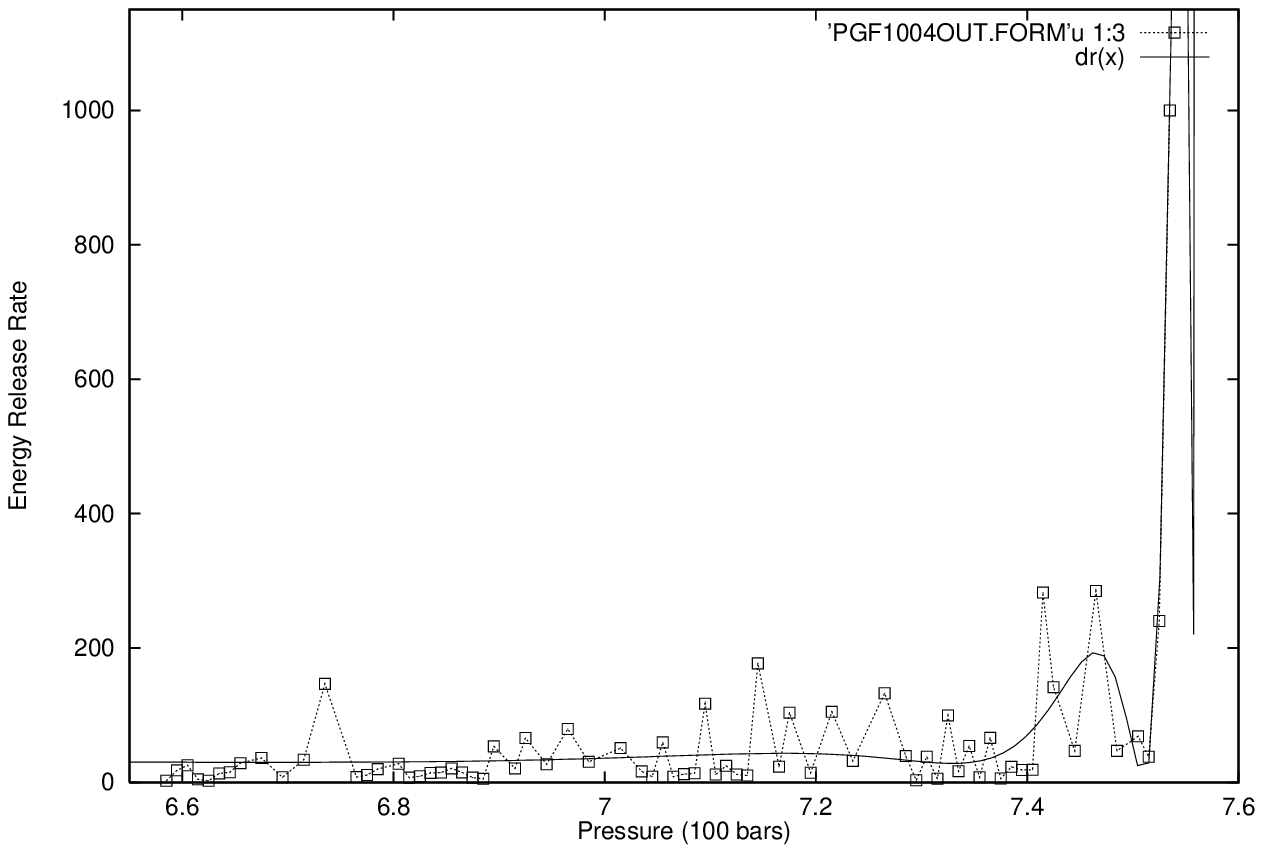,width=0.45\textwidth}}
\parbox[l]{0.45\textwidth}{
\epsfig{file=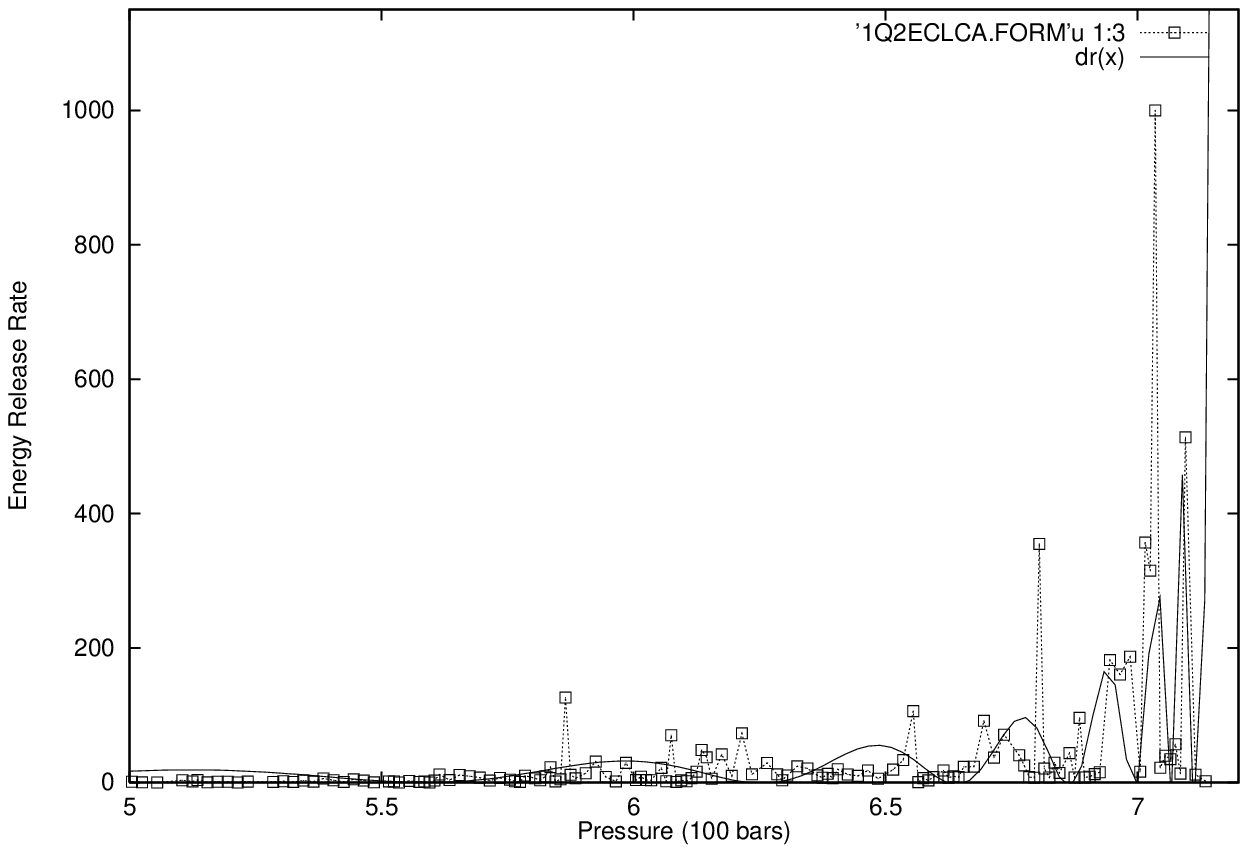,width=0.45\textwidth}}
\parbox[r]{0.45\textwidth}{
\epsfig{file=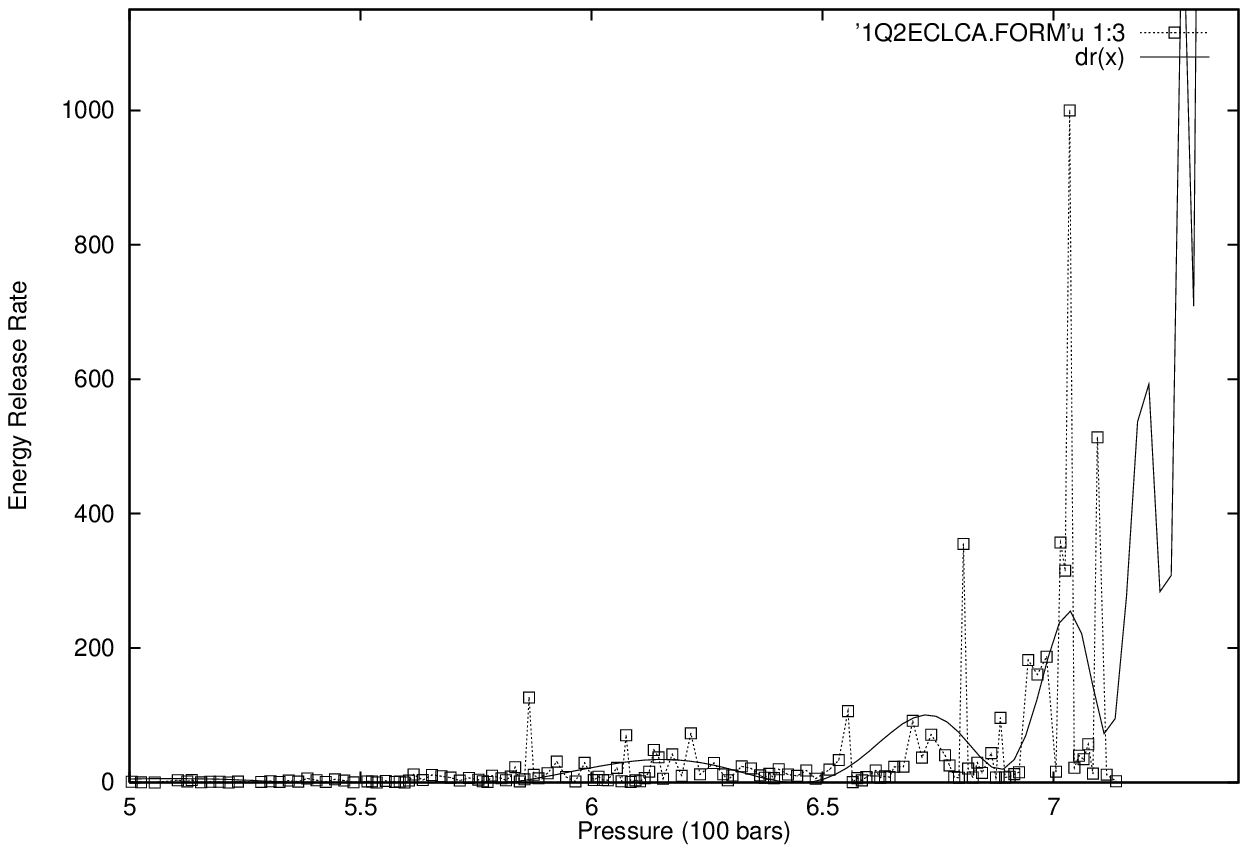,width=0.45\textwidth}}
\parbox[l]{0.45\textwidth}{
\epsfig{file=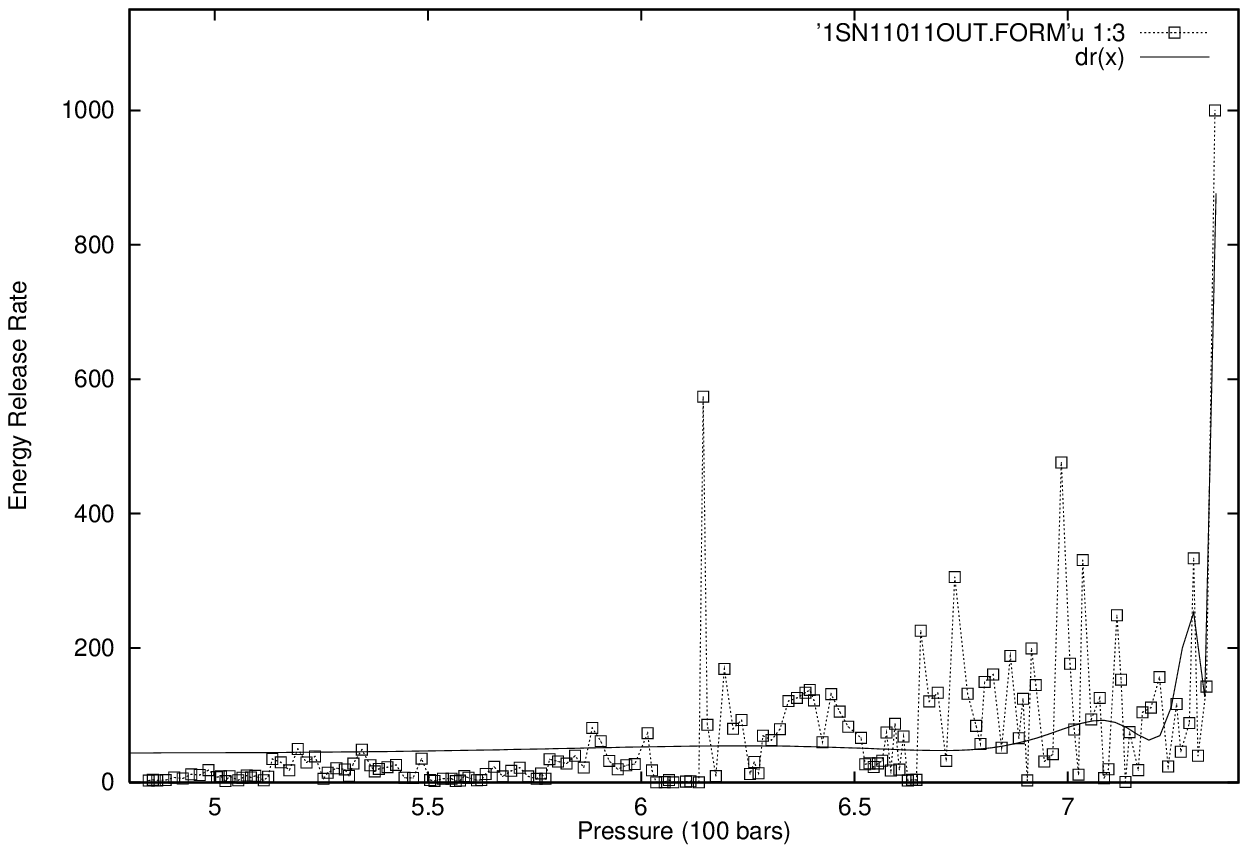,width=0.45\textwidth}}
\parbox[r]{0.45\textwidth}{
\hspace{15mm}
\epsfig{file=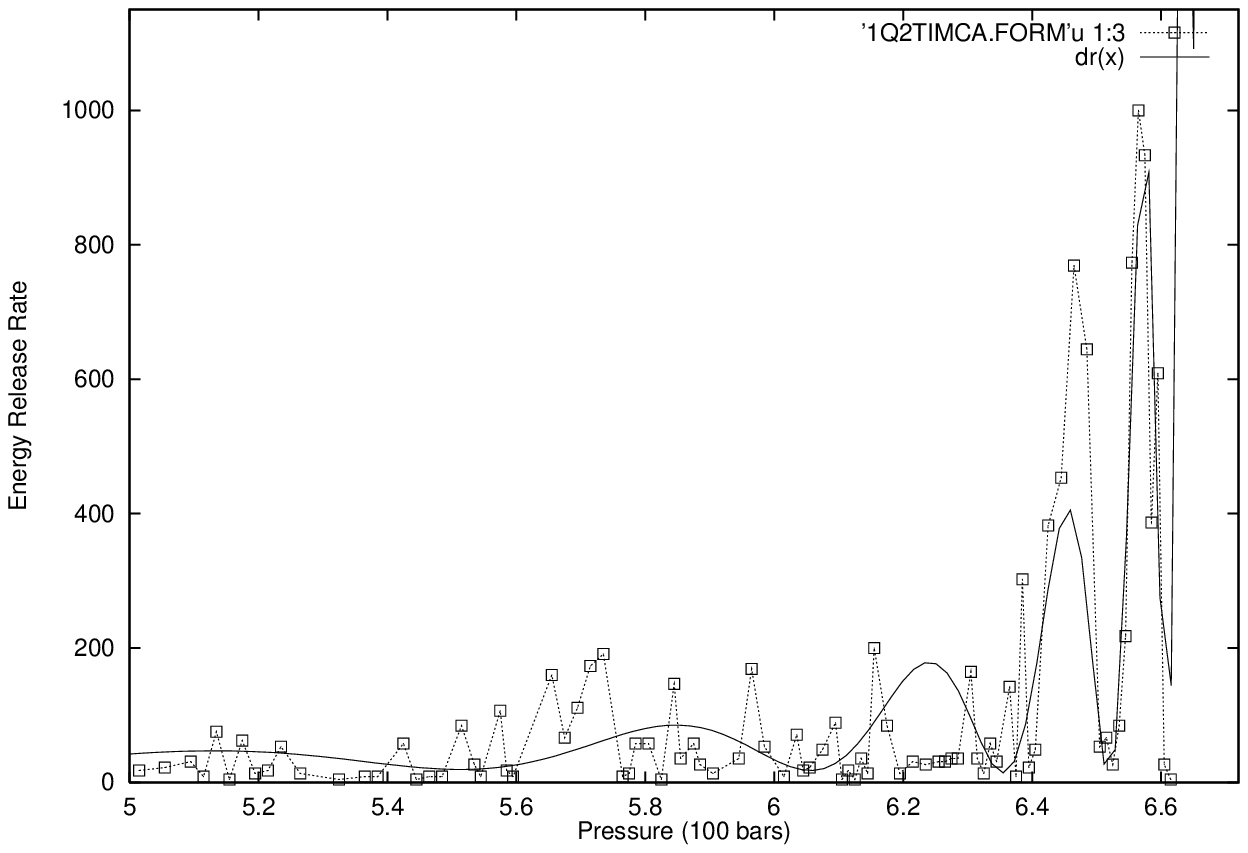,width=0.45\textwidth}}
\caption{\protect\label{anarate}Starting from the upper left corner, we
show the best
fit of the energy release rate with eq.~(\ref{logpera}) for data set 1,2,3,4,
5 (best and second best),6 and 7. Notice how the log-periodic oscillations
allow to
account for an accelerating rate of bursts on the approach of the rupture.}
\end{figure}

The results were encouraging for all data sets.
In table \ref{tabrate}, we see the values of the physical
parameters for the best fit of each data set with equation (\ref{logpera}). Use
of equation (\ref{eq1sa}) for the energy release rate is unreliable due to the
huge fluctuations shown in figure \ref{anarate}.
The fits were performed using the ``amoeba-search'' algorithm \cite{numerecipe}
minimizing the variance of the fit to the data. We stress that all
three linear variables $A$, $B$ and $C$ are slaved to the other nonlinear
variables by imposing the condition that, at a local minimum, the variance
has zero first
derivative with respect these variables. Hence, they should not be regarded as
free parameters, but are calculated solving three linear equations using
standard techniques including pivoting. Note, in addition, that the phase
$\phi$
in  (\ref{logpera}) is just a (pressure) unit and the coefficients $A$, $B$ and
$C$ have all dimensions of energy.
The key physical variables are thus $p_c$, $z$ and $\omega$.

The corresponding plots are shown in
figure \ref{anarate}.  The best fit is defined as the fit with the
lowest r.m.s. (root-mean-square) as well as reasonable values for $\omega$.
Specifically, this
means that we do not consider solutions with $\omega \stackrel{<}{\sim} 1$ and
$\omega \stackrel{>}{\sim} 14$. The reason is that too large values for
$\omega$ indicates noise-fitting. This means that for data set 5 and 7, the
minimum with the lowest r.m.s. was discarded because $\omega =25$ and
$\omega =17$, respectively. For data set 5, we list the two best minima.
Too small values for $\omega$ mean that the fit is not truly log-periodic
with less
than one oscillation. The corresponding log-periodic
correction to the pure power law is thus not valid.

Table \ref{tabrate} shows that the log-angular frequencies $\omega$ cluster
around two values  $\omega \approx 5$ or $\omega \approx 10$,
corresponding to a frequency multiplexing (doubling), as observed
also in diffusion-limited-aggregation
(frequency doubling)
\cite{DLA} and in 2D-freely decaying turbulence (frequency tripling)
\cite{2Dturbl}.
Another noticeable feature is
that the exponent is rather well defined at $z \approx -1.4 \pm 0.7$,
notwithstanding
the well-known difficulties in estimating critical exponents, especially in
such noisy data as analysed here. The values obtained for the critical 
pressure $p_c$
are all rather close to the last point in the data sets with the exception
of the second minimum for data set 5. This means that eq. (2) does
a good job of parameterising the data in a consistent manner for all seven
data sets. However, it does not provide the correct value for $p_c$ when
the last data point is far away from $p_c$ indicating that the energy
release rate might not be the best quantity to analyze in order to obtain
a predictive power. We will hence switch to the cumulative distribution in
order to investigate this aspect as well as further test the critical point
concept.

\section{Analysis of the cumulative energy release}

\subsection{Power laws}

Due to the noisy nature of data in general and especially the acoustic
emission data analyzed here, a power law fit is not always numerically stable.
The reason is the following: if the data exhibits rather large fluctuation in
the end part, the search algorithm used in the optimization process of the fit
will not necessarily find a local minimum for any choice of $p_c$ larger than
the last point $p_{last}$, driving the search towards a choice $p_c <
p_{last}$ and thus creating a numerical instability. Of the 7 data sets
considered here only 4 data sets, sets 1,2,5,7 achieved (a single)
numerically stable power law fit for the entire data interval (except the
part below 100 bars that has been omitted as previously mentioned) as shown in
figure \ref{anacumu1}.
Data sets 3 and 4 had to be
truncated in the lower end at a point where a change of regime could be
identified as a ``kink'' in the curve. Data set 6 could not be parameterised
by a power law due to a kink in the last part of the data.  In table
\ref{tabpow},
the values for the physical parameters $p_c$ and $z$ are listed.
We see  that only for data sets 3 and 4 the power law does a good job of
estimating $p_c$ whereas it overshoots for data sets 1 and 2. For data
sets 5 and 7 where the last point is far away from $p_c$, only in the first
case do we get a reasonable estimate of $p_c$.

Figure \ref{nonpara} provides
 a non-parametric visual test of the critical point concept. We show
the logarithm of the cumulative energy release as a function of the
logarithm in base $10$ of the distance $\lp p_c - p\rp /p_c$ to the critical 
rupture pressure $p_c$
determined from the fits shown in figure \ref{anacumu1}. While the power law
regime qualified by a straight line in this representation does not extend
to many
decades, the plots shown in figure \ref{nonpara} are nevertheless suggestive
of a critical point. The relative limited range of pressure-to-failure
$\lp p_c - p\rp /p_c$ defining the critical regime is defined by the 
resolution in pressure which is
here no better than one bar and thus limits the investigation of the
energy release rate closer to the rupture.

The over-all conclusion of the analysis presented here is that a pure power
law does a reasonable job of parameterising the cumulative data but does
not seem to
provide for a predictive tool: truncating the data in the upper
end will only increase the over-shooting. We hence have to move beyond pure
power laws in order to be able to use the cumulative energy release
for prediction purposes. It is not very surprising that a
pure power law fails to capture essential features of the data. Whether one
believes in log-periodic oscillations or not, the analysis presented in the
previous section clearly shows that the energy release rate is quite
intermittent. The low-pass filtering performed in calculating the cumulative
energy release have reduced these fluctuations to a large extent but clearly
not enough. Secondly, it is clear that a truncation of the data in the lower
end is necessary in order to identify the transition point to a power law
acceleration. In fact, the identification of this transition point between
random and cooperative behavior may very well be {\em the crux} when it comes
to predictability.

We will thus investigate whether eq.'s (2) and eq. (3) do a better job of 
parameterising the cumulative data. The rationale behind the last extension 
to eq. (3) is that we cannot hope for a prediction capability if we cannot 
to a reasonable extent capture the features of the full data set which
contains a transition in the acceleration of the energy release rate.

\subsection{Beyond Pure Power Laws}

\subsubsection{Fit with equation \ref{logpera}}

It is well-known that calculating the cumulative of some quantity effectively
corresponds to performing a low-pass filtering thus diminishing fluctuations
in the data. However, it will not completely remove them and thus a residue
of the oscillations found in section \ref{rate} should still be present in
the data. Furthermore, adding an extra degree of freedom in the equation to
be fitted will remove the problem with the numerical instability, since
by the very nature of the experiments the fluctuation around some average
behaviour will be slower for low pressures and more rapid for higher pressures.
Eq.'s (2) and (3) exactly takes such a behaviour into account. This means that
from a purely technical perspective, eq.'s (2) and (3) offers significant 
advantages to eq. (1).

\begin{figure}
\parbox[l]{0.45\textwidth}{
\epsfig{file=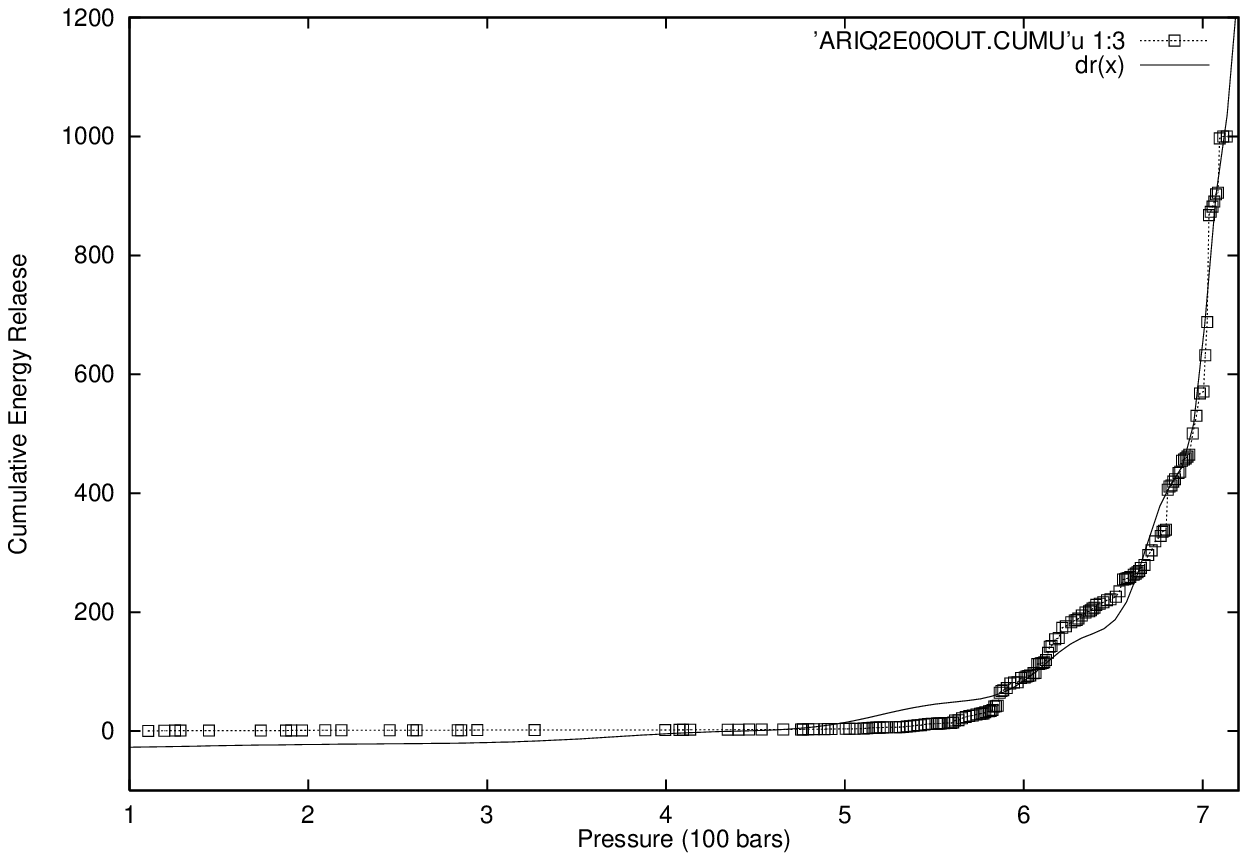,width=0.45\textwidth}}
\parbox[r]{0.45\textwidth}{
\epsfig{file=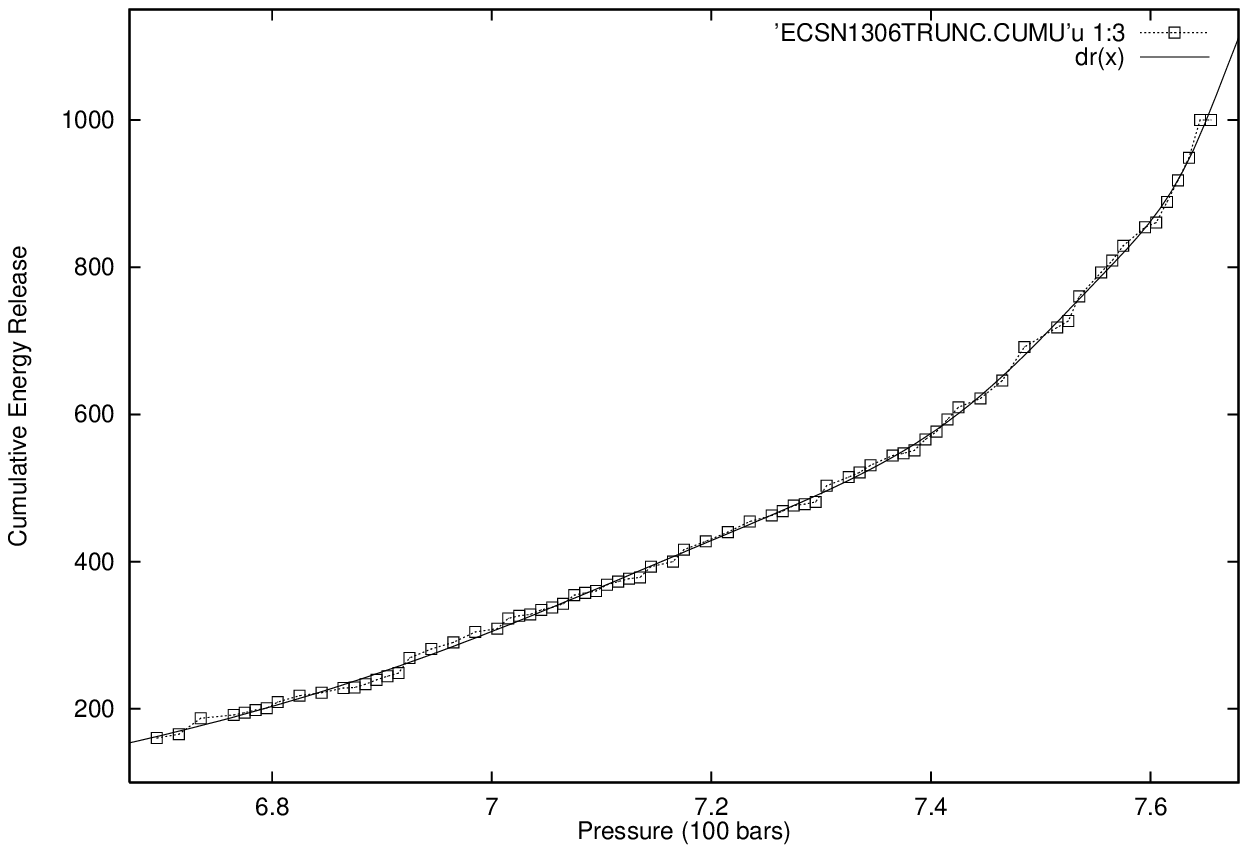,width=0.45\textwidth}}
\parbox[l]{0.45\textwidth}{
\epsfig{file=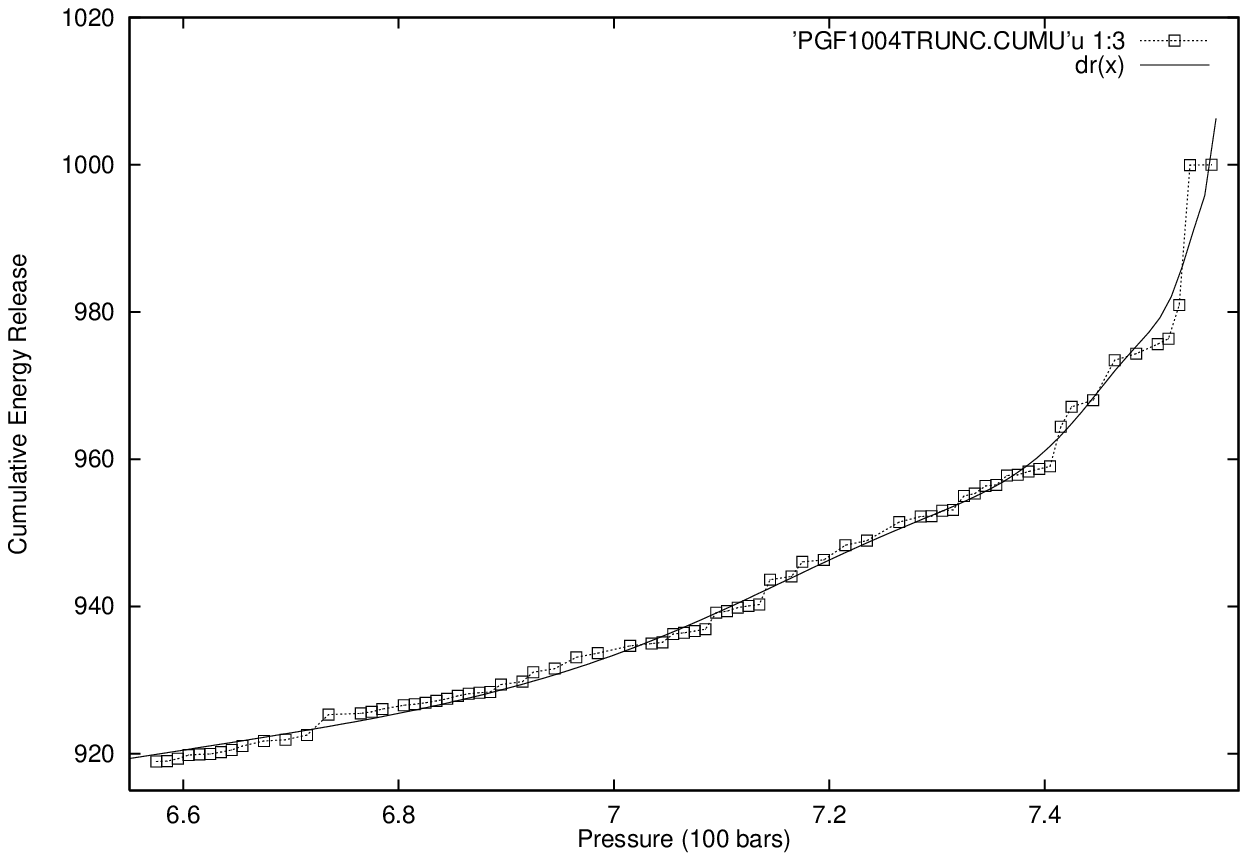,width=0.45\textwidth}}
\parbox[r]{0.45\textwidth}{
\epsfig{file=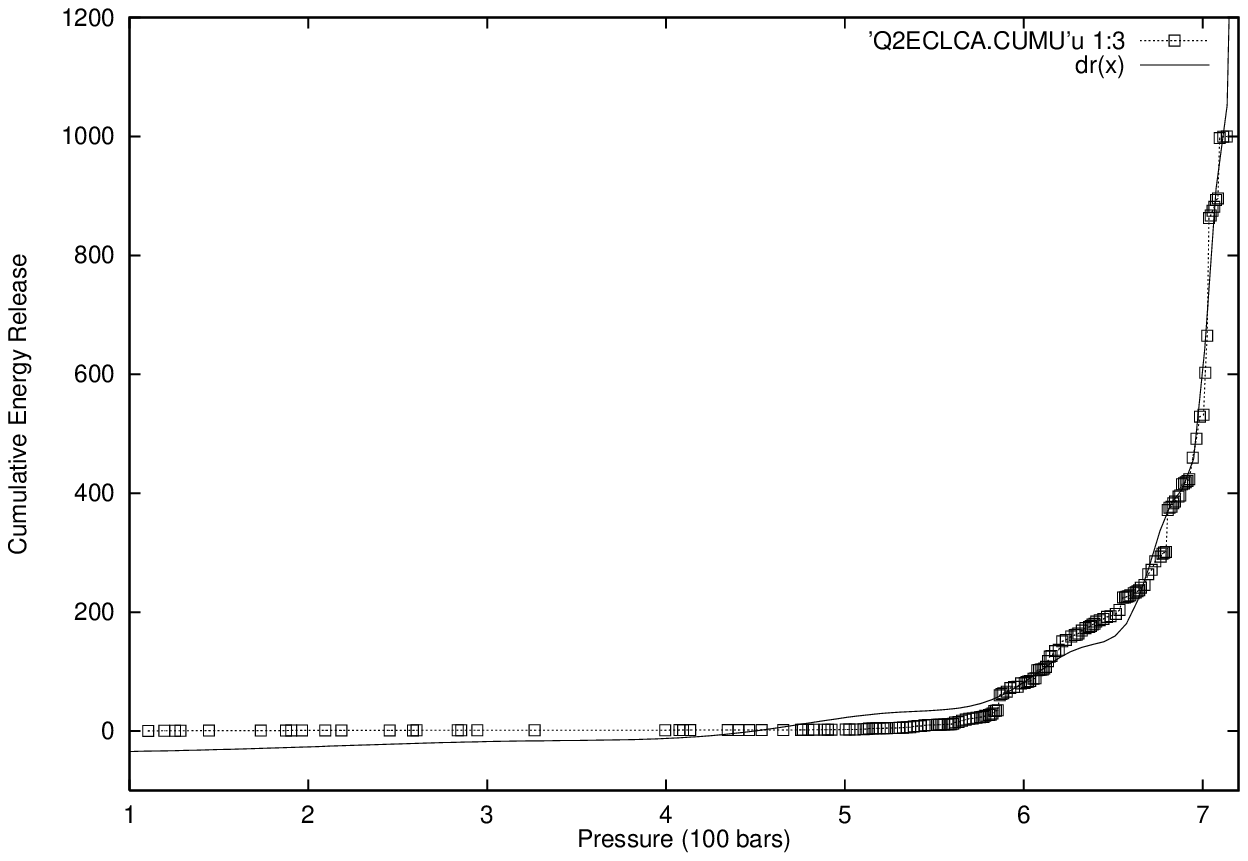,width=0.45\textwidth}}
\parbox[l]{0.45\textwidth}{
\epsfig{file=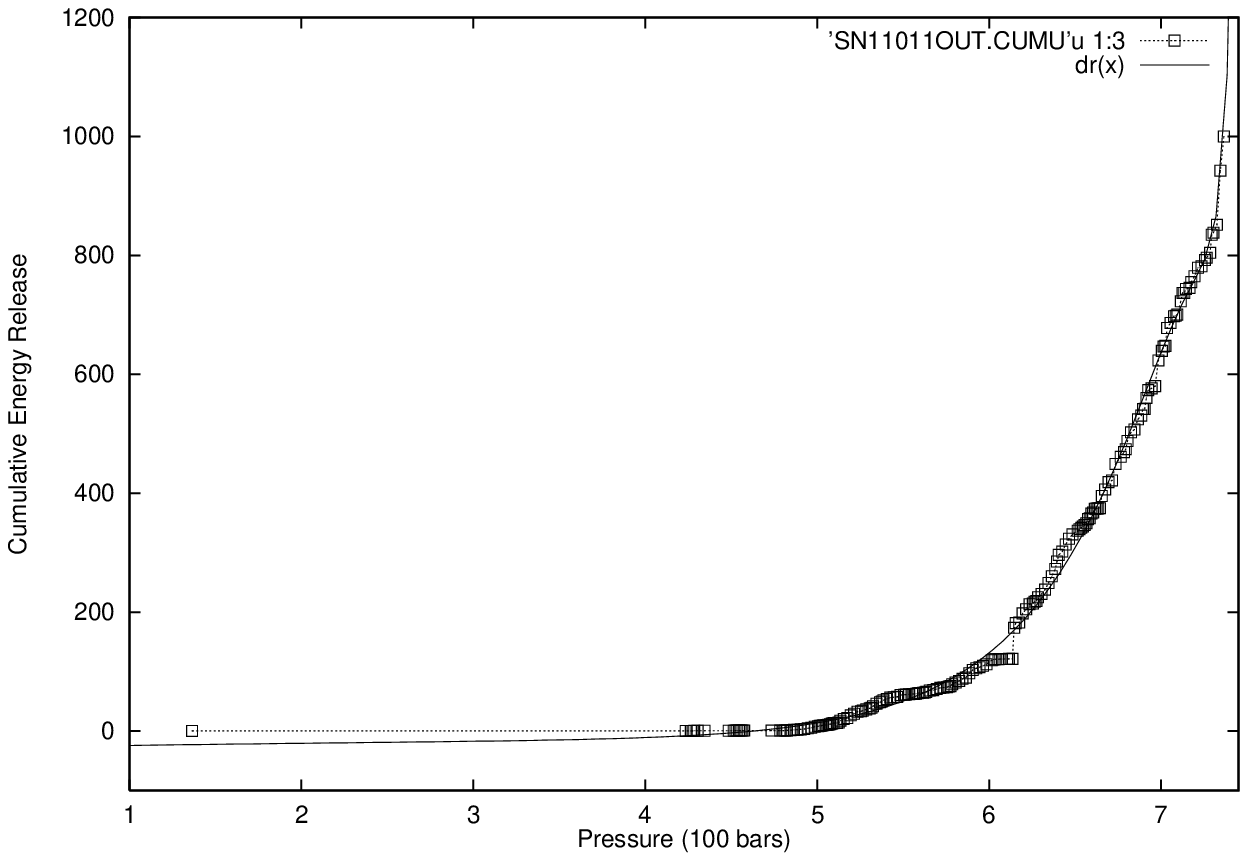,width=0.45\textwidth}}
\hspace{15mm}
\parbox[r]{0.45\textwidth}{
\epsfig{file=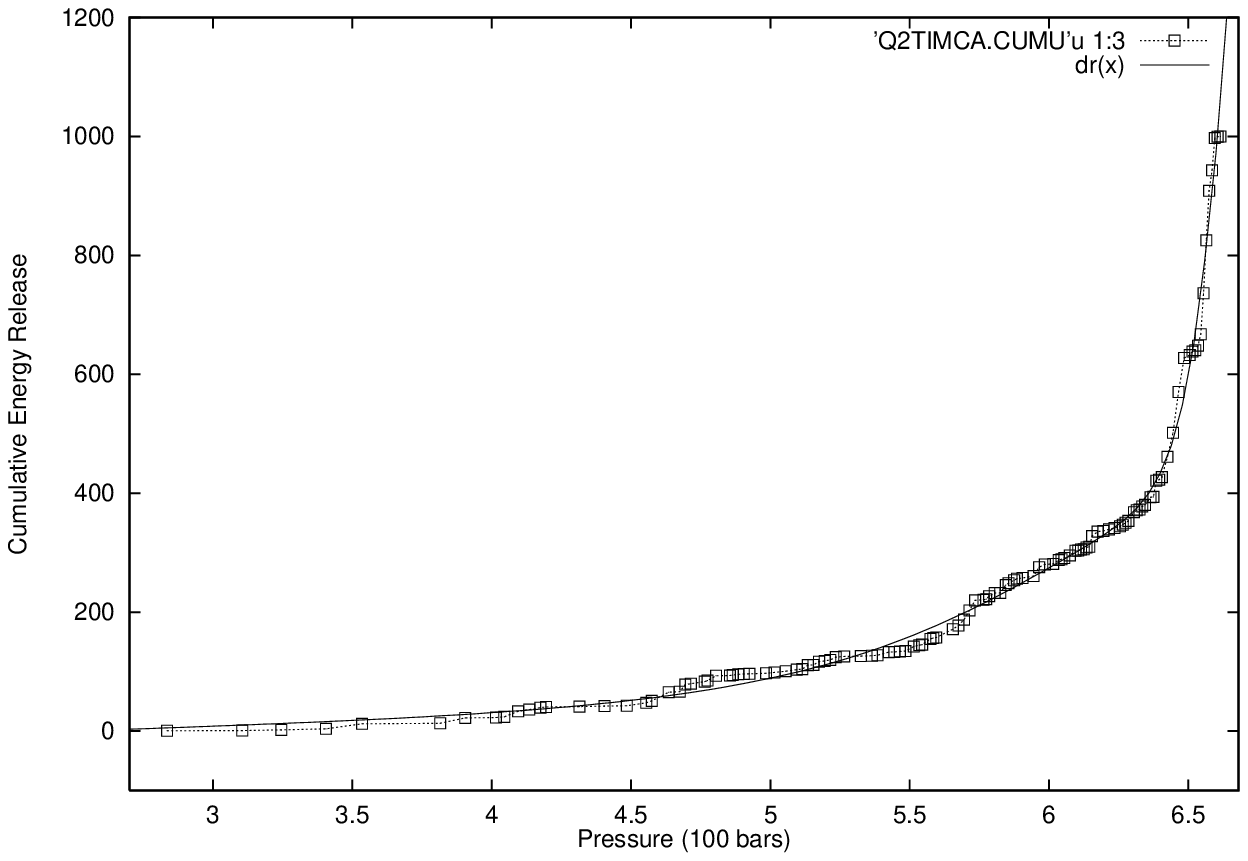,width=0.45\textwidth}}
\parbox[l]{0.45\textwidth}{
\epsfig{file=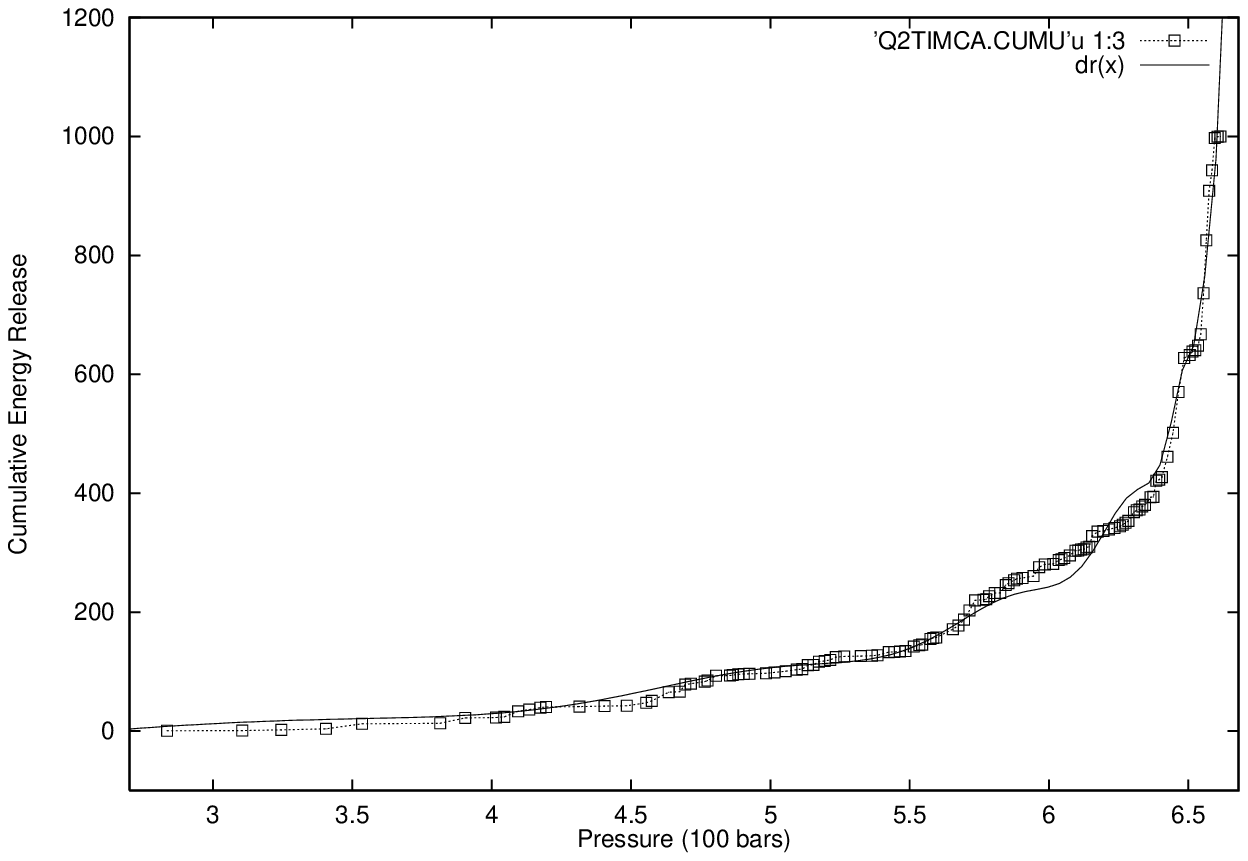,width=0.45\textwidth}}
\hspace{15mm}
\parbox[r]{0.45\textwidth}{}

\caption{\protect\label{anacumu2} Cumulative energy release fitted with
eq.~(2).
Starting from the upper left corner, we have data set 1,3,4,5,6,7 (first
minimum) and 7(second minimum). Data sets 3 and 4 has been truncated in the
lower end, taking as the first point the point where the acceleration in the
acoustic emission begins to be similar to the other data sets. }
\end{figure}

In figure \ref{anacumu2}, we see the fits of the 7 data sets with eq. (2).
Whereas the fits with eq. (1) in all cases only provided us with a single
fit (or in the case of data set 6 none), we now have several solutions per
time series. The fits shown in figure \ref{anacumu2} and the corresponding 
parameter values listed in table \ref{tablp} are those of the best fit 
which full-fills the criteria previously given. Data set 2 did not give any 
such fits, since $\omega \ll 1$ for all of them.

\subsubsection{Fit with equation \ref{fmkakka}}

As previously mentioned, a transition point exists where the acceleration
in the energy release rate increases significantly. Standard theoretical
arguments
from critical phenomena \cite{Goldenfeld} suggest that the
acceleration is approximately exponential before the transition point and
goes to a power law after the transition point up to the critical point, 
hence defining the so-called critical region. As a justification of eq.(3), 
we stress that such a cross-over has already been studied in detail in a 
numerical model of rupture \cite{Andersen2}. We thus propose that eq. (3) 
might provide a better fit of the data sets without the
need of truncation as was the case in section \ref{rate}. By using
eq.(3), we are introducing an additional parameter, the typical width $\tau$ of
the critical region, and a better fit is thus
expected. However, if we also get a better estimate of $p_c$ and a better
predictive power, we can argue that this transition in the acceleration is
indeed captured by eq. (3).

Figures \ref{anacumu3a} and \ref{anacumu3b} show the fits of the seven data
sets with
equation \ref{fmkakka}. Sets 1-4 and 6 have one acceptable solution while
two solutions
are given for data sets 5 and 7. There are more solutions but most can be
discarded
or aggregated. The reason for the large number of minima for data sets 3, 4
and 7 is
due to a degeneracy with respect to the new parameter $\tau$, when the
number of
data points is small (72, 73 and 119 respectively). Thus, in these data
sets, $\tau$ is not
constrained well. If one insists that two solutions are identical if
they have approximately the same values for $\omega$ and $z$,
then the number of minima are reduced to approximately 15, 11
and 5. For data set 5, the best solution is shown together with the
solution which had $\omega$ closest to $2\pi$, a value that has been found
repeatedly
in previous works \cite{Huang,otherlogperiodic,DLA}
and argued to be close to the universal mean field value
\cite{logperiodicreport}. For data set 7, the best
solution is shown together with the only solution which did not have $p_c$
close to $p_{last}$.

Comparing the results presented in table \ref{tablp} and \ref{tabtanh},
a major improvement is obtained for data sets 1 and 2 by using eq. (3)
compared to eq. (2). For data sets 3, 5 and 7, the improvement is minor while,
for data sets 4 and 6, we get the same solutions.
Hence, the overall conclusion is that eq. (3) better captures
important features in the data and supports the idea of a transition in the
acceleration of the energy release rate from exponential to power law.

\section{Prediction of the critical pressure of rupture}

Armed with these empirical tests of the concept of critical rupture,
prediction should in principle be possible by extrapolation of
the acoustic emission data using the theoretical formulas. This scheme is
similar to that proposed by Voight to describe and predict rate-dependent
material failure \cite{Voight}, based on the use of an empirical power law
relation
supposed to be followed by an observable variable. However, Voight's procedure
is impractical due to the narrowness of the domain of validity of the pure
power law, preventing a realistic implementation of the prediction. This
will be
confirmed by our tests presented below with equation (1). In order to extend
the domain of validity beyond the rather narrow critical region, we are
going to
test for the predictive merits of eq.'s (2) and (3).

The  question we now address is whether we can use eq.'s (1), (2) and (3)
in order to predict from the value of $p_c$ the approximate value of the
pressure at rupture. Data sets 5 and 7 being incomplete, {\it i.e.}, having 
acoustic emissions recorded up to a value $p_{\rm last}$ far
from the pressure at rupture ($11\%$ below $p_c$ for data set 5 and
 $17\%$ below $p_c$ for data set 7), this prediction question has
already been answered to some extent. For data set 5, eq. (1) gave just 
as good result as eq. (3), the result with eq. (2) being slightly worse. 
Choosing the second solution for eq. (3), the estimate of
$p_c$ in these three cases was obtained with an error of
 $6\%$, $9 \%$ and $7 \%$, respectively. For
data set 7, eq. (1) and (2) only gave solutions with $p_c \approx 
p_{\rm last}$, the usual signature that the prediction is not reliable 
because it is controlled by
the very last acceleration. In contrast, eq. (3) produces a better
 solution with an error of $12\%$ which can be compared with
the difference of $17\%$ between  $p_c$ and $p_{\rm last}$.

Since the results for data set 7 using the entire data set are not very 
good (presumably because $p_{last}$ is so far away from $p_c$), it does 
not seem reasonable to include this data set in in a prediction scheme. 
Of the remaining data sets, the first prediction attempt for data set 1 
was made for $p \approx 676$, for data set 2 it was $629$, for data set 3 
it was $742$, for data set 4 it was $733$ and for data set 6 (which had 
no solutions for eq. (1)) it was $688$. These truncations were based on 
purely numerical 
consideration, {\it i.e.}, how many data points can one afford to remove 
without severely increasing the degeneracy of the cost-function used in 
the optimization of the fit. As a consequence, a maximum of 30 points were 
removed from the larger data sets ($\approx 170$ points for data sets 1,2,5 
and 6) and 20 points from the smaller data sets ($\approx 70$ for data sets 
3 and 4).

As we can see from table \ref{powpred}, the prediction performance of eq (1) 
is quite bad and only data set 5 gives something interesting. 

In tables \ref{lppred1} to \ref{lppred6}, we see the corresponding results
using eq. (2). Again the the prediction performance is not good. 

In tables \ref{tanhpred1} to \ref{tanhpred6} we see the corresponding results
using eq. (3). When two solutions are given, the first is the best fit and
the second is the best fit with $\omega$ closest to $2\pi$. If two fits have
$\omega$'s approximately at the same distance from $2\pi$ then both are listed.
If only one fit is listed, then $\omega$ of this fit was also closest to
$2\pi$. We also demand that $p_c$ is not very close to $p_{last}$. The reason 
for including these additional fits is to illustrate whether one always get a 
solution with a $p_c$ close to the true $p_c$ or not.

As mentioned, numerical degeneracy of the cost-function used in the 
optimization of the fit with eq.'s (2) and (3) can be a problem when 
the number of data points is not large. Hence, we have recorded the 
predicted $p_c$ as a function of $p_{last}$ for all fits with eq.'s (2) 
and (3) obeying the constraints on $\omega$ previously mentioned. The results 
are shown in figures \ref{lppred} and \ref{tanhpred}. Whereas no pattern
can be identified using eq. (2) we do see a clustering around the true
$t_c$ using eq. (3) for all data sets except 6.

\section{Conclusion}

\begin{figure}
\parbox[l]{0.45\textwidth}{
\epsfig{file=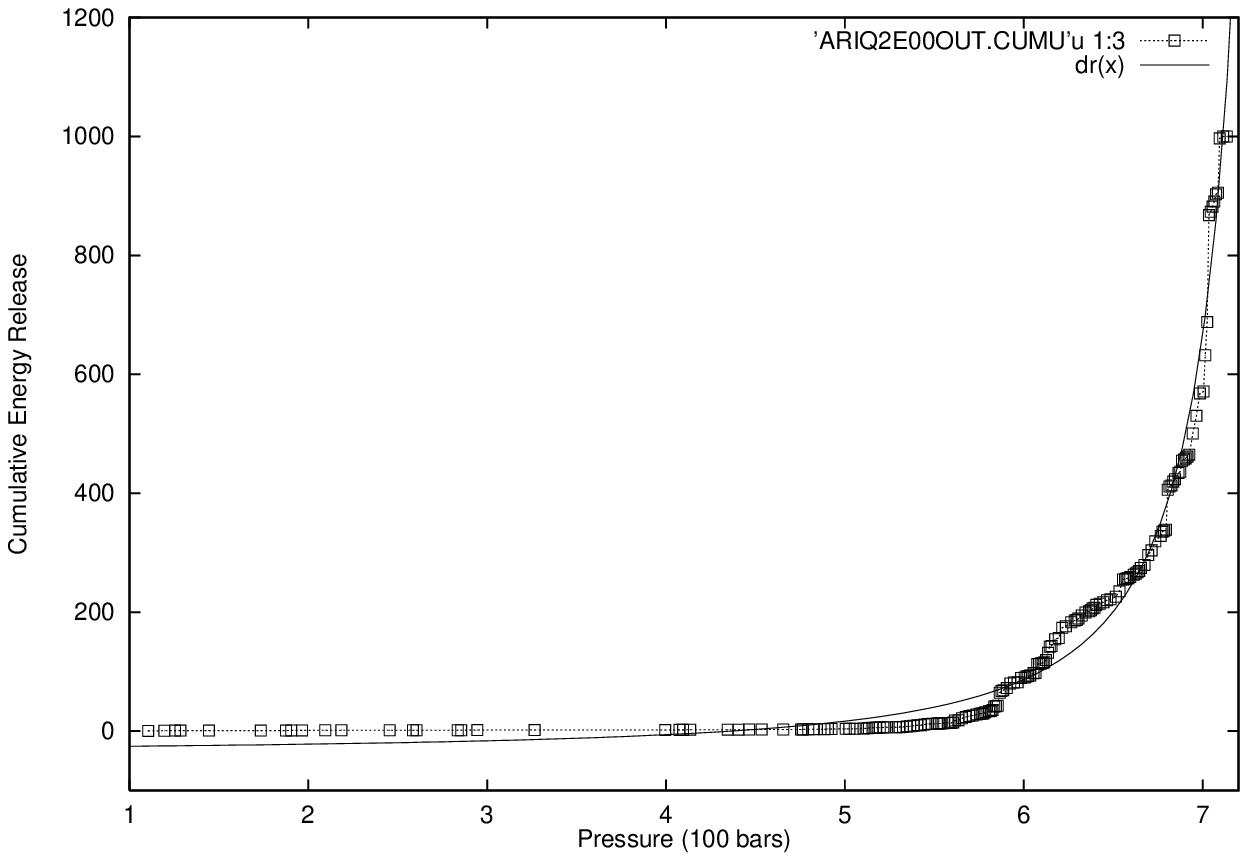,width=0.45\textwidth}}
\parbox[r]{0.45\textwidth}{
\epsfig{file=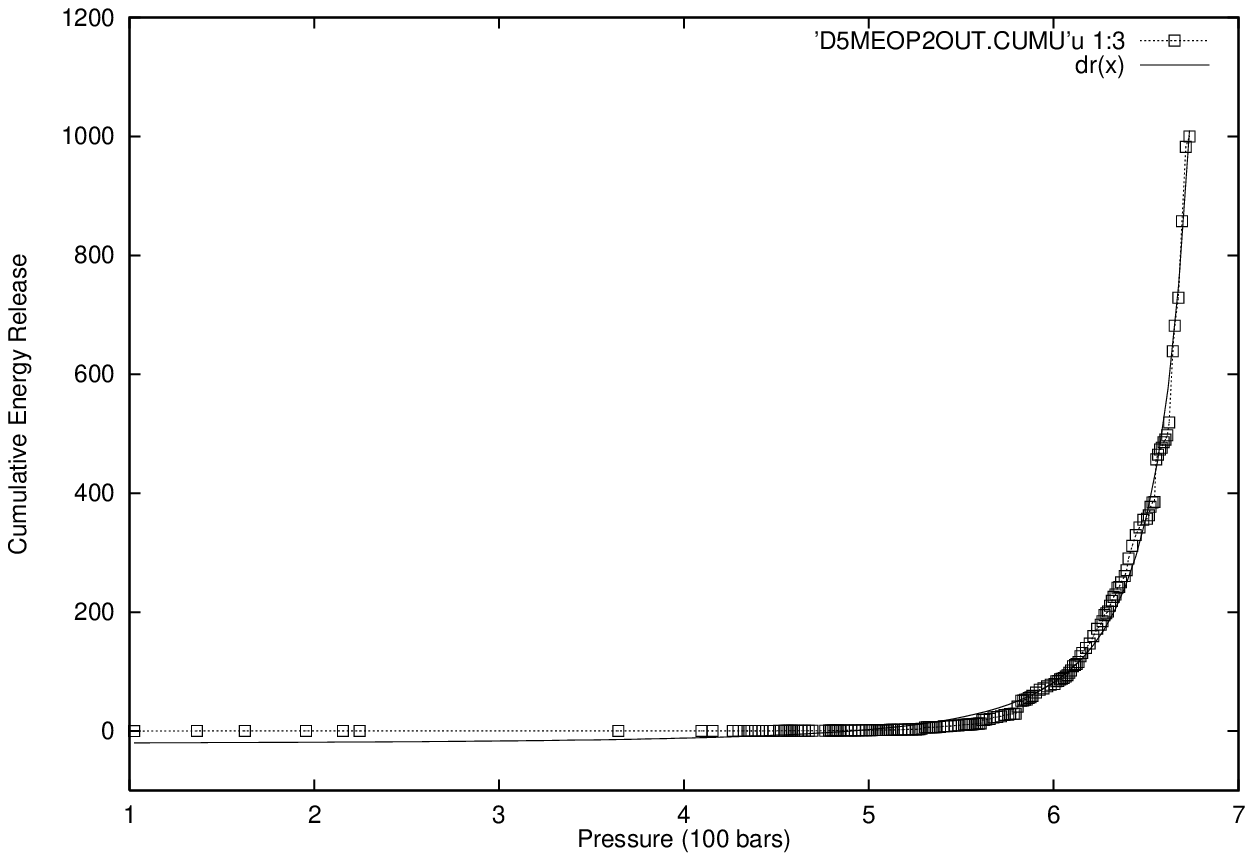,width=0.45\textwidth}}
\parbox[l]{0.45\textwidth}{
\epsfig{file=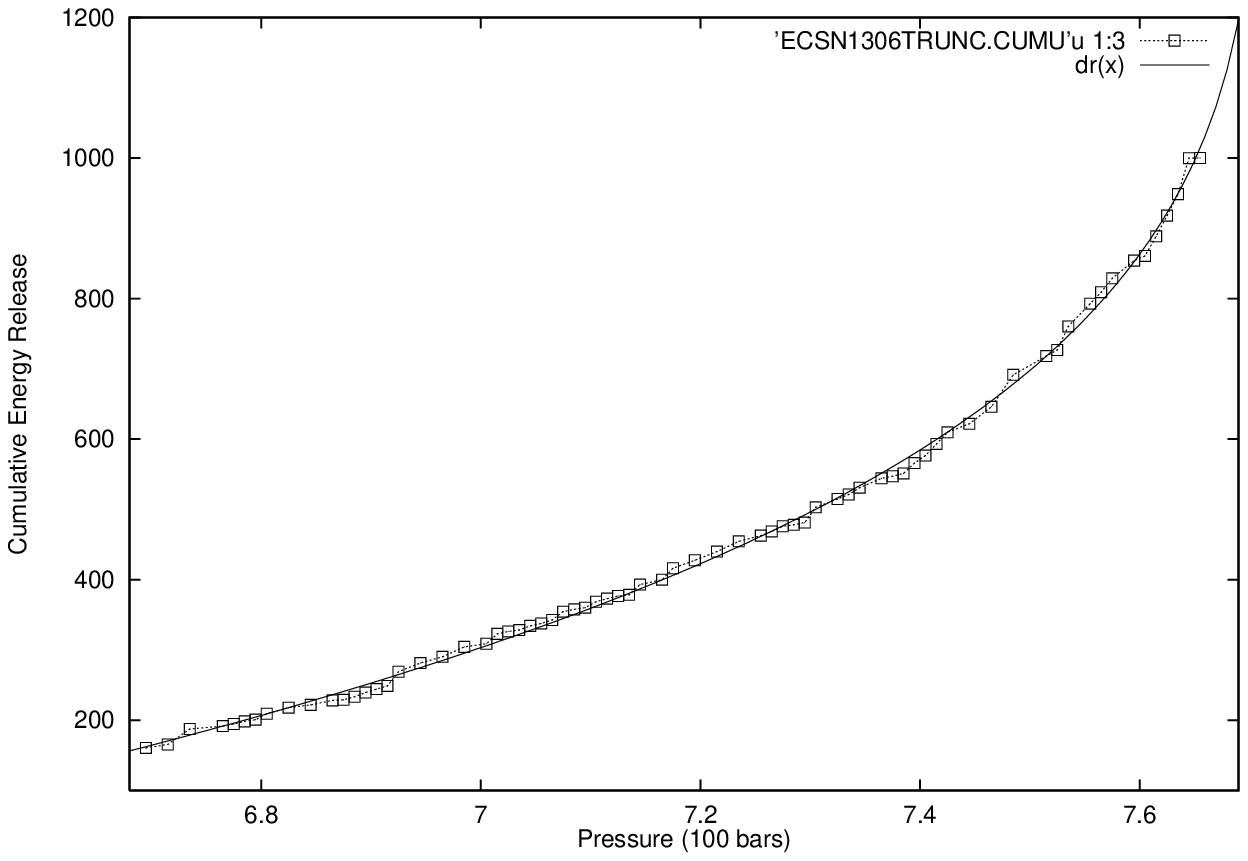,width=0.45\textwidth}}
\parbox[r]{0.45\textwidth}{
\epsfig{file=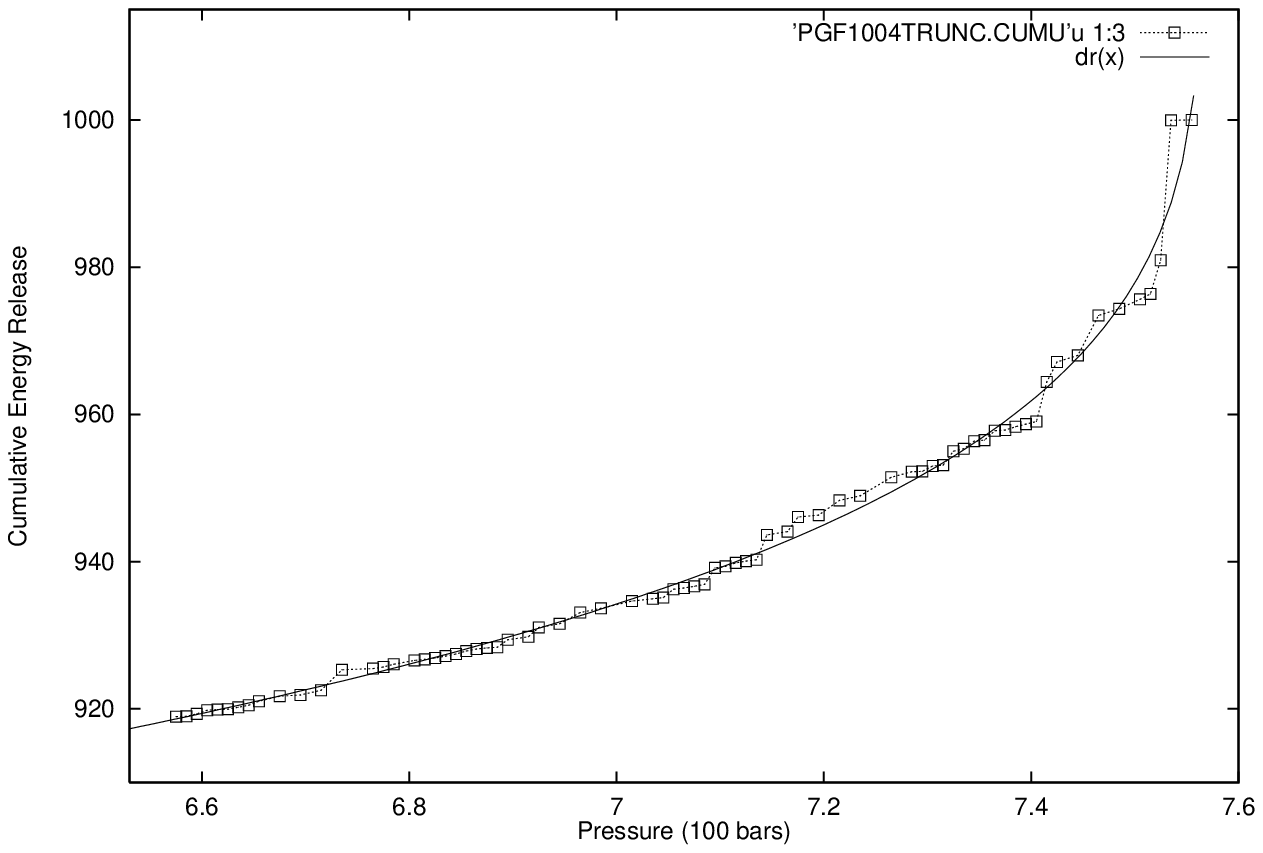,width=0.45\textwidth}}
\parbox[l]{0.45\textwidth}{
\epsfig{file=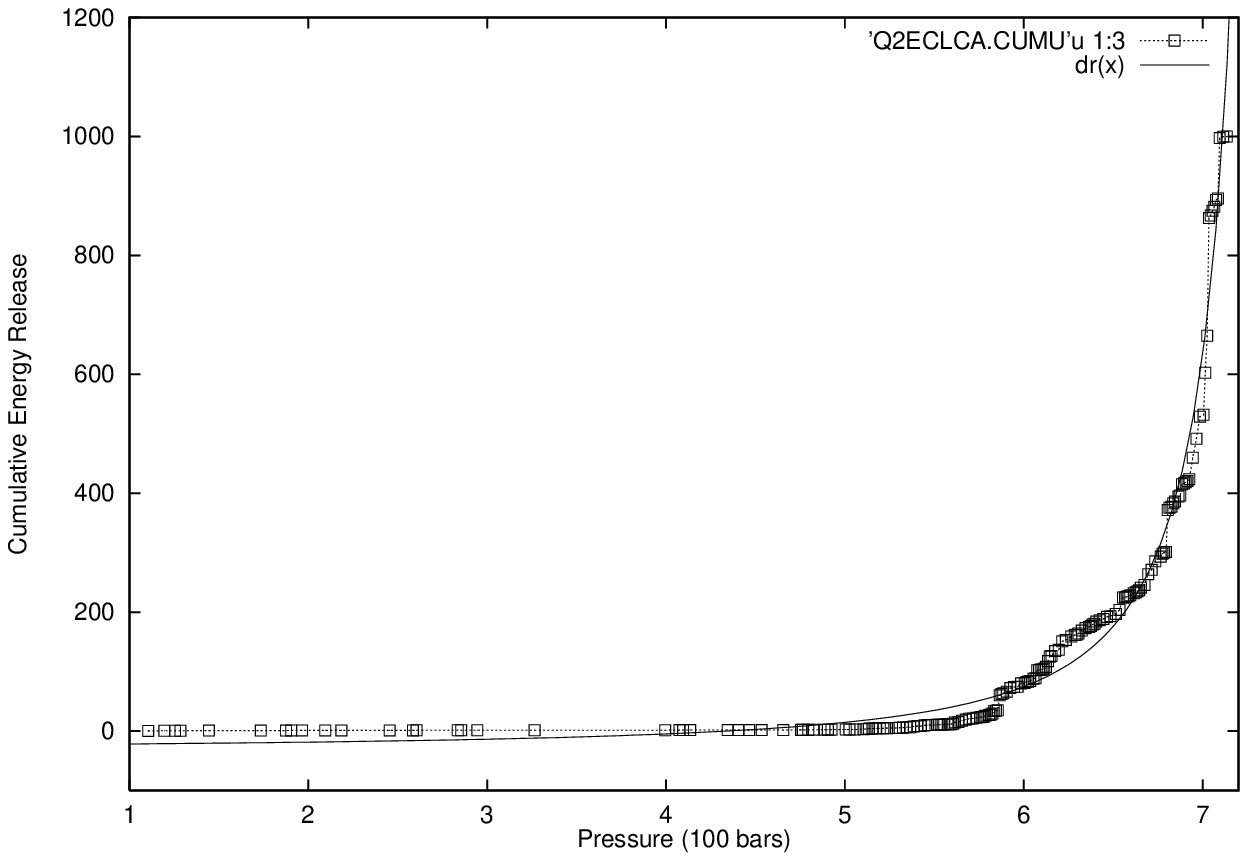,width=0.45\textwidth}}
\hspace{15mm}
\parbox[r]{0.45\textwidth}{
\epsfig{file=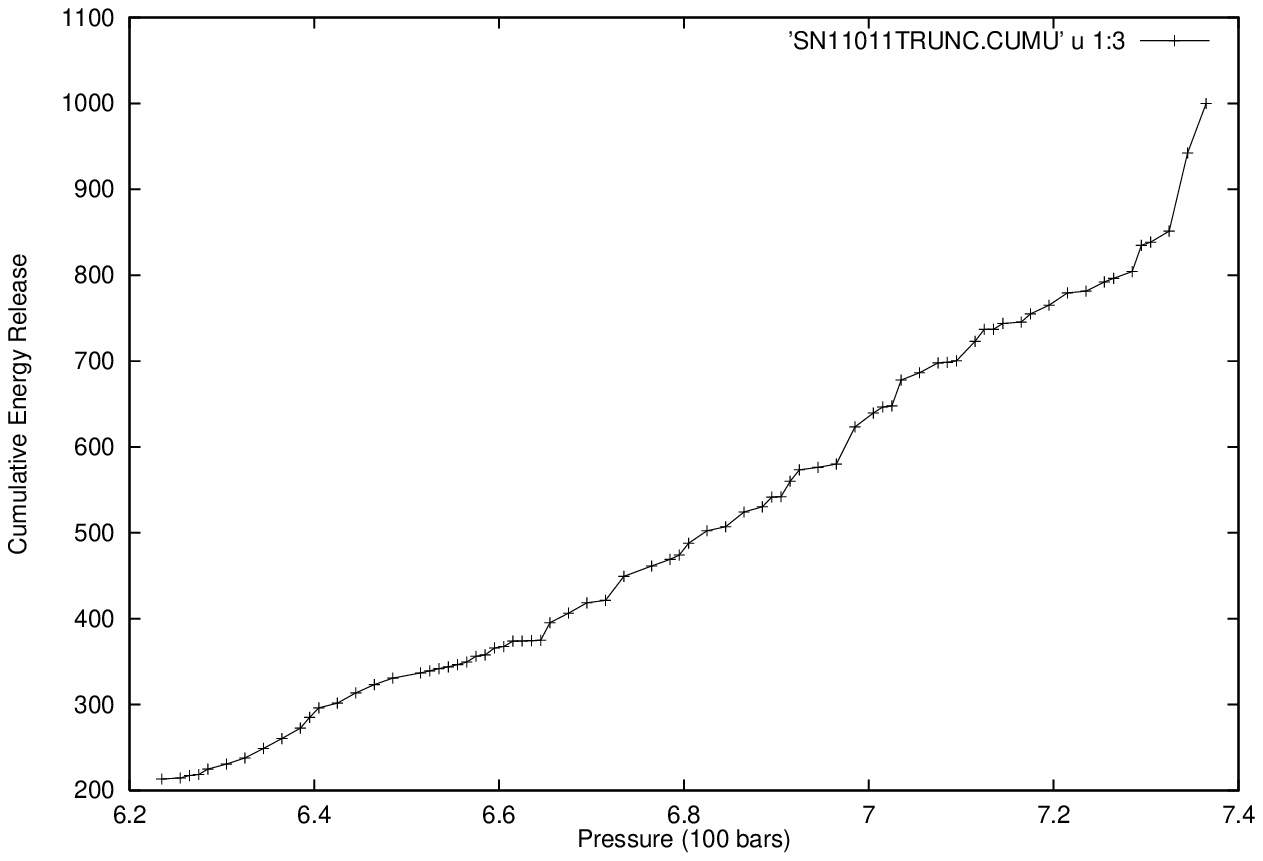,width=0.45\textwidth}}
\parbox[l]{0.45\textwidth}{
\epsfig{file=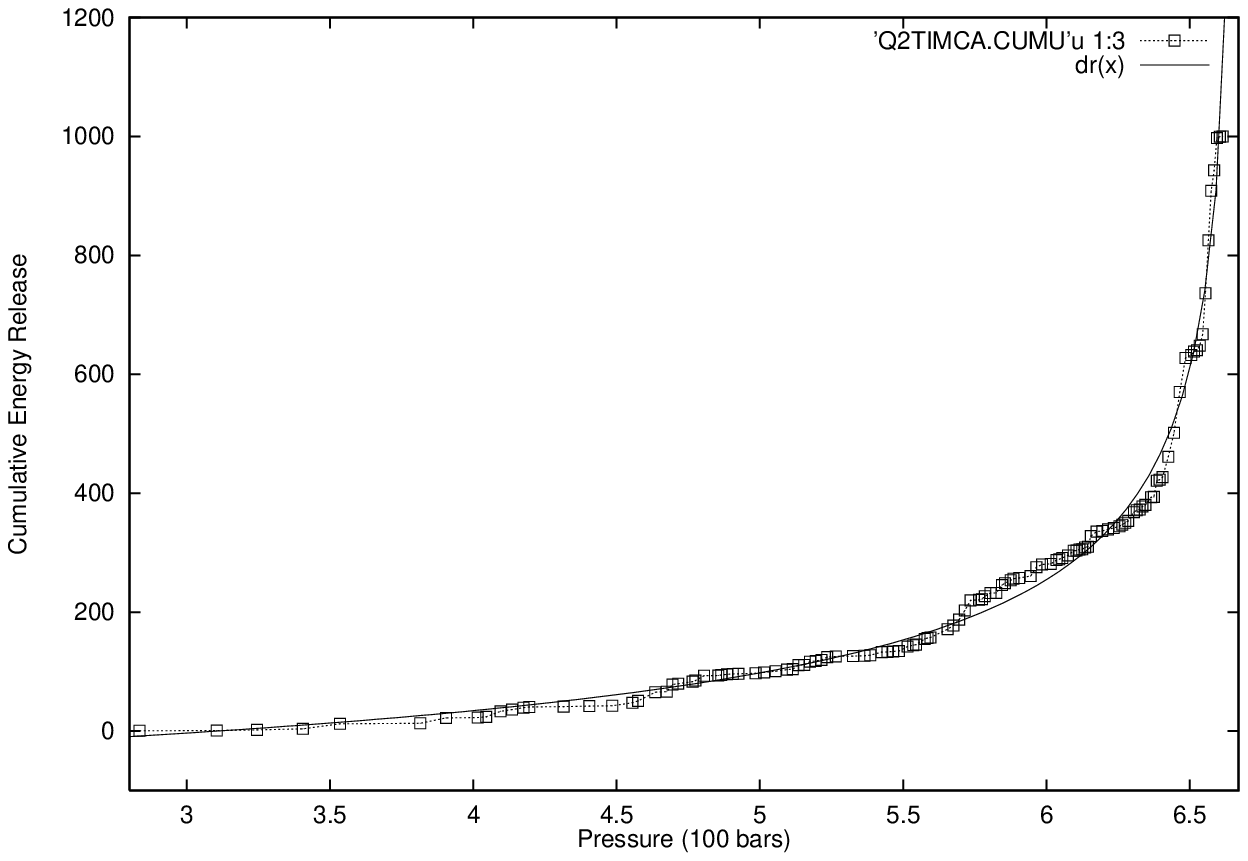,width=0.45\textwidth}}
\hspace{15mm}
\parbox[r]{0.45\textwidth}{}
\caption{\protect\label{anacumu1} Cumulative energy release fitted with eq.
(1). Starting from the upper left corner, we have data set 1,2,3, ...7.
Data sets
3 and 4 has been truncated in the lower end, taking as the first point the
point where the acceleration in the acoustic emission begins in a similar
way as for the other data sets. }
\end{figure}

\begin{figure}
\parbox[l]{0.45\textwidth}{
\epsfig{file=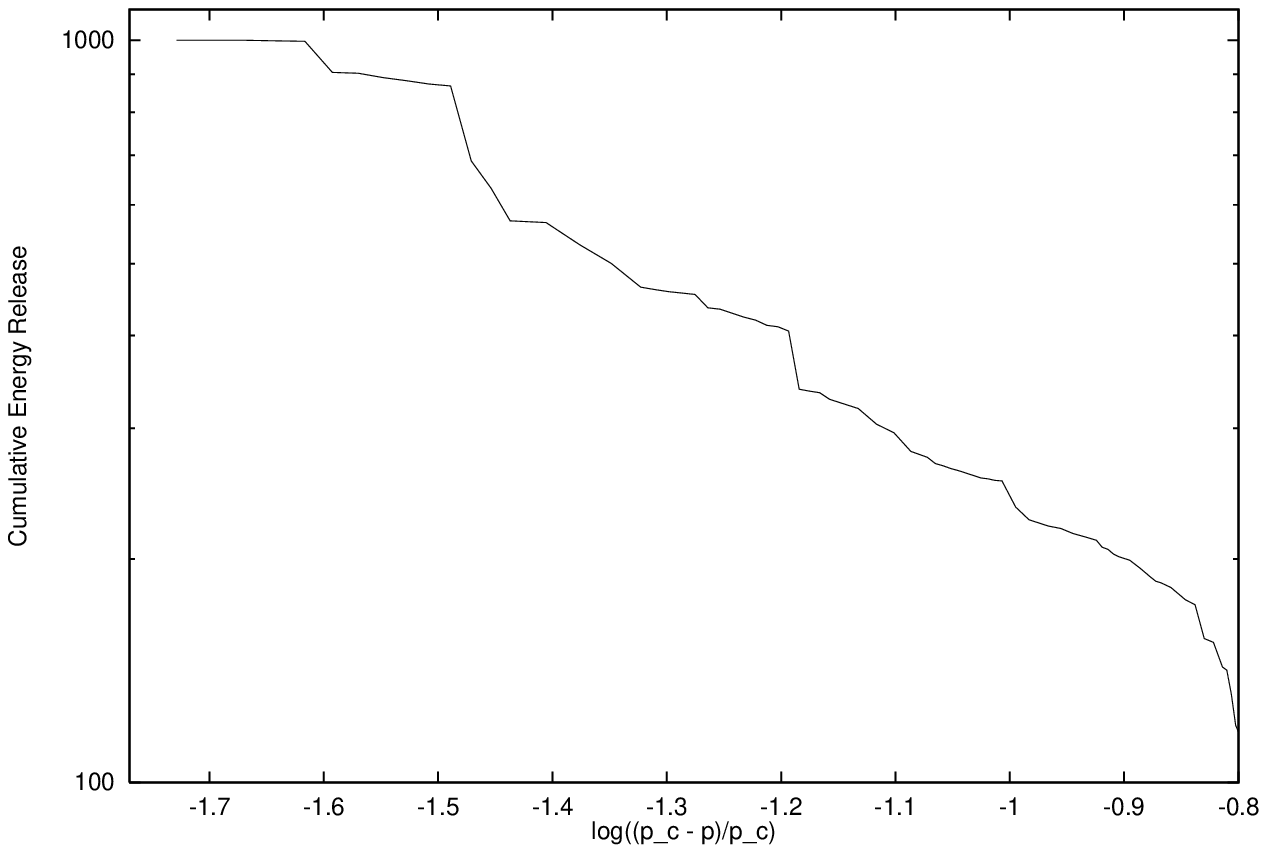,width=0.45\textwidth}}
\parbox[r]{0.45\textwidth}{
\epsfig{file=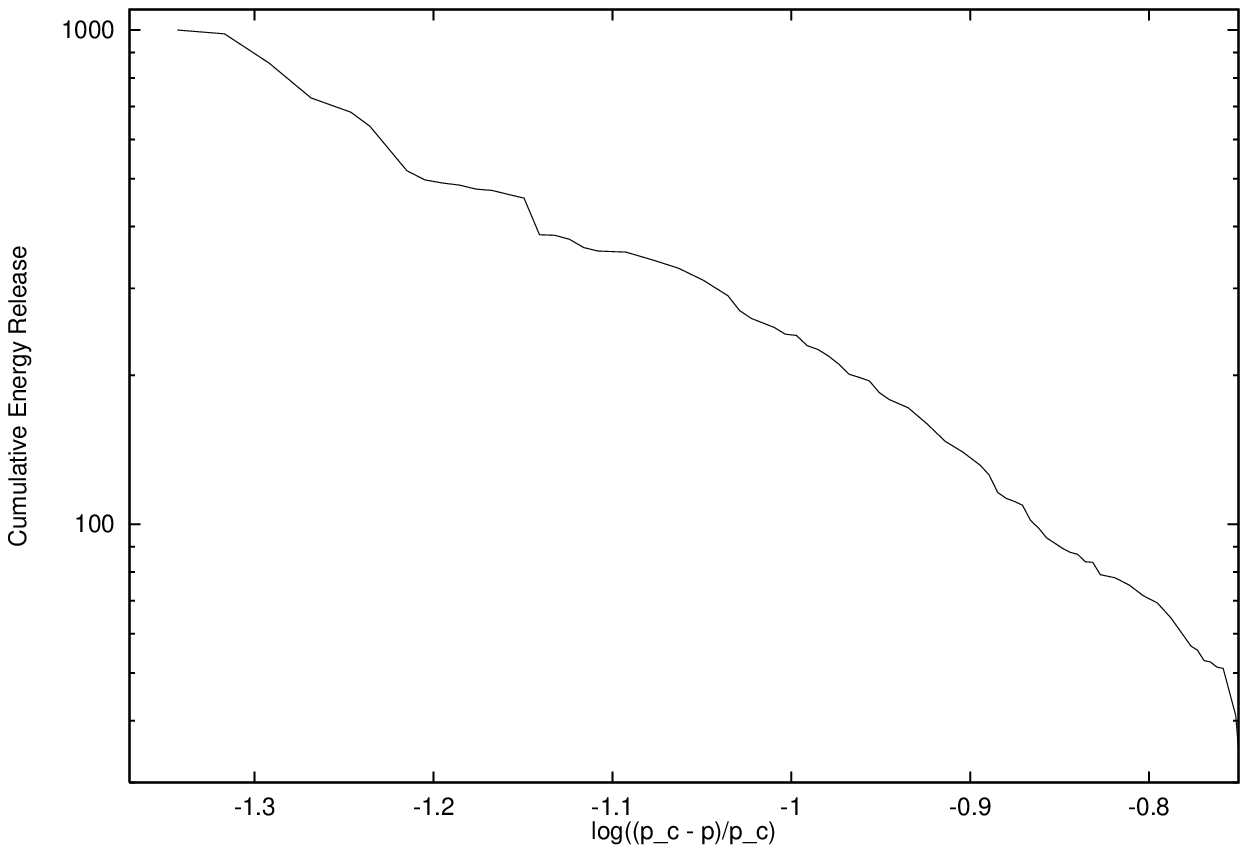,width=0.45\textwidth}}
\parbox[l]{0.45\textwidth}{
\epsfig{file=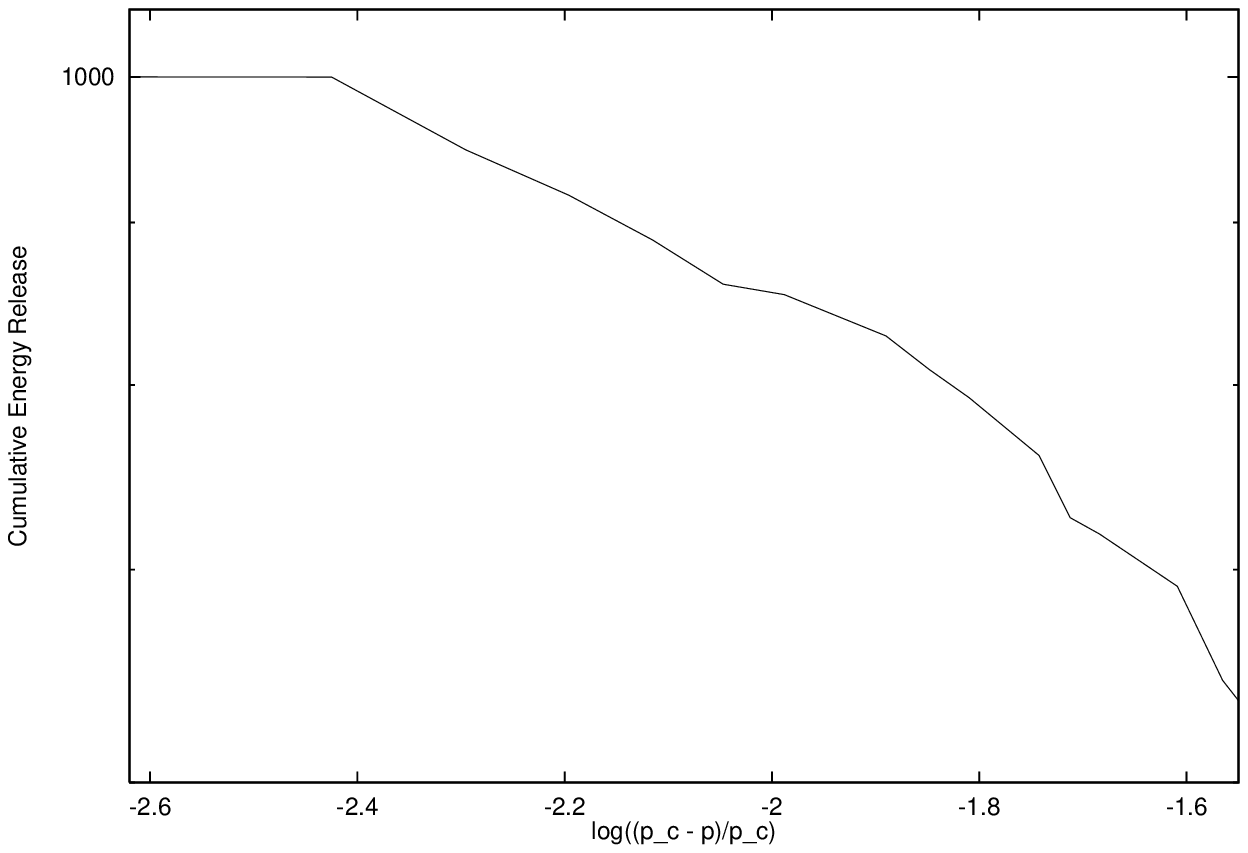,width=0.45\textwidth}}
\parbox[r]{0.45\textwidth}{
\epsfig{file=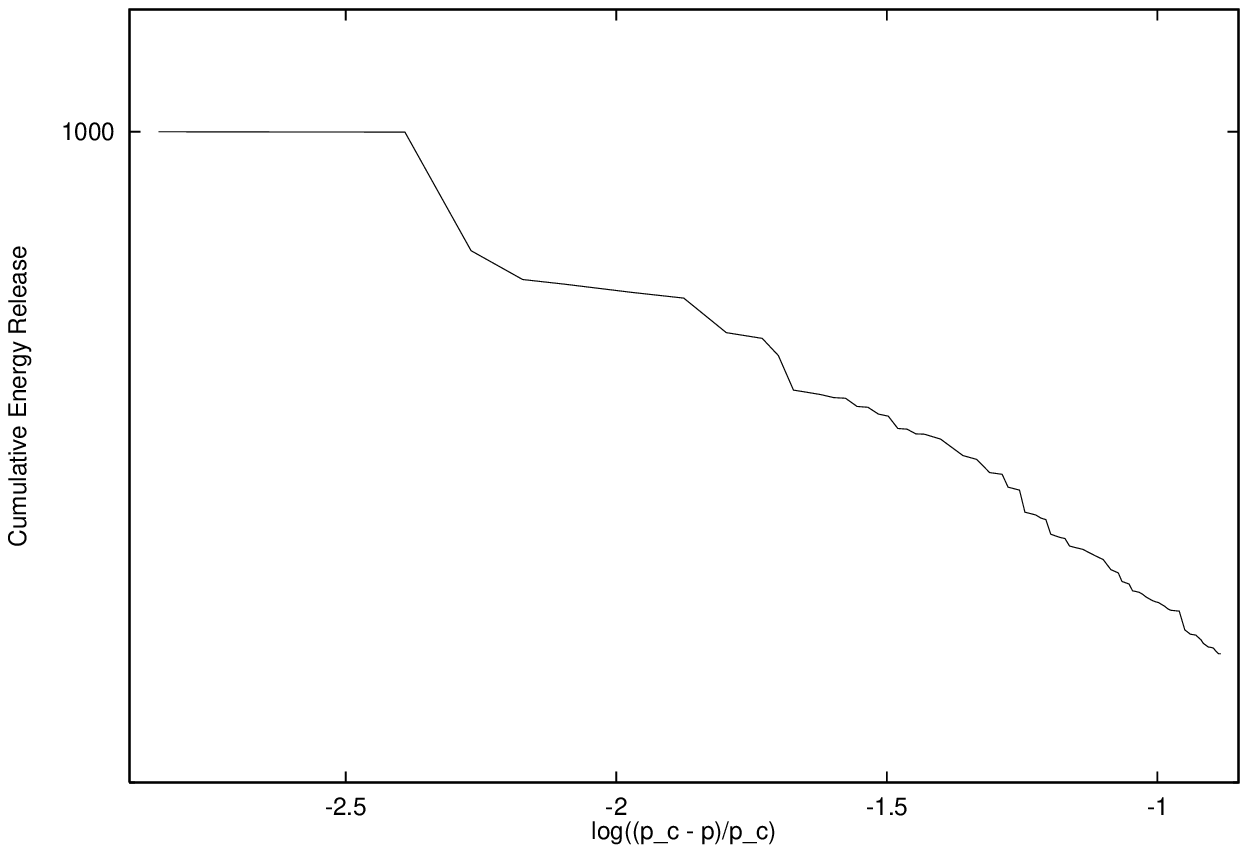,width=0.45\textwidth}}
\parbox[l]{0.45\textwidth}{
\epsfig{file=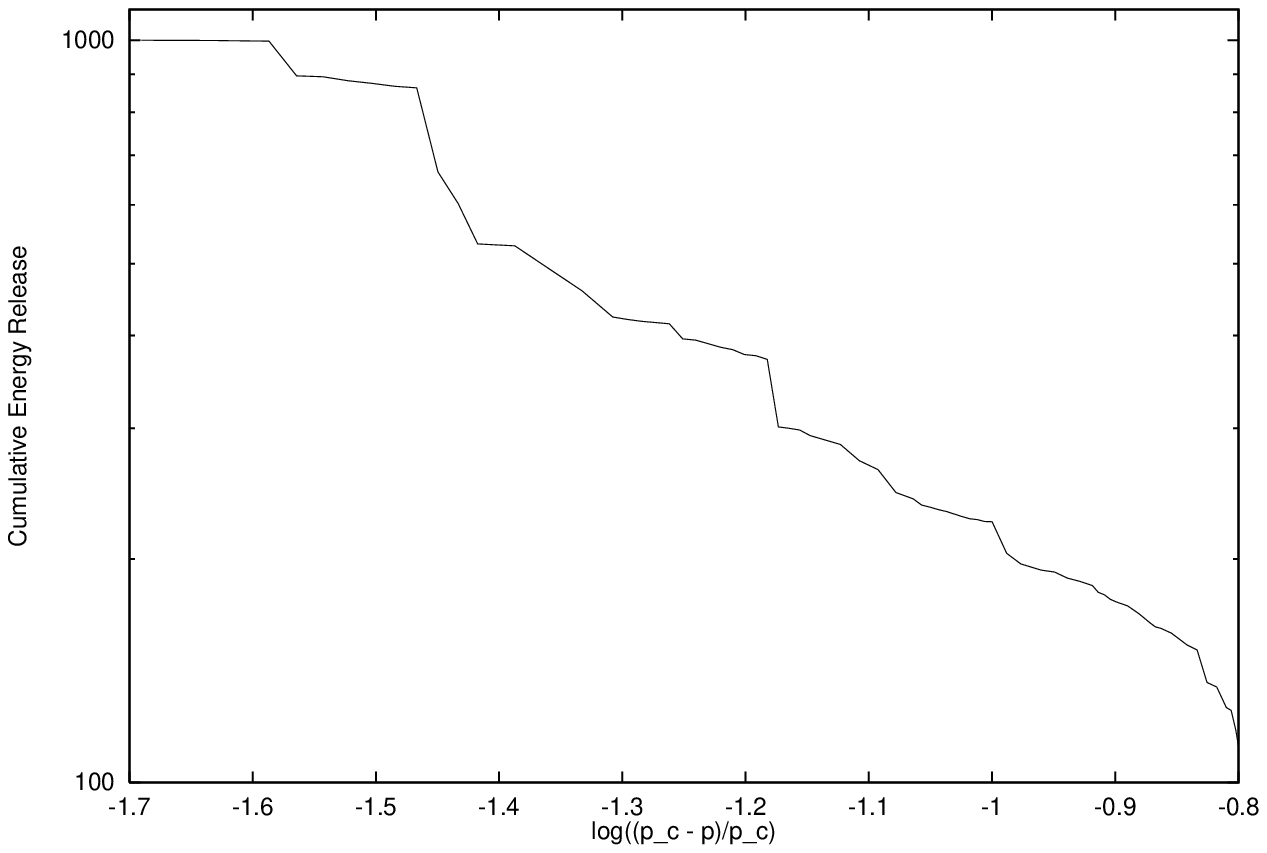,width=0.45\textwidth}}
\hspace{15mm}
\parbox[r]{0.45\textwidth}{
\epsfig{file=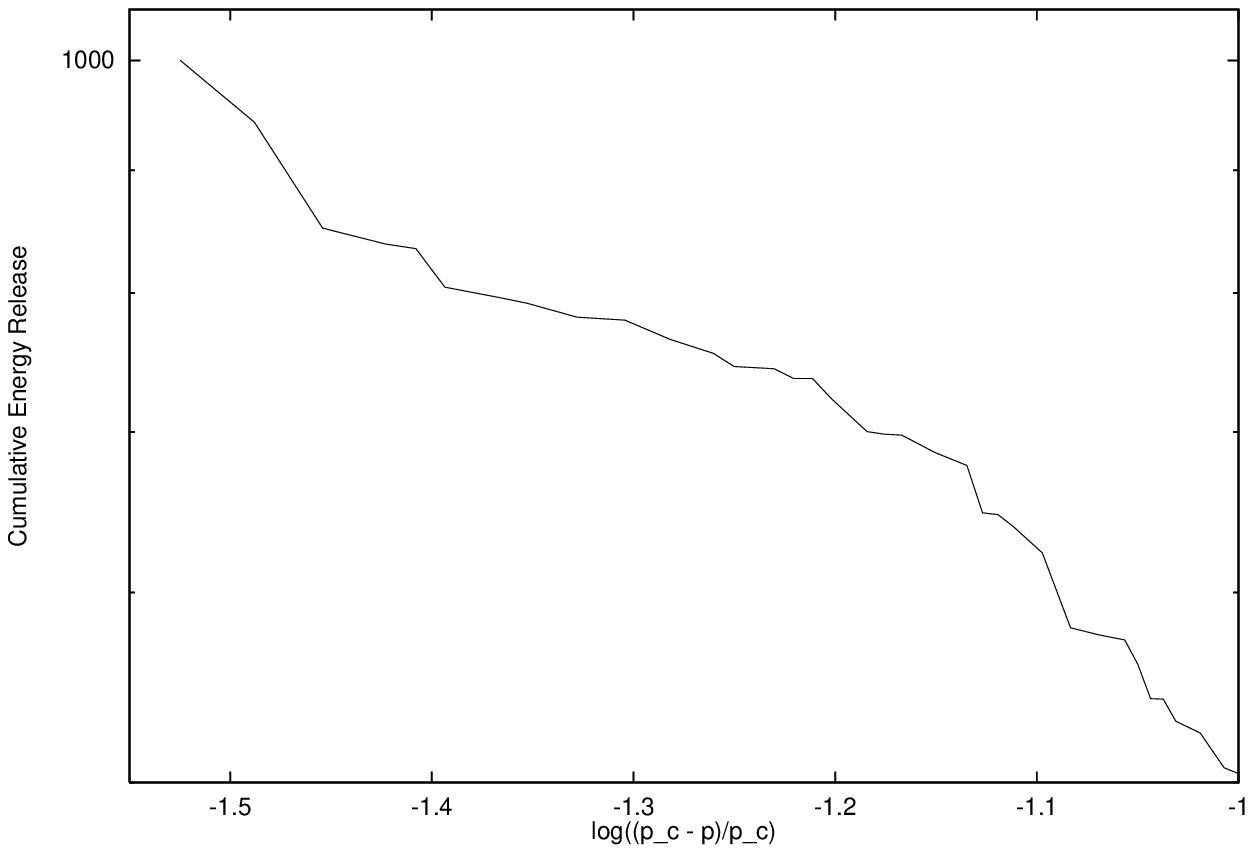,width=0.45\textwidth}}
\parbox[l]{0.45\textwidth}{
\epsfig{file=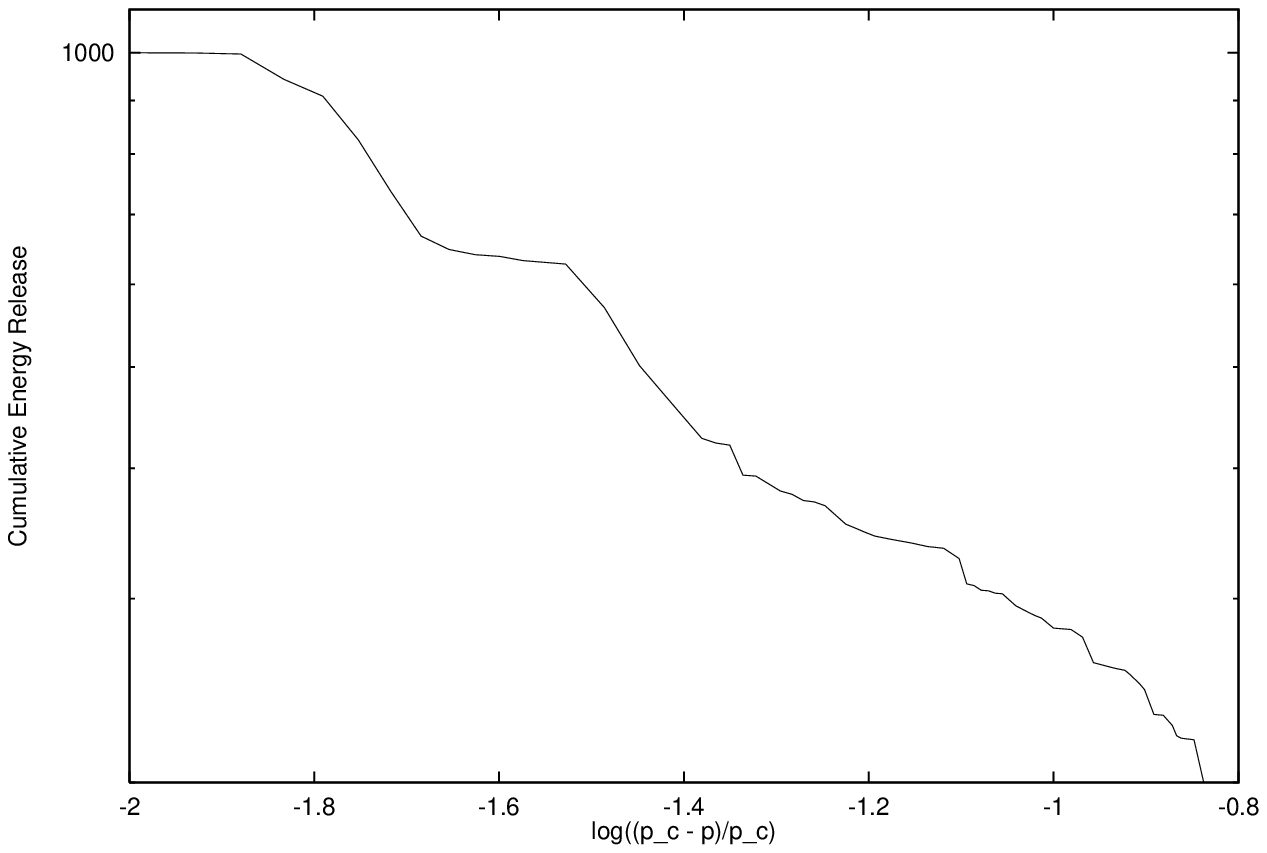,width=0.45\textwidth}}
\caption{\protect\label{nonpara} Cumulative energy release (in $\ln_{10}$
scale)
as a function of the logarithm (base 10) of the distance $(p_c-p)/p_c$ to the
critical
rupture pressure $p_c$ determined from the fits shown in figure \ref{anacumu1}.
Starting from the upper left corner, we have data set 1,2,3, ...7. We show
the last ``critical region'' close to rupture which is suggestive of a
power law
qualified by a straight line in this representation.
}
\end{figure}

\begin{figure}
\parbox[l]{0.45\textwidth}{
\epsfig{file=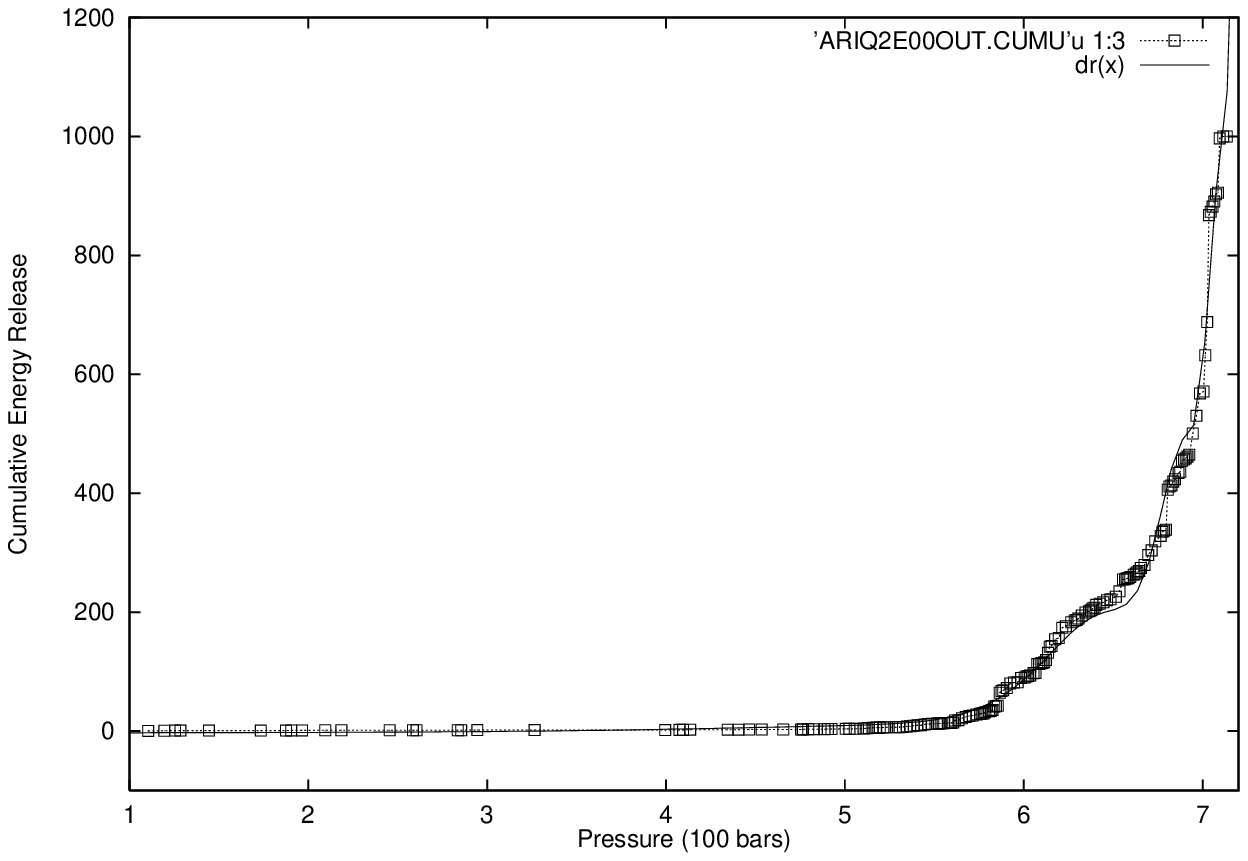,width=0.45\textwidth}}
\parbox[r]{0.45\textwidth}{
\epsfig{file=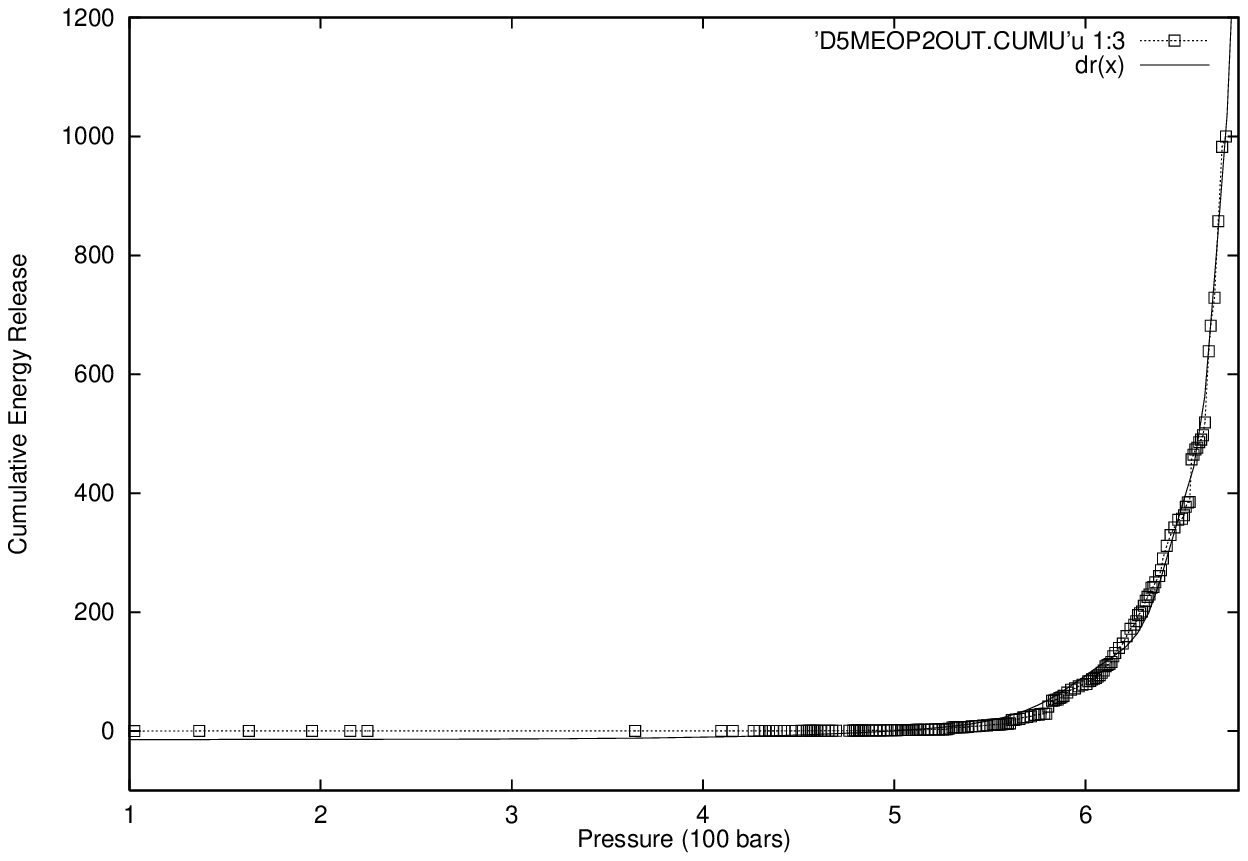,width=0.45\textwidth}}
\parbox[l]{0.45\textwidth}{
\epsfig{file=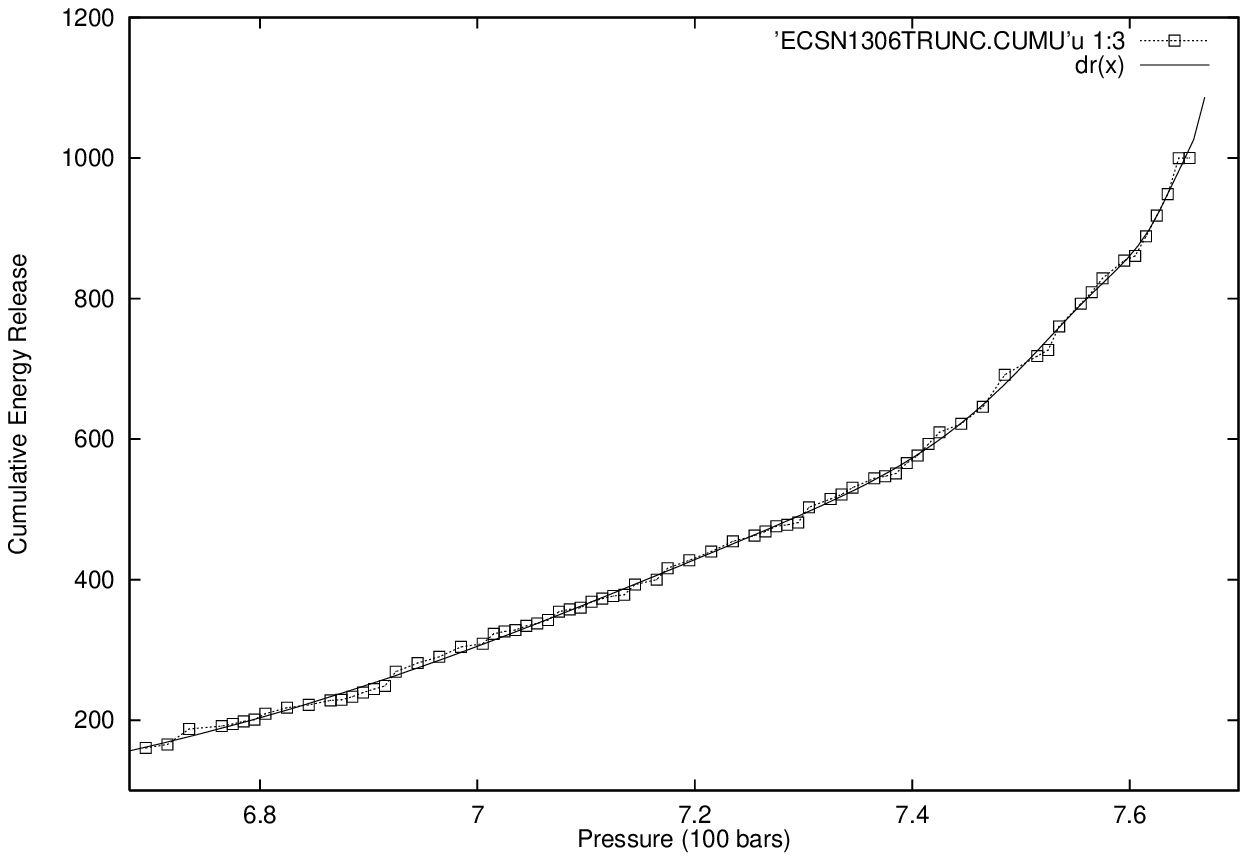,width=0.45\textwidth}}
\parbox[r]{0.45\textwidth}{
\epsfig{file=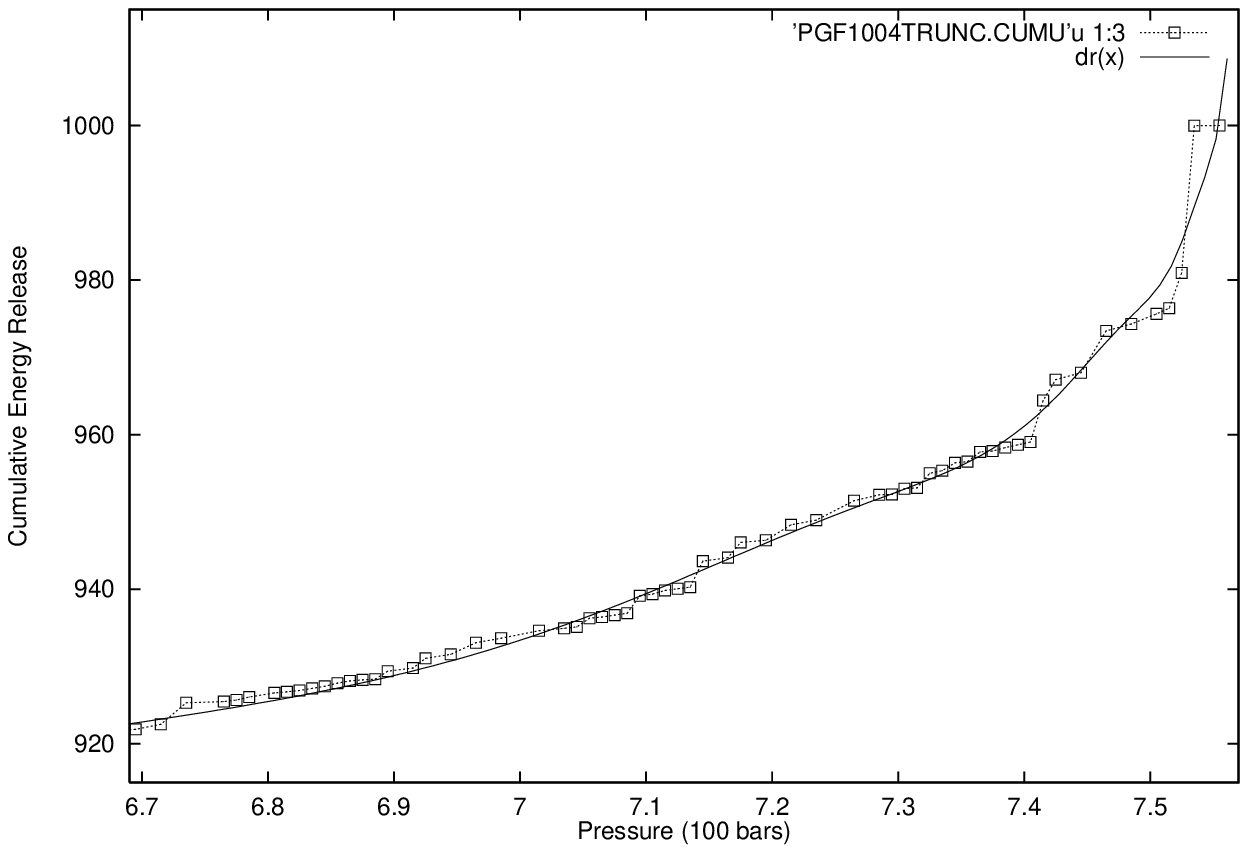,width=0.45\textwidth}}
\parbox[l]{0.45\textwidth}{
\epsfig{file=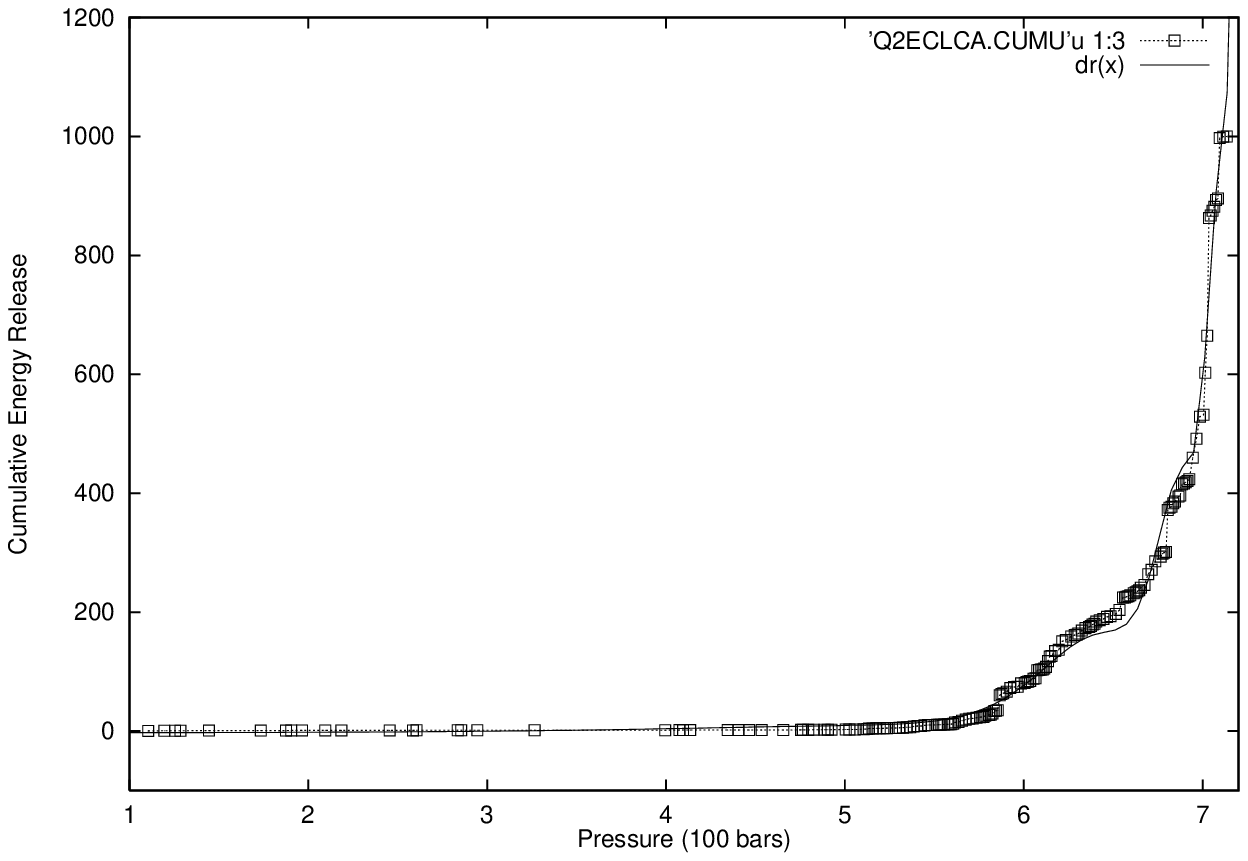,width=0.45\textwidth}}
\hspace{15mm}
\parbox[r]{0.45\textwidth}{
\epsfig{file=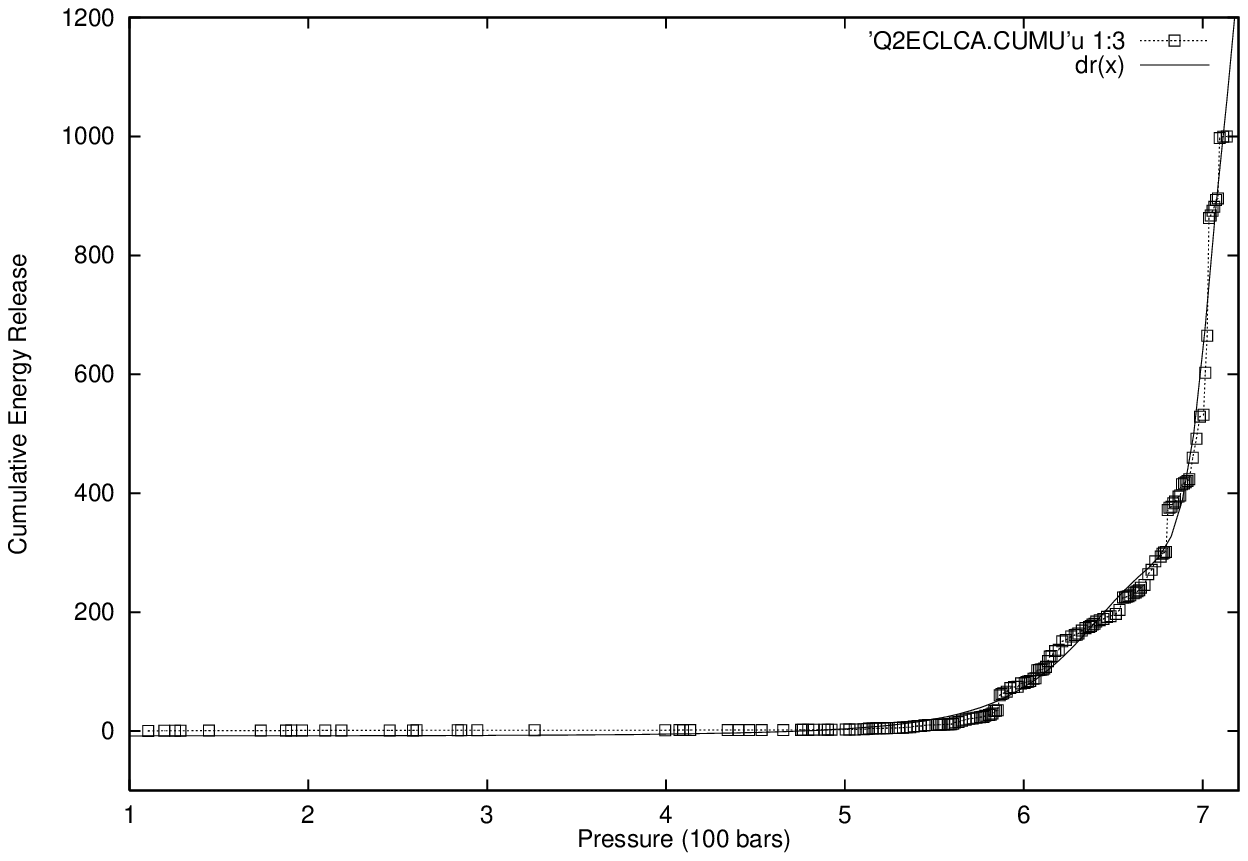,width=0.45\textwidth}}
\parbox[l]{0.45\textwidth}{
\epsfig{file=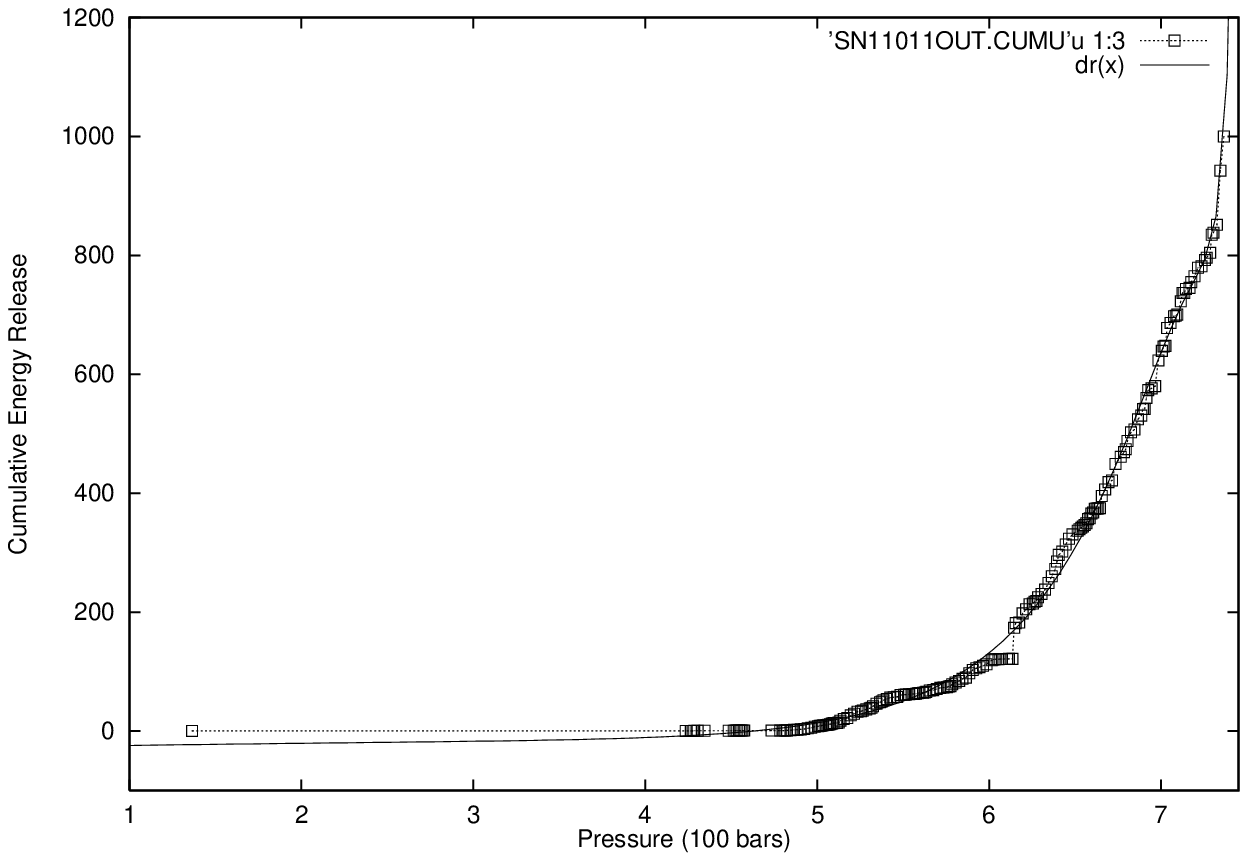,width=0.45\textwidth}}
\hspace{15mm}
\parbox[r]{0.45\textwidth}{
\epsfig{file=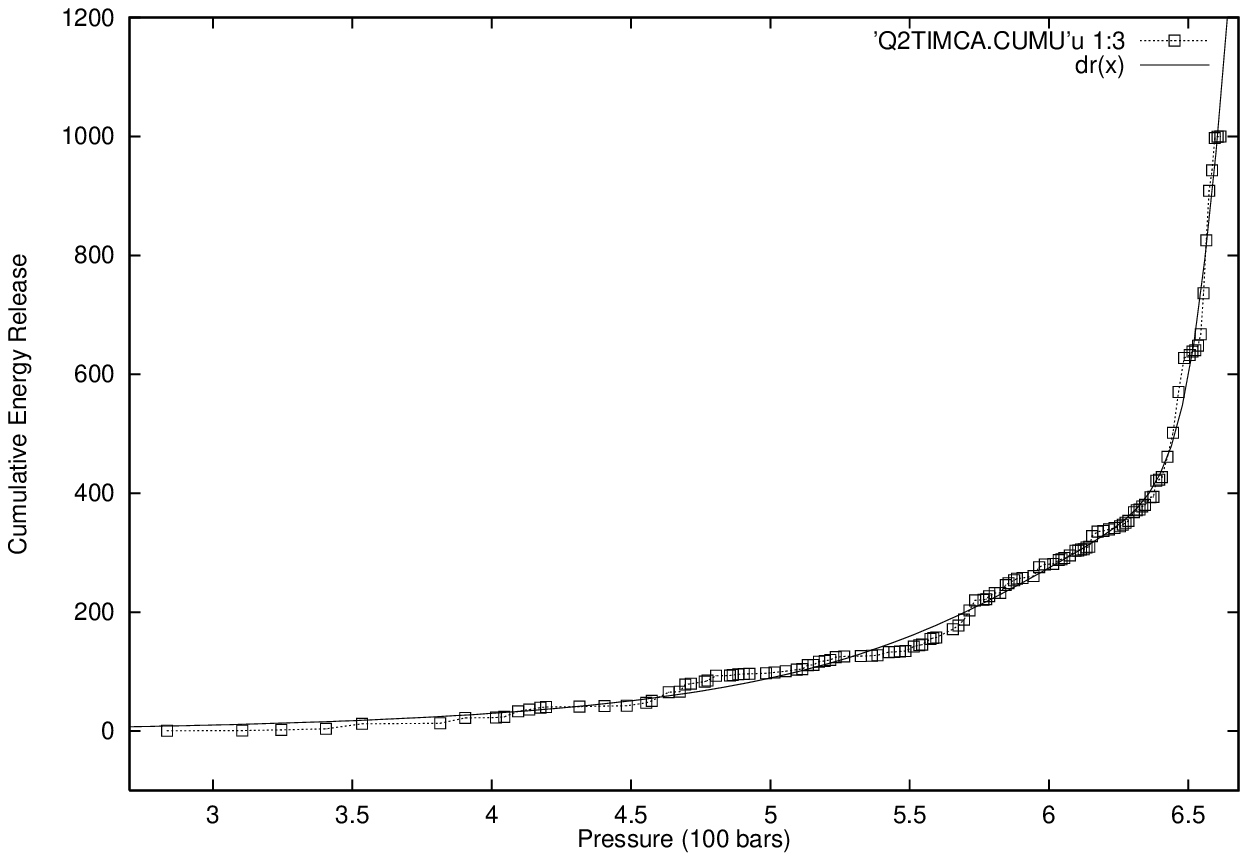,width=0.45\textwidth}}
\caption{\protect\label{anacumu3a} Cumulative energy release fitted with
eq.~(3).
Starting the from upper left corner, we have data set 1,3,4,5,6,7 (first
minimum). Data sets 3 and 4 has been truncated in the
lower end, taking as the first point the point where the acceleration in the
acoustic emission begins to be similar to the other data sets.}
\end{figure}

\begin{figure}
\parbox[l]{0.45\textwidth}{
\epsfig{file=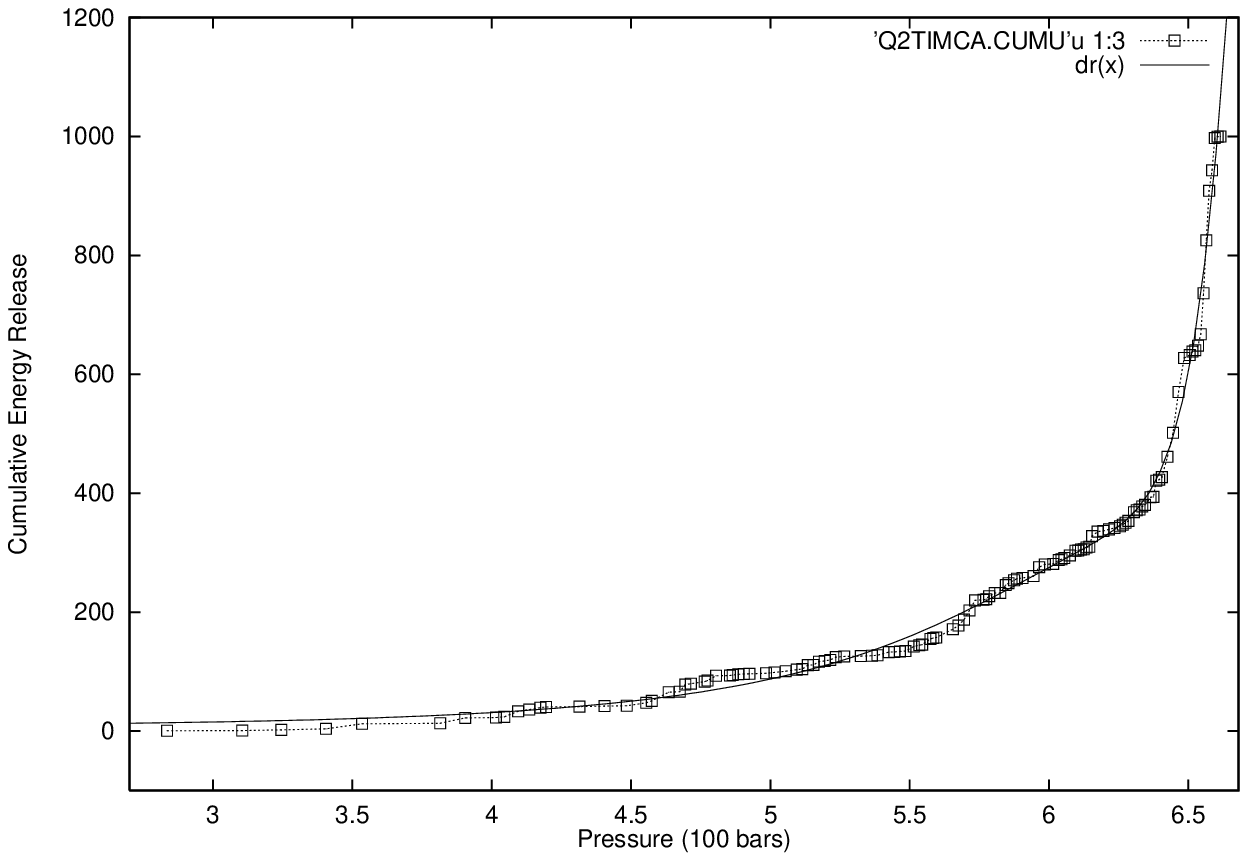,width=0.45\textwidth}}
\caption{\protect\label{anacumu3b} Second solution of the
cumulative energy release of data set 7 fitted with eq.~(3). }
\parbox[l]{0.45\textwidth}{
\epsfig{file=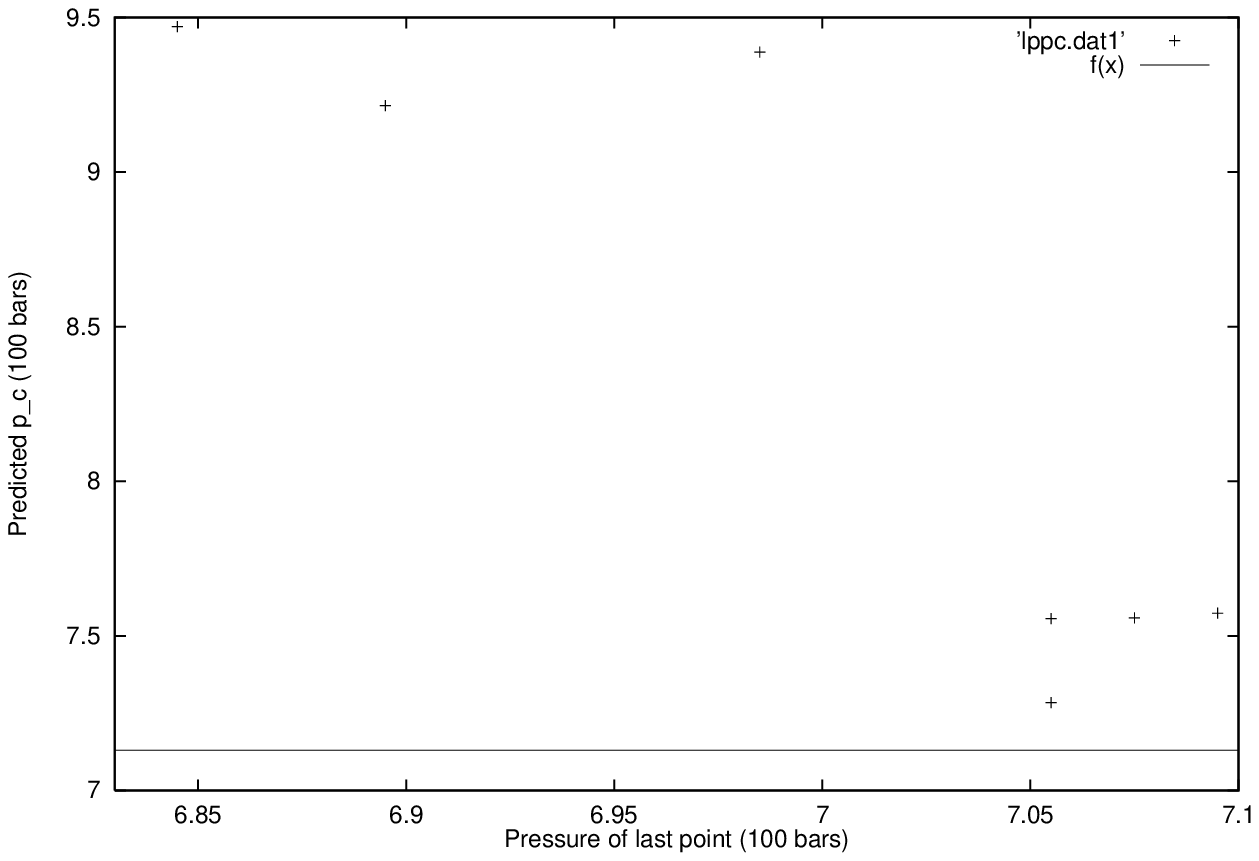,width=0.45\textwidth}}
\parbox[r]{0.45\textwidth}{
\epsfig{file=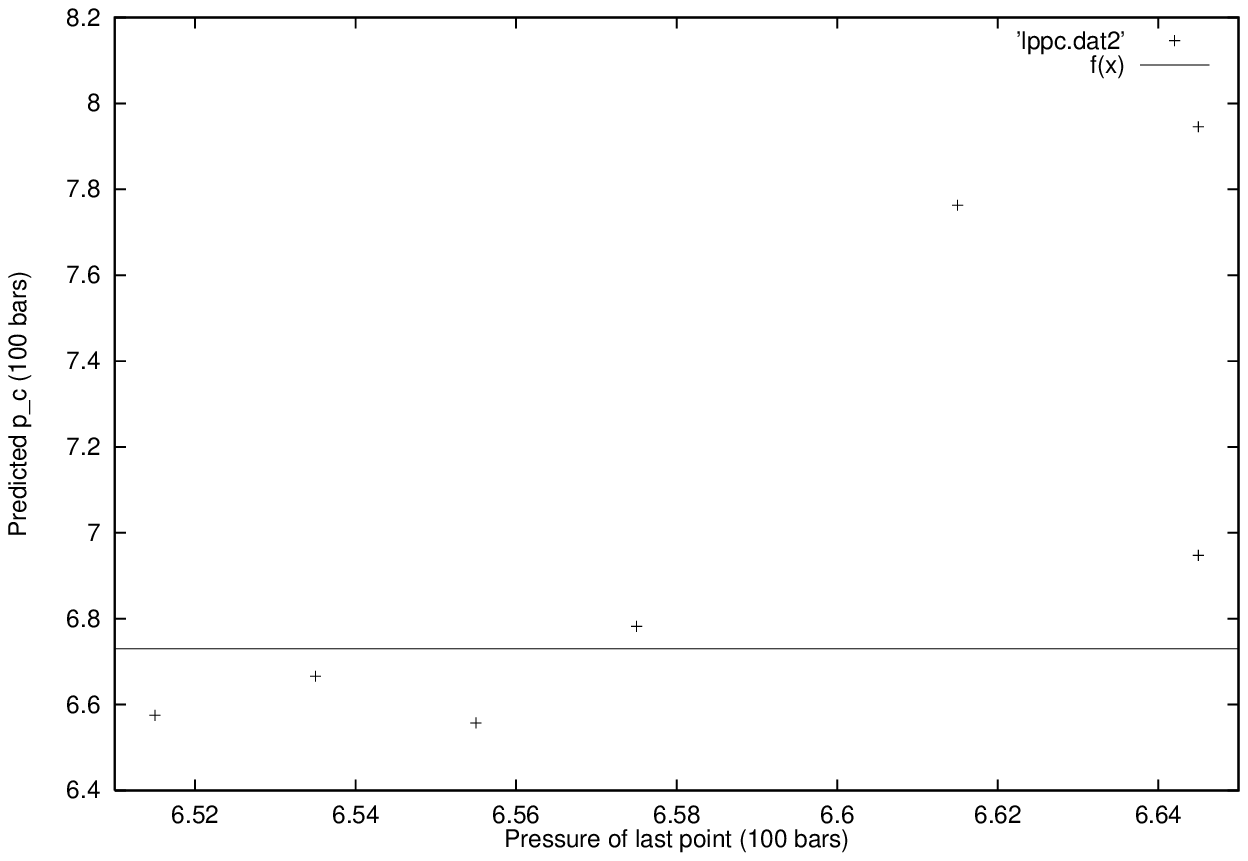,width=0.45\textwidth}}
\parbox[l]{0.45\textwidth}{
\epsfig{file=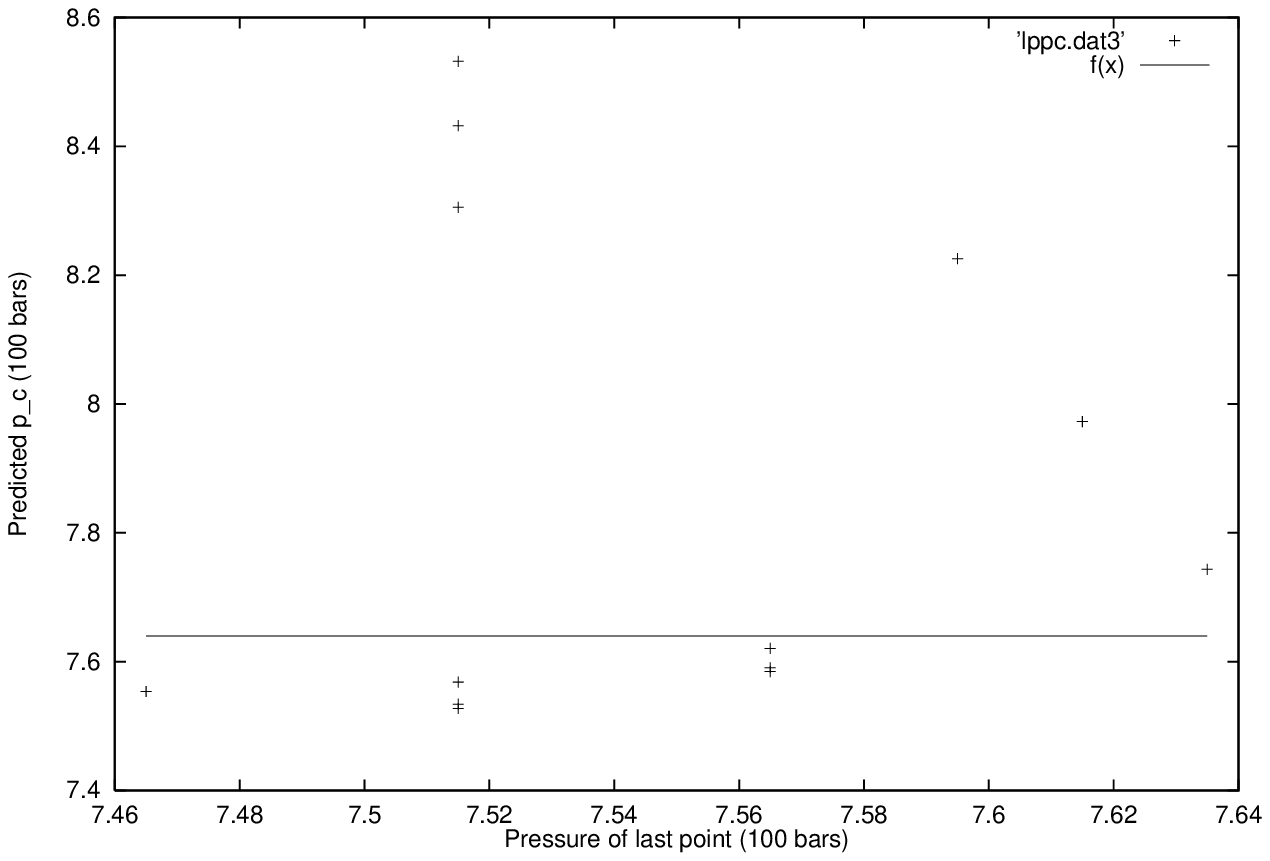,width=0.45\textwidth}}
\parbox[r]{0.45\textwidth}{
\epsfig{file=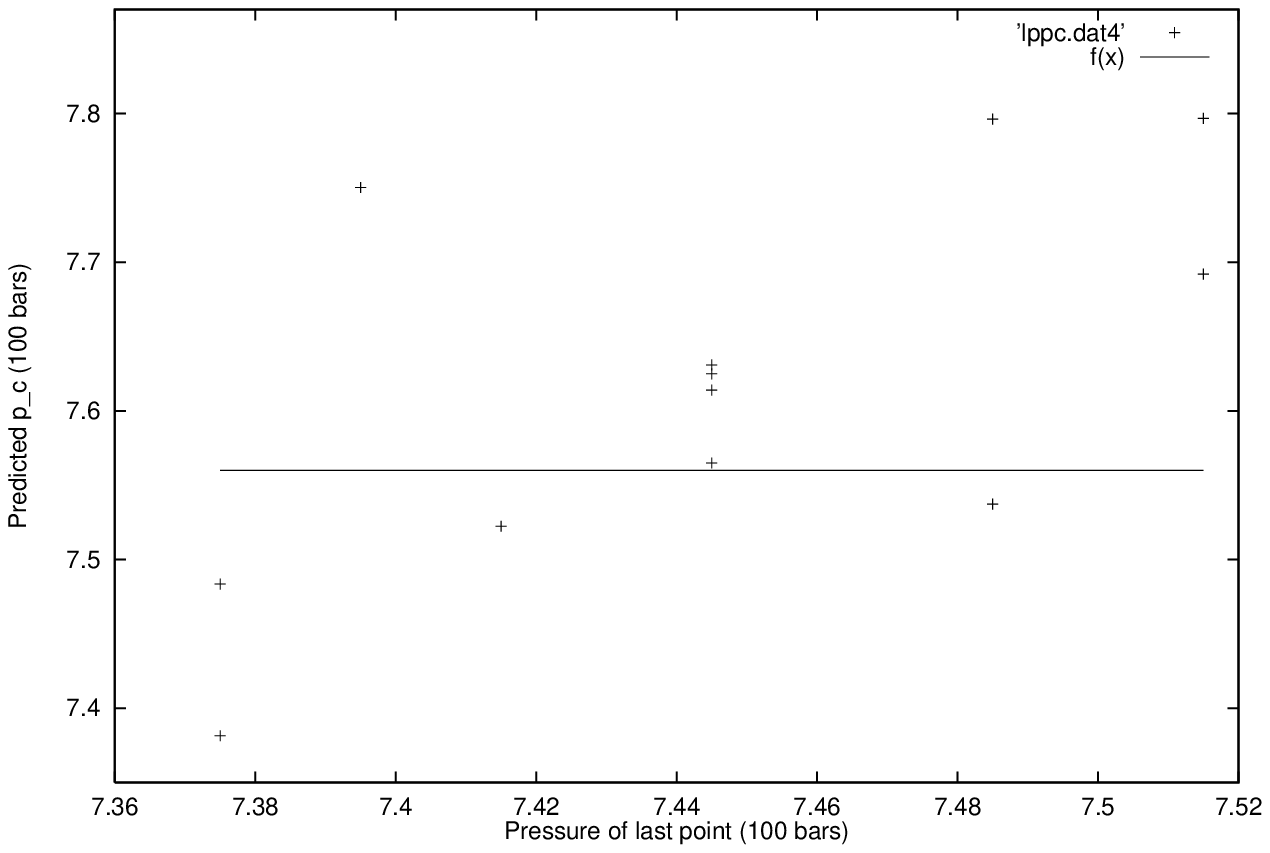,width=0.45\textwidth}}
\parbox[l]{0.45\textwidth}{
\epsfig{file=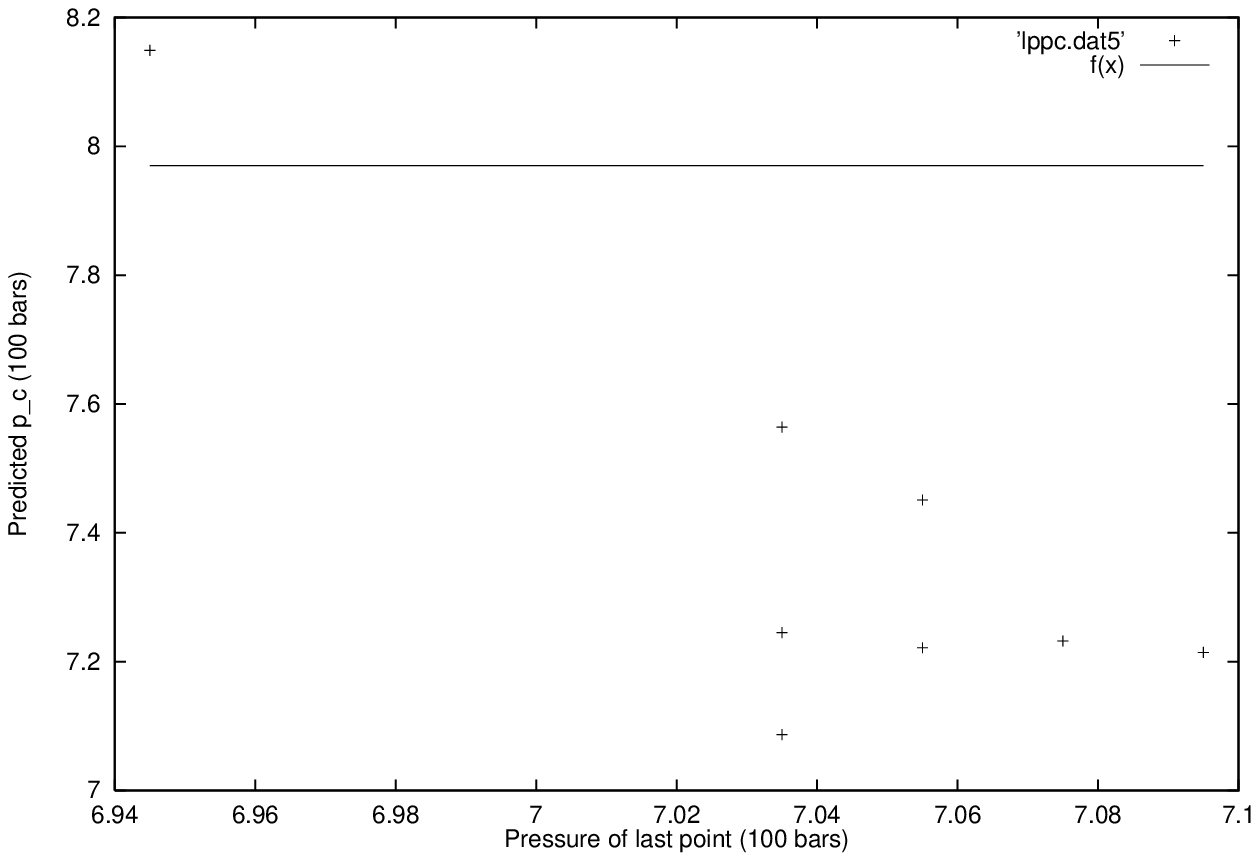,width=0.45\textwidth}}
\hspace{15mm}
\parbox[r]{0.45\textwidth}{
\epsfig{file=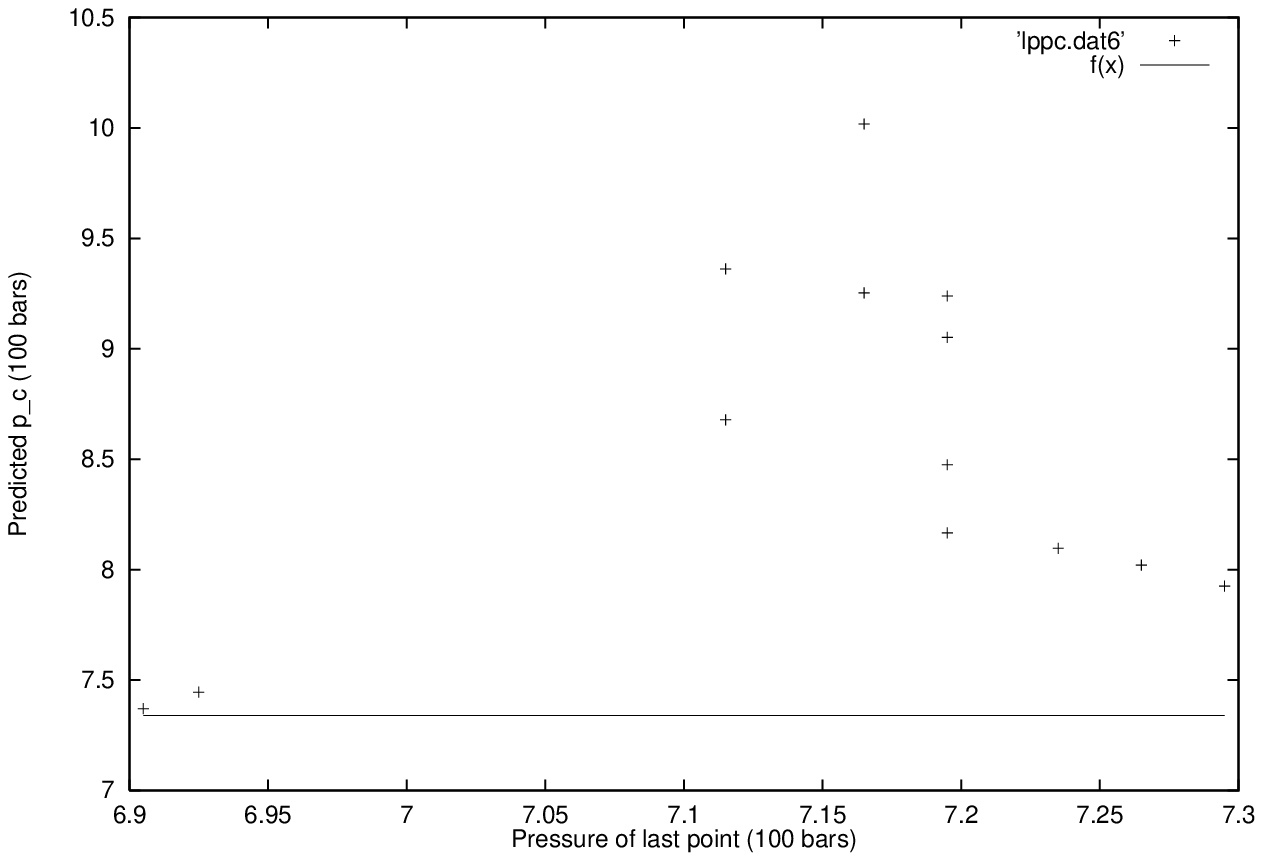,width=0.45\textwidth}}
\caption{\protect\label{lppred} Compilation of all predicted $p_c$
as a function of $p_{last}$ using eq. (2). The straight line represent
the true $p_c$.}
\end{figure}

Considering the limited quality of the data recorded in a sub-optimal
industrial environment, it is quite interesting that a reasonably clear picture
has emerged from the analysis presented here. Beginning with the log-periodic
analysis of the energy release rate, it is remarkable that the
parameter values for the exponent $z$ and log-periodic angular
frequency $\omega$ obtained from the fits with eq. (2)
to the 7 first data set actually agree on $z \approx -1.4 \pm 0.7$ and
$\omega \approx 5$ or $\omega \approx 10$ corresponding to a frequency
doubling as seen in table \ref{tabpow}. Furthermore, from figures
\ref{anacumu1} and \ref{nonpara},
it is clear that at least 6 of the data sets exhibit an average power law
acceleration as $p \rightarrow p_c$. The consistent results obtained for the
energy release rate with respect to log-periodic oscillations is reasonably
confirmed by the results obtained with eq. (3), see table \ref{tabtanh} and
figures \ref{anacumu3a} and \ref{anacumu3b}. Comparing these results with those
obtained with eq. (2) suggest that the cumulative energy release does exhibit
a transition from an approximately exponential increase to that of a power
law, in agreement with the numerical simulations of Sornette and Andersen
\cite{Andersen2}.
Additional support for such a transition comes from the better predictions
results obtained using eq. (3) instead of eq. (2).

The general results for the predictive power of eq. (1), (2) and (3) is that
the first two equation does not perform at all. The results with eq. (3) are
more positive. For data set 1, we get a suggestion for $p_c \approx
720$ bars already at $p_{last} \approx 682$ bars. For data set 2, we start
to get a reasonable stable estimate $p_c \approx 680-90$ bars for $p_{\rm
last} \approx 651$ bars with some prior indications down to $p_{\rm last} 
\approx 621$ bars. For data set 3, the results are not very convincing, but 
we do get solutions close to the true $p_c$. It is interesting to note that if
we {\it only} consider solutions with $\omega \approx 2\pi$, the we get a good 
estimate for  $p_{\rm last} \approx 752$ bars. Remarkably, the same is true for
data set 4 all the way down to the lowest pressure of $\approx 733$ bars used. 
Data set 5 is as mentioned incomplete, the last point being $\approx 85$ bars 
away from $p_c$. For this data set, we get a reasonable estimate on $p_c$ if we
focus on solutions with $\omega \approx 2\pi$ for $p_{\rm last}$ in the
range of $689-701$. For higher pressures, the fits lock on $p_c$ close to 
the last point in the untruncated data set. The results for data set 7
are very mixed without any clear pattern.

\begin{figure}
\parbox[l]{0.45\textwidth}{
\epsfig{file=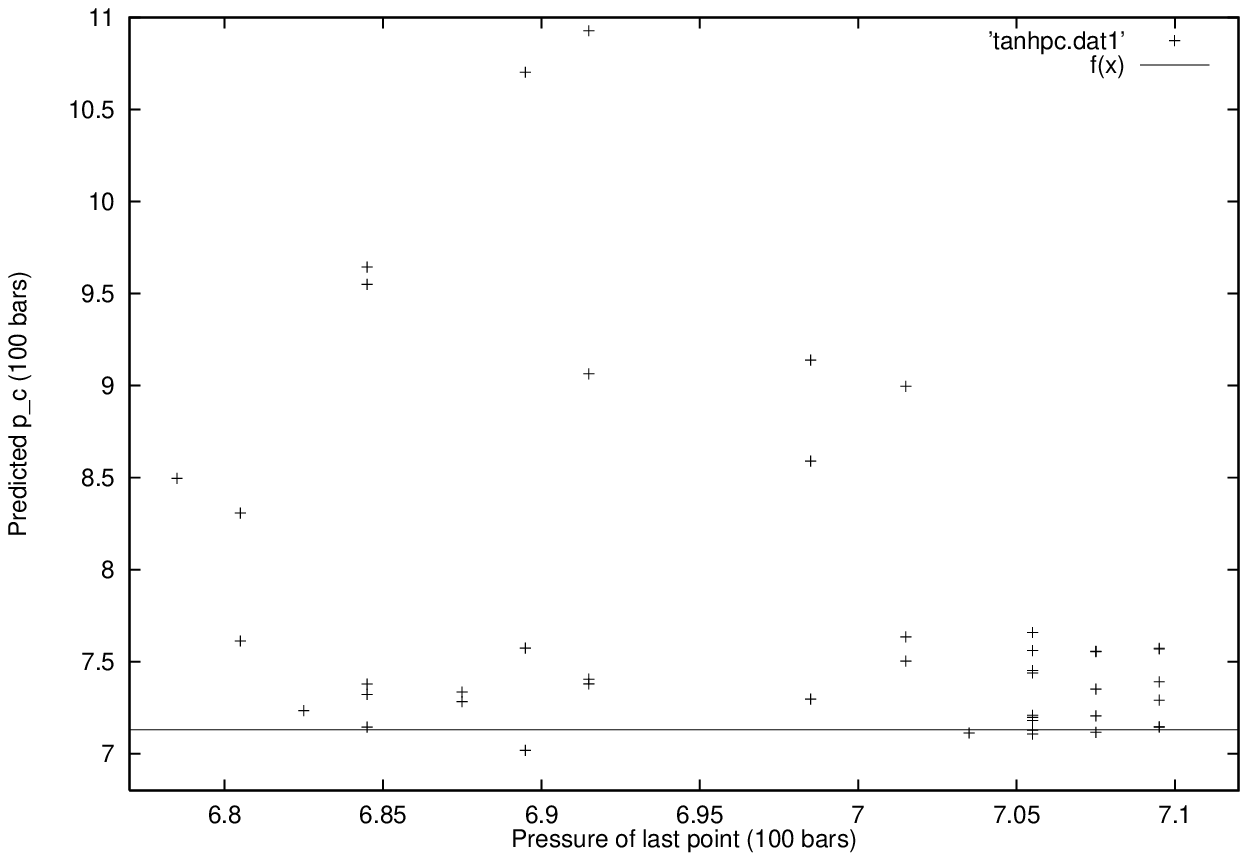,width=0.45\textwidth}}
\parbox[r]{0.45\textwidth}{
\epsfig{file=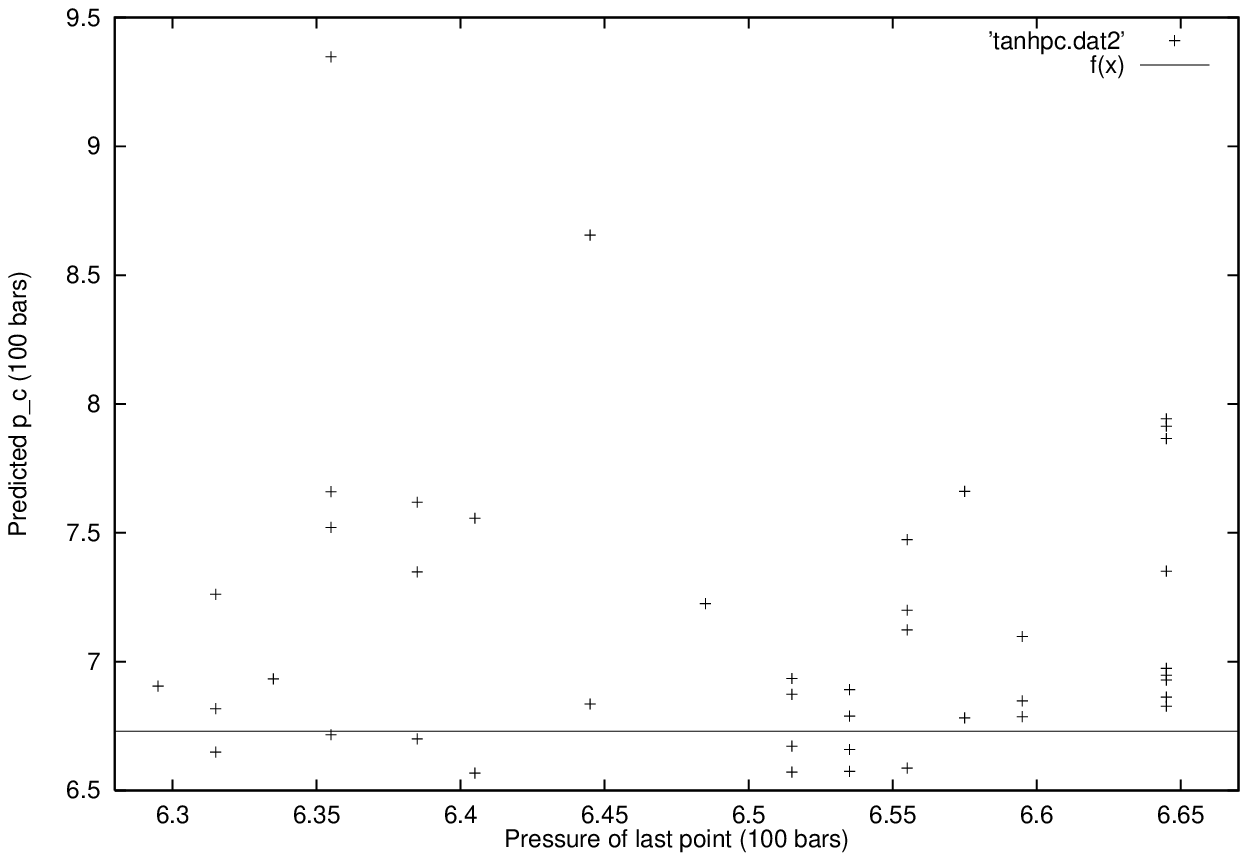,width=0.45\textwidth}}
\parbox[l]{0.45\textwidth}{
\epsfig{file=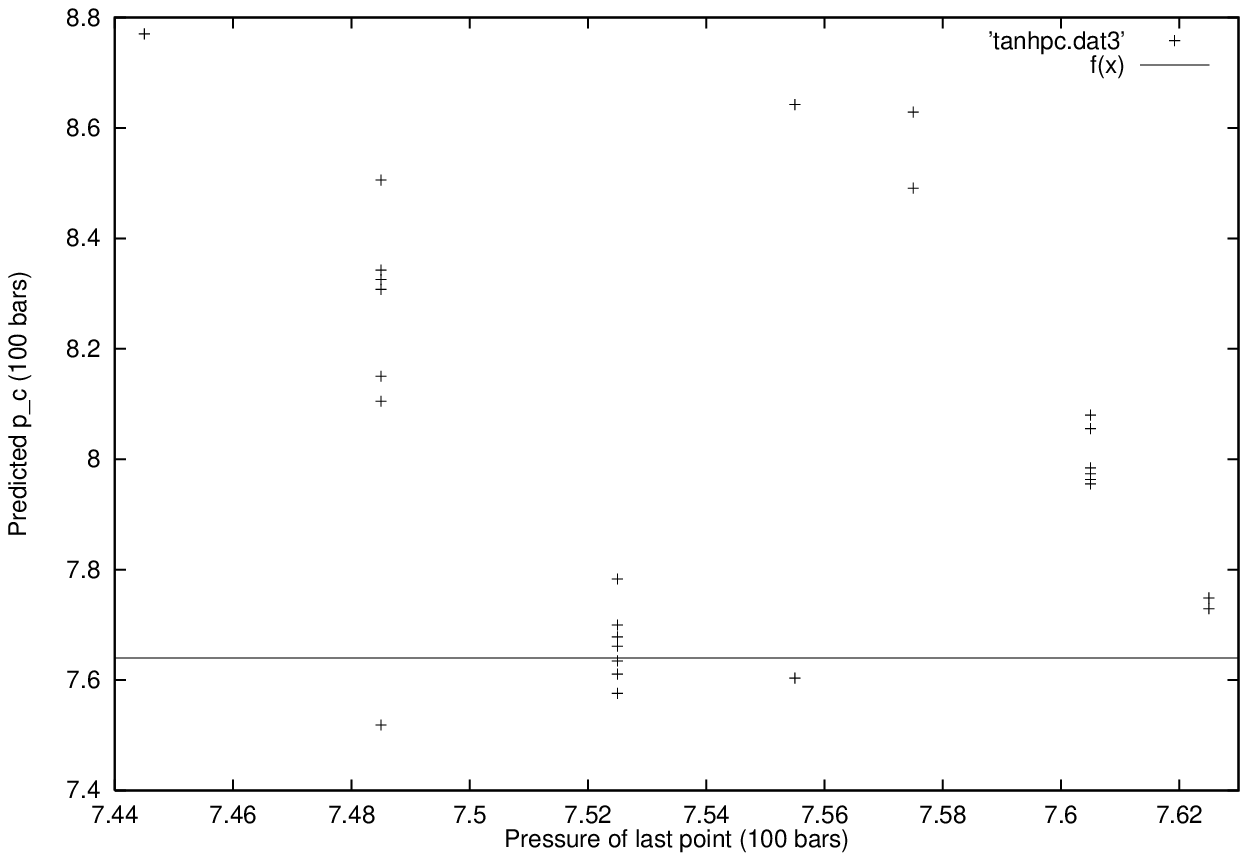,width=0.45\textwidth}}
\parbox[r]{0.45\textwidth}{
\epsfig{file=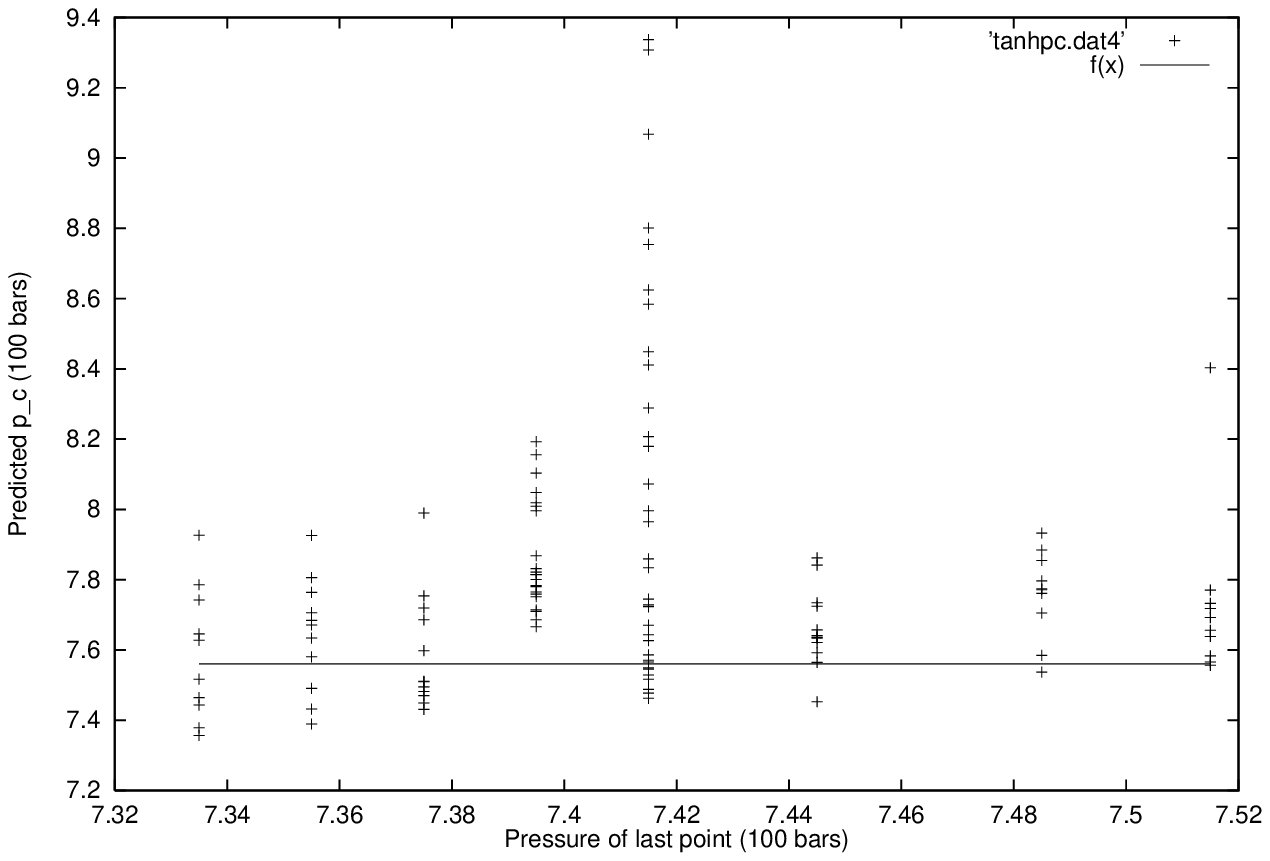,width=0.45\textwidth}}
\parbox[l]{0.45\textwidth}{
\epsfig{file=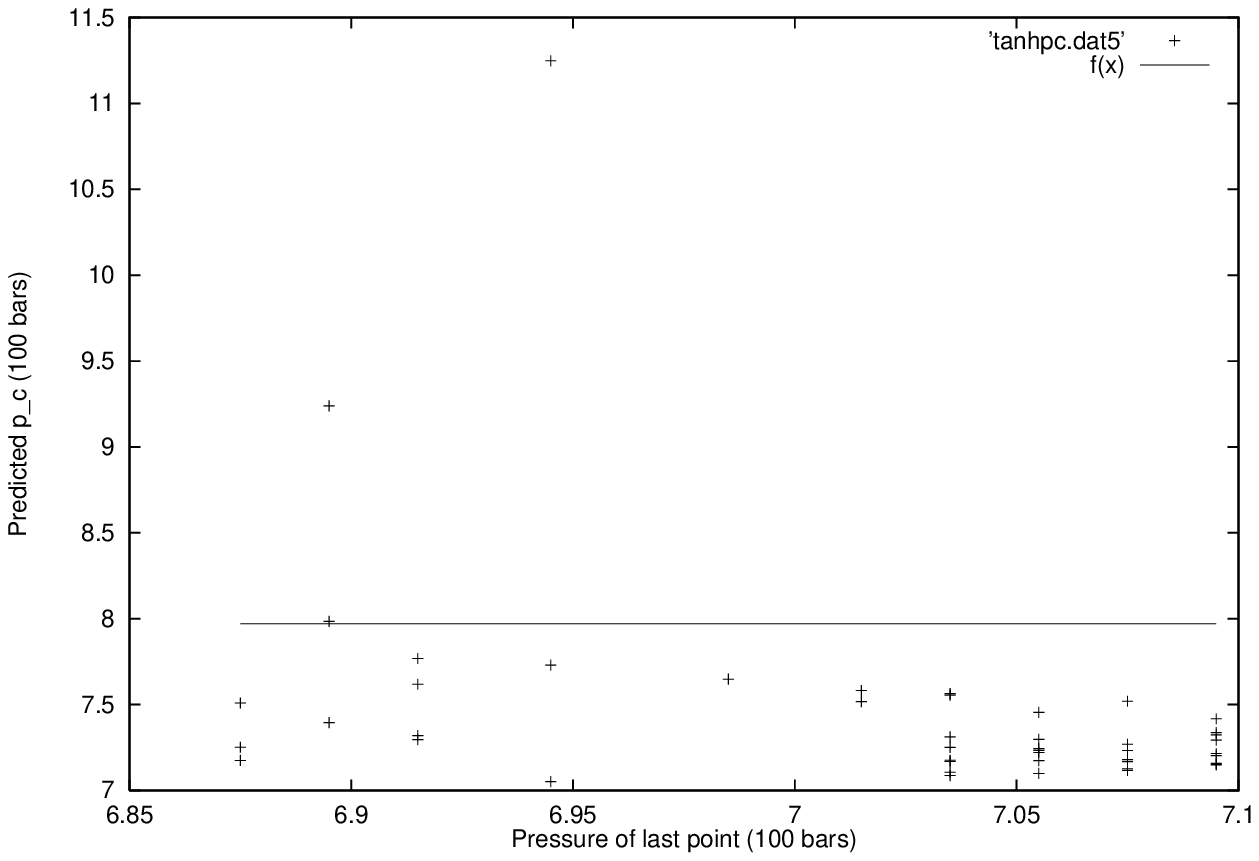,width=0.45\textwidth}}
\hspace{15mm}
\parbox[r]{0.45\textwidth}{
\epsfig{file=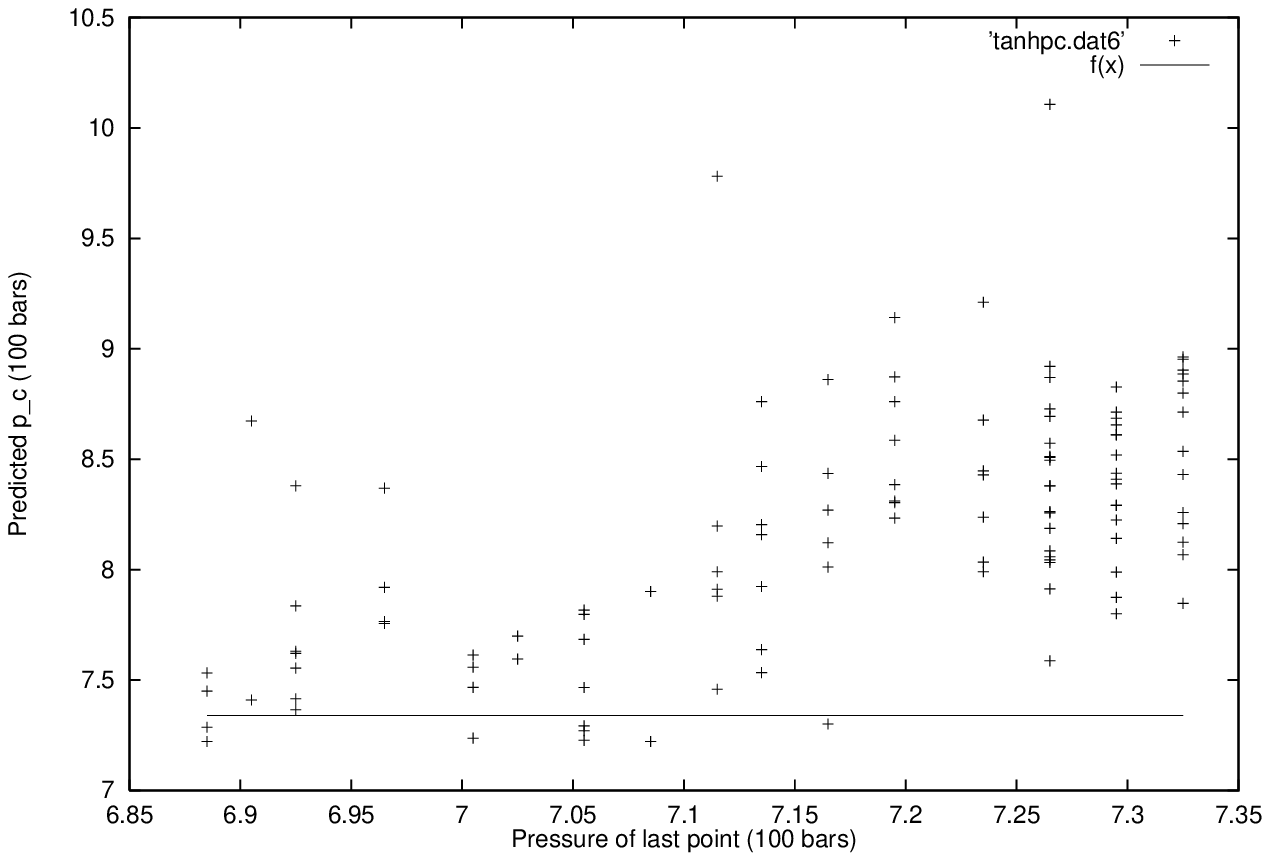,width=0.45\textwidth}}
\caption{\protect\label{tanhpred} Compilation of all predicted $p_c$
as a function of $p_{last}$ using eq. (3). The straight line represent
the true $p_c$.}
\end{figure}

As seen in figure \ref{tanhpred}, using all fits shows that a predictive 
potential of eq. (3) exists since a clustering of the predicted $p_c$'s 
occurs around the true $p_c$ for all data sets except one (data set 7). This
suggests that for a better controlled experimental situation, a reliable
prediction procedure based on eq. (3) can be achieved.

\vskip 0.5cm
{\bf Acknowledgements:} We thank J.-C. Anifrani from A\'erospatiale-Matra
Inc., Bordeaux,
France, for sharing the data with us. This research was supported by NSF
under the
grant NSF-DMR99-71475.

\mbox{}
\newpage

\begin{table}
\begin{center}
\begin{tabular}{|c|c|c|c|c|c|c|c|c|} \hline
Set \# & \# points & \# fits & $p_{\rm min}$ & fit $p_c$  & true $p_c$ &
$p_{last}$ & $z$ & $\omega$ \\ \hline
1 & $\approx 170$ & $1$ & $110.5$ &  $707$  & $713$ & $703$ & $-1.4$ & $5.4$
\\ \hline
2 & $\approx 170$ & $3$  & $102.5$ & $674$  & $673$ & $669$ & $-2.1$ &
$10.5$ \\ \hline
3 & $\approx 70$ & $1$  & $538.5$  & $772$  & $764$ & $764$ & $-0.7$ & $4.5$
\\ \hline
4 & $\approx 70$ & $1$  & $334.5$ & $758$  & $756$ & $753$ & $-2.0$ &
$4.9$\\ \hline
5 & $\approx 170$ & $2$  & $110.5$ & $717,738$  & $797$ & $713$ &
$-1.4,-1.0$ & $10.1,11.5$  \\ \hline
6 & $\approx 180$ & $1$  & $136.5$ & $738$  & $734$ & $734$ & $-1.1$ & $4.7$
\\ \hline
7 & $\approx 120$ & $1$  & $283.5$ & $672$  & $797$ & $661$ & $-1.5$ &
$10.5$ \\ \hline
\end{tabular}
\end{center}
\caption{\label{tabrate} Parameter values for fits with eq. (2) to
the energy release rate. The fits are shown in figure
\protect\ref{anarate}. All pressure
tanks are made of Kevlar composite except tank 2 which is a carbon composite.
}
\end{table}

\begin{table}
\begin{center}
\begin{tabular}{|c|c|c|c|c|} \hline
Set \# & fit $p_c$ & $z$ &  $t_{last}$ & true $p_c$ \\ \hline
1  & $756$  &  $-1.7$  & $713$ & $713$  \\ \hline
2  & $718$  &  $-2.4$  & $673$ & $673$   \\ \hline
3  & $770$  &  $0.26$  & $765$ & $764$ \\ \hline
4  & $756$  &  $0.25$  & $753$ & $756$  \\ \hline
5  & $747$ & $-1.6$ &  $713$ & $797$   \\ \hline
7  & $666$ & $-0.33$ &  $661$ & $797$  \\ \hline
8  & $718$ & $-2.4$ &  $673.5$ & $673$  \\ \hline \hline
\end{tabular}
\end{center}
\caption{\label{tabpow}Parameter values for fits with eq. (1) to
the cumulative energy released. The fits are shown in figure
\protect\ref{anacumu1}.}
\end{table}

\begin{table}
\begin{center}
\begin{tabular}{|c|c|c|c|c|c|c|} \hline
Set \#  & \# fits & fit $p_c$  & true $p_c$ & $t_{last}$ & $z$ & $\omega$
\\ \hline
1 & $2$  &  $760$  & $713$ & $713$ & $-1.7$ & $13.0$ \\ \hline
3 & $2$  &  $774$  & $764$ & $765$ & $0.13$ & $6.0$ \\ \hline
4 & $3$  &  $757$  & $756$ & $755$ & $0.23$ & $5.0$\\ \hline
5 & $3$  &  $734$  & $797$ & $713$ & $-1.2$ & $8.9$  \\ \hline
6 & $4$  &  $759$  & $734$ & $736$ & $-0.98$ & $2.4$ \\ \hline
7 & $2$  &  $669,668$  & $797$ & $661$ & $-0.42,-0.43$ & $3.1,8.6$ \\ \hline
8 & $4$  &  $549$ & $548$  & $548$ & $-0.06$ & $1.3$  \\ \hline \hline
\end{tabular}
\end{center}
\caption{\label{tablp} Parameter values for fits with eq. (2) to
the cumulative energy released. The fits are shown in figure
\protect\ref{anacumu2}.}
\end{table}

\begin{table}
\begin{center}
\begin{tabular}{|c|c|c|c|c|c|c|c|} \hline
Set \#  & \# fits & fit $p_c$  & true $p_c$ & $t_{last}$ & $z$ & $\omega$
& $\tau$ \\ \hline
1 & $3$  &  $727$  & $713$ & $713$ & $-0.57$ & $8.6$  & $1.8$ \\ \hline
2 & $1$  &  $705$  & $673$ & $673$ & $-1.7$  & $13.4$ & $1.4$   \\ \hline
3 & $33$  &  $767$  & $764$ & $765$ & $0.52$ & $4.5$  & $1.0$ \\ \hline
4 & $15$  &  $757$  & $756$ & $755$ & $0.23$ & $5.0$  & $428$ \\ \hline
5 & $4$  &  $728,743$  & $797$ & $713$ & $-0.8,-1.1$ & $8.7,6.7$ &
$1.9,1.9$ \\ \hline
6 & $4$  &  $759$  & $734$ & $736$ & $-0.98$ & $2.4$ & $124$ \\ \hline
7 & $17$  &  $668,699$  & $797$ & $661$ & $-0.33,-1.3$ & $3.0,4.4$ &
$2.3,1.9$ \\ \hline
8 & $4$  &  $639$ & $548$  & $548$ & $-1.8$ & $7.8$ & $7.4$ \\ \hline \hline
\end{tabular}
\end{center}
\caption{\label{tabtanh} Parameter values for fits with eq. (3) to
the cumulative energy released. The fits are shown in figures
\protect\ref{anacumu3a} and \protect\ref{anacumu3b}.}
\end{table}

\begin{table}[t]
\begin{center}
\begin{tabular}{|c|c|c|c|c|c|} \hline
Set \#  &  fit $p_c$  & true $p_c$ & $t_{last}$ & $z$  \\ \hline
1 & $749$  &  $713$  & $705$ & $-1.5$  \\ \hline
1 & $739$  &  $713$  & $707$ & $-1.3$  \\ \hline
1 & $739$  &  $713$  & $709$ & $-1.3$  \\ \hline
3 & $819$  &  $764$  & $742$ & $-0.02$  \\ \hline
3 & $750$  &  $764$  & $746$ & $0.67$  \\ \hline
3 & $753$  &  $764$  & $751$ & $0.60$  \\ \hline
3 & $755$  &  $764$  & $753$ & $0.58$  \\ \hline
3 & $758$  &  $764$  & $756$ & $0.50$  \\ \hline
3 & $764$  &  $764$  & $759$ & $0.39$  \\ \hline
3 & $769$  &  $764$  & $761$ & $0.29$  \\ \hline
3 & $770$  &  $764$  & $763$ & $0.27$  \\ \hline
4 & $809$  &  $756$  & $744$ & $-0.64$  \\ \hline
4 & $754$  &  $756$  & $748$ & $0.30$  \\ \hline
4 & $772$  &  $756$  & $751$ & $-0.05$  \\ \hline
\end{tabular}
\end{center}
\caption{\label{powpred} Summary of the predicted critical pressures
and comparison with the true pressure at rupture using eq.(1) on the
seven pressure tanks. }
\end{table}

\begin{table}
\begin{center}
\begin{tabular}{|c|c|c|c|c|c|c|c|} \hline
Set \#  & \# fits & fit $p_c$  & true $p_c$ & $t_{last}$ & $z$  & $\omega$
\\ \hline
1 & $1$ & $947$  &  $713$  & $684$ & $-1.6$  & $5.1$  \\ \hline
1 & $1$ & $922$  &  $713$  & $689$ & $-0.75$ & $3.1$  \\ \hline
1 & $1$ & $939$  &  $713$  & $698$ & $-0.48$ & $1.4$  \\ \hline
1 & $2$ & $728$  &  $713$  & $705$ & $-0.96$ & $7.8$  \\ \hline
1 & $1$ & $756$  &  $713$  & $707$ & $-1.7$  & $12.2$ \\ \hline
1 & $1$ & $757$  &  $713$  & $709$ & $-1.7$  & $12.4$ \\ \hline
\end{tabular}
\end{center}
\caption{\label{lppred1} Same as table \ref{powpred} with equation (2) on
the pressure tank 1.}
\end{table}

\begin{table}
\begin{center}
\begin{tabular}{|c|c|c|c|c|c|c|c|} \hline
Set \#  & \# fits & fit $p_c$  & true $p_c$ & $t_{last}$ & $z$  & $\omega$
\\ \hline
2 & $1$ & $658$  &  $673$  & $651$ & $-0.03$ & $1.5$  \\ \hline
2 & $3$ & $667$  &  $673$  & $653$ & $-0.43$ & $1.8$  \\ \hline
2 & $1$ & $656$  &  $673$  & $655$ & $-0.22$ & $1.4$  \\ \hline
2 & $1$ & $668$  &  $673$  & $657$ & $-2.1$  & $2.6$  \\ \hline
2 & $1$ & $668$  &  $673$  & $657$ & $-2.1$  & $2.6$  \\ \hline
2 & $1$ & $776$  &  $673$  & $661$ & $-2.5$  & $3.3$  \\ \hline
2 & $2$ & $795$  &  $673$  & $664$ & $-2.7$  & $3.9$  \\ \hline
\end{tabular}
\end{center}
\caption{\label{lppred2} Same as table \ref{powpred} with equation (2) on
the pressure tank 2.}
\end{table}

\begin{table}
\begin{center}
\begin{tabular}{|c|c|c|c|c|c|c|c|} \hline
Set \#  & \# fits  & fit $p_c$  & true $p_c$ & $t_{last}$ & $z$  & $\omega$
\\ \hline
3 & $3$ & $755$  &  $764$  & $746$ & $0.62$  & $13.0$ \\ \hline
3 & $5$ & $757$  &  $764$  & $751$ & $0.63$  & $12.6$ \\ \hline
3 & $5$ & $757$  &  $764$  & $753$ & $0.54$  & $13.8$ \\ \hline
3 & $3$ & $762$  &  $764$  & $756$ & $0.37$  & $5.0$  \\ \hline
3 & $1$ & $823$  &  $764$  & $759$ & $-0.77$ & $10.6$  \\ \hline
3 & $1$ & $797$  &  $764$  & $761$ & $-0.29$ & $8.4$  \\ \hline
3 & $1$ & $774$  &  $764$  & $761$ & $0.12$  & $6.0$  \\ \hline
\end{tabular}
\end{center}
\caption{\label{lppred3} Same as table \ref{powpred} with equation (2) on
the pressure tank 3.}
\end{table}

\begin{table}
\begin{center}
\begin{tabular}{|c|c|c|c|c|c|c|c|} \hline
Set \#  & \# fits  & fit $p_c$  & true $p_c$ & $t_{last}$ & $z$  & $\omega$
\\ \hline
4 & $2$ & $738,748$  &  $756$  & $737$ & $0.85,0.59$  & $3.3,4.8$  \\ \hline
4 & $1$ & $775$  &  $756$  & $739$ & $0.20$  & $8.1$  \\ \hline
4 & $1$ & $752$  &  $756$  & $741$ & $0.57$  & $5.1$  \\ \hline
4 & $4$ & $756$  &  $756$  & $744$ & $0.40$  & $6.1$  \\ \hline
4 & $2$ & $780$  &  $756$  & $748$ & $-0.16$ & $10.4$ \\ \hline
4 & $2$ & $769$  &  $756$  & $751$ & $-0.09$ & $8.5$  \\ \hline
\end{tabular}
\end{center}
\caption{\label{lppred4} Same as table \ref{powpred} with equation (2) on
the pressure tank 4.}
\end{table}

\begin{table}
\begin{center}
\begin{tabular}{|c|c|c|c|c|c|c|} \hline
Set \#  & \# fits & fit $p_c$  & true $p_c$ & $t_{last}$ & $z$ & $\omega$
\\ \hline
5 & $1$  &  $815$  & $797$ & $694$ & $0.31$ & $11.1$  \\ \hline
5 & $3$  &  $709,724$  & $797$ & $703$ & $-0.71,-0.92$ & $1.8,7.3$  \\ \hline
5 & $2$  &  $745$  & $797$ & $705$ & $-2.4$ & $3.2$  \\ \hline
5 & $1$  &  $723$  & $797$ & $707$ & $-0.85$ & $7.3$  \\ \hline
5 & $2$  &  $721$  & $797$ & $709$ & $-0.79$ & $7.1$  \\ \hline
\end{tabular}
\end{center}
\caption{\label{lppred5} Same as table \ref{powpred} with equation (2) on
the pressure tank 5.}
\end{table}

\begin{table}
\begin{center}
\begin{tabular}{|c|c|c|c|c|c|c|c|} \hline
Set \#  & \# fits  & fit $p_c$  & true $p_c$ & $t_{last}$ & $z$  & $\omega$
\\ \hline
6 & $1$ & $737$  &  $734$  & $690$ & $-1.9$  & $2.9$  \\ \hline
6 & $1$ & $744$  &  $734$  & $692$ & $-2.4$  & $2.9$  \\ \hline
6 & $2$ & $936$  &  $734$  & $711$ & $-2.4$  & $4.5$  \\ \hline
6 & $2$ & $925$  &  $734$  & $716$ & $-2.0$  & $4.3$  \\ \hline
6 & $4$ & $924$  &  $734$  & $719$ & $-2.7$  & $4.6$  \\ \hline
6 & $1$ & $810$  &  $734$  & $723$ & $-0.88$ & $2.9$  \\ \hline
6 & $1$ & $802$  &  $734$  & $726$ & $-0.92$ & $2.8$  \\ \hline
6 & $1$ & $792$  &  $734$  & $729$ & $-0.60$ & $2.7$  \\ \hline
\end{tabular}
\end{center}
\caption{\label{lppred6} Same as table \ref{powpred} with equation (2) on
the pressure tank 6.}
\end{table}

\begin{table}
\begin{center}
\begin{tabular}{|c|c|c|c|c|c|c|c|} \hline
Set \#  & \# fits & fit $p_c$  & true $p_c$ & $t_{last}$ & $z$ & $\omega$
& $\tau$ \\ \hline
1 & $1$  &  $850$  & $713$ & $678$ & $-2.4$ & $5.6$  & $530$ \\ \hline
1 & $2$  &  $761,831$  & $713$ & $680$ & $-1.4,-1.7$ & $11.4,5.6$ &
$1.7,5.8$ \\ \hline
1 & $1$  &  $723$  & $713$ & $682$ & $-0.45$& $6.9$  & $1.7$ \\ \hline
1 & $5$  &  $732,714$  & $713$ & $684$ & $-0.07,-0.09$& $9.4,6.8$  &
$1.4,1.5$ \\ \hline
1 & $2$  &  $733,728$  & $713$ & $687$ & $0.14,0.15$& $10.3,9.3$  &
$1.4,1.4$ \\ \hline
1 & $3$  &  $702,1070$  & $713$ & $689$ & $0.19,-2.4$& $5.1,5.4$  &
$1.6,24$ \\ \hline
1 & $4$  &  $738,1093$  & $713$ & $691$ & $0.10,-2.2$& $11.1,5.1$  &
$1.4,32$ \\ \hline
1 & $3$  &  $859,730$  & $713$ & $698$ & $0.75,-1.6$& $1.0,2.8$  & $6.8,10$
\\ \hline
1 & $3$  &  $763$  & $713$ & $701$ & $-2.4$ & $4.3$  & $10.0$ \\ \hline
1 & $1$  &  $711$  & $713$ & $703$ & $-2.4$& $5.7$  & $1.8$ \\ \hline
1 & $9$  &  $711,713$  & $713$ & $705$ & $-0.14,-0.17$ & $5.3,5.8$  &
$2.0,1.7$ \\ \hline
1 & $5$  &  $712$  & $713$ & $707$ & $-0.17$&$5.5$  & $2.0$ \\ \hline
1 & $6$  &  $715$  & $713$ & $709$ & $-0.24$&$6.2$  & $1.9$ \\ \hline
\end{tabular}
\end{center}
\caption{\label{tanhpred1} Same as table \ref{powpred} with equation (3) on
the pressure tank 1.}
\end{table}

\begin{table}
\begin{center}
\begin{tabular}{|c|c|c|c|c|c|c|c|} \hline
Set \#  & \# fits & fit $p_c$  & true $p_c$ & $t_{last}$ & $z$ & $\omega$
& $\tau$ \\ \hline
2 & $1$  &  $690$  & $673$ & $629$ & $-2.6$  & $2.6$ & $6.8$   \\ \hline
2 & $3$  &  $726,682$  & $673$ & $631$ & $0.51,-1.5$  & $1.3,2.5$ &
$2.4,8.9$   \\ \hline
2 & $1$  &  $693$  & $673$ & $633$ & $-1.1$  & $2.5$ & $6.2$   \\ \hline
2 & $4$  &  $752,672$  & $673$ & $635$ & $0.49,-2.5$  & $1.3,3.2$ &
$2.4,7.5$   \\ \hline
2 & $3$  &  $735,762$  & $673$ & $638$ & $0.26,-2.9$  & $2.7,3.5$ &
$2.3,44$   \\ \hline
2 & $2$  &  $756$  & $673$ & $640$ & $-2.4$  & $3.3$ & $16.0$  \\ \hline
2 & $2$  &  $866,683$  & $673$ & $644$ & $-1.7,-2.8$  & $2.2,3.4$ &
$3.1,9.5$   \\ \hline
2 & $1$  &  $722$  & $673$ & $648$ & $-1.7$  & $2.7$ & $13$  \\ \hline
2 & $4$  &  $657,696$  & $673$ & $651$ & $0.08,-2.1$  & $1.5,2.5$ & $18,13$
\\ \hline
2 & $4$  &  $657,689$  & $673$ & $653$ & $0.50,-1.6$  & $1.5,2.5$ &
$2.8,9.0$  \\ \hline
2 & $4$  &  $659,747$  & $673$ & $655$ & $-0.43,-2.7$  & $1.8,3.9$ &
$3.5,5.2$  \\ \hline
2 & $2$  &  $678,766$  & $673$ & $657$ & $-2.1,-2.2$   & $2.6,3.1$ &
$887,63$  \\ \hline
2 & $3$  &  $685$  & $673$ & $659$ & $-1.8$   & $3.2$ & $3.5$  \\ \hline
2 & $9$  &  $695,697$  & $673$ & $664$ & $-2.6,2.8$   & $3.0,3.1$ & $67,69$
\\ \hline
\end{tabular}
\end{center}
\caption{\label{tanhpred2} Same as table \ref{powpred} with equation (3) on
the pressure tank 2.}
\end{table}

\begin{table}
\begin{center}
\begin{tabular}{|c|c|c|c|c|c|c|c|} \hline
Set \#  & \# fits & fit $p_c$  & true $p_c$ & $t_{last}$ & $z$ & $\omega$
& $\tau$ \\ \hline
3 & $8$  &  $898,825$  & $764$ & $741$ & $-2.7,-2.1$ & $13.3,6.8$  &
$227,7.2$ \\ \hline
3 & $13$  &  $877,803$  & $764$ & $744$ & $-2.1,-1.0$ & $13.4,6.4$  &
$572,8.1$ \\ \hline
3 & $8$  &  $752,807$  & $764$ & $748$ & $0.62,-1.2$ & $12.0,6.6$  &
$120,4.7$ \\ \hline
3 & $8$  &  $758,770$  & $764$ & $752$ & $0.63,0.44$ & $4.0,6.2$  &
$1.0,1.3$ \\ \hline
3 & $2$  &  $760,782$  & $764$ & $755$ & $0.41,0.09$ & $4.9,6.7$  &
$4.7,1.8$ \\ \hline
3 & $2$  &  $863,784$  & $764$ & $757$ & $-1.60.12$ & $13.2,6.3$  & $951$
\\ \hline
3 & $8$  &  $796,784$  & $764$ & $760$ & $-0.03,-0.04$& $9.0,6.9$  &
$1.6,6.5$ \\ \hline
3 & $2$  &  $772,776$  & $764$ & $762$ & $0.26,0.09$ & $5.6,6.1$  &
$1.7,27$ \\ \hline
\end{tabular}
\end{center}
\caption{\label{tanhpred3} Same as table \ref{powpred} with equation (3) on
the pressure tank 3.}
\end{table}

\begin{table}
\begin{center}
\begin{tabular}{|c|c|c|c|c|c|c|c|} \hline
Set \#  & \# fits & fit $p_c$  & true $p_c$ & $t_{last}$ & $z$ & $\omega$
& $\tau$ \\ \hline
4 & $13$  &  $736,752$  & $756$ & $733$ & $0.86,0.42$ & $3.1,5.9$  &
$2.4,474$ \\ \hline
4 & $11$  &  $739,758$  & $756$ & $735$ & $0.78,0.32$ & $3.5,6.7$  &
$52,165$ \\ \hline
4 & $12$  &  $751,760$  & $756$ & $737$ & $0.63,0.34$ & $5.6,6.7$  &
$1.8,87$ \\ \hline
4 & $22$  &  $781,767$  & $756$ & $739$ & $0.16,0.36$ & $9.5,7.0$  &
$3.3,4.5$ \\ \hline
4 & $32$  &  $763,759$  & $756$ & $741$ & $0.79,0.40$ & $8.6,6.2$  &
$1.1,373$ \\ \hline
4 & $12$  &  $756,759$  & $756$ & $744$ &  $0.40,0.35$  & $6.1,6.4$ &
$428,354$ \\ \hline
4 & $10$  &  $780,758$  & $756$ & $748$ & $-0.16,0.30$ & $10.3,6.3$  &
$924,16$ \\ \hline
4 & $10$  &  $769,758$  & $756$ & $751$ & $0.09,0.30$ & $10.3,6.3$  &
$924,16$ \\ \hline
\end{tabular}
\end{center}
\caption{\label{tanhpred4} Same as table \ref{powpred} with equation (3) on
the pressure tank 4.}
\end{table}

\begin{table}
\begin{center}
\begin{tabular}{|c|c|c|c|c|c|c|c|} \hline
Set \#  & \# fits & fit $p_c$  & true $p_c$ & $t_{last}$ & $z$ & $\omega$
& $\tau$ \\ \hline
5 & $3$  &  $751,717$  & $797$ & $687$ & $-1.3,0.06$ & $13.6,7.5$ &
$1.4,1.5$ \\ \hline
5 & $3$  &  $739,798$  & $797$ & $689$ & $-0.08,-2.9$ &$11.2,8.6$ &
$1.4,148$ \\ \hline
5 & $4$  &  $732,762$  & $797$ & $691$ & $-0.03,-2.5$ & $9.8,5.2$ &
$1.4,11$ \\ \hline
5 & $3$  &  $705,773$  & $797$ & $694$ & $0.09,-2.4$ & $5.6,7.2$ &
$1.6,2.9$ \\ \hline
5 & $1$  &  $765$  & $797$ & $698$ & $-2.0$ &  $6.5$ &  $2.8$ \\ \hline
5 & $2$  &  $752,758$  & $797$ & $701$ & $-0.61,-2.1$ & $11.1,4.7$ &
$1.5,4.6$ \\ \hline
5 & $10$  &  $710,718$  & $797$ & $703$ & $-1.8,-0.57$ &  $5.5,6.2$ &
$2.0,2.3$ \\ \hline
5 & $7$  &  $710,722$  & $797$ & $705$ & $-0.22,-0.85$ &  $5.1,7.0$ &
$2.1,12$ \\ \hline
5 & $7$  &  $712,713$  & $797$ & $707$ & $-0.28,0.31$ &  $5.5,5.9$ &
$2.1,1.9$ \\ \hline
5 & $9$  &  $715$  & $797$ & $709$ & $-0.38$ &  $6.2$ &  $2.0$ \\ \hline
\end{tabular}
\end{center}
\caption{\label{tabtanh5} Same as table \ref{powpred} with equation (3) on
the pressure tank 5}
\end{table}

\begin{table}
\begin{center}
\begin{tabular}{|c|c|c|c|c|c|c|c|} \hline
Set \#  & \# fits & fit $p_c$  & true $p_c$ & $t_{last}$ & $z$ & $\omega$
& $\tau$ \\ \hline
6 & $4$  &  $722,745$  & $734$ & $688$ & $1.3,-1.5$ & $13.7,5.3$ &
$1.1,1.5$ \\ \hline
6 & $2$  &  $741,867$  & $734$ & $690$ & $-2.3,-2.8$ & $2.8,5.0$ & $35,897$
\\ \hline
6 & $7$  &  $763,762$  & $734$ & $692$ & $-1.4,-2.0$ & $6.9,6.1$ &
$1.6,1.6$ \\ \hline
6 & $4$  &  $776$  & $734$ & $696$ & $-2.0$ & $6.7$ & $1.9$ \\ \hline
6 & $4$  &  $724,761$  & $734$ & $700$ & $0.94,-2.2$ & $12.2,3.4$ & $1.2$
\\ \hline
6 & $2$  &  $770$  & $734$ & $702$ & $-2.8$ & $3.3$ & $7.5$ \\ \hline
6 & $7$  &  $729,782$  & $734$ & $705$ & $0.81,-2.4$ & $12.4,4.9$ &
$1.2,3.0$ \\ \hline
6 & $2$  &  $722,790$  & $734$ & $708$ & $0.76,-2.0$ & $10.6,5.4$ &
$1.3,3.5$ \\ \hline
6 & $6$  &  $746,978$  & $734$ & $711$ & $0.66,-2.6$ & $11.1,6.4$ &
$1.3,6.1$ \\ \hline
6 & $7$  &  $753,820$  & $734$ & $713$ & $0.67,-2.7$ & $12.8,5.2$ &
$1.3,5.6$ \\ \hline
6 & $6$  &  $730,812$  & $734$ & $716$ & $0.86,-2.3$ & $12.9,4.7$ &
$1.2,5.5$ \\ \hline
6 & $8$  &  $839,914$  & $734$ & $719$ & $2.2,-2.0$ & $3.3,5.0$ & $2.6,6.2$
\\ \hline
6 & $7$  &  $921$  & $734$ & $723$ & $-1.8$ & $6.0$ & $4.5$ \\ \hline
6 & $20$  &  $759,892$  & $734$ & $726$ & $1.0,-2.7$ & $2.0,4.4$ &
$2.9,5.6$ \\ \hline
6 & $13$  &  $883$  & $734$ & $729$ & $-1.8$ & $4.9$ & $4.0$ \\ \hline
6 & $13$  &  $785,889$  & $734$ & $732$ & $1.6,-0.03$ & $2.4,6.1$ &
$2.5,3.0$ \\ \hline
\end{tabular}
\end{center}
\caption{\label{tanhpred6} Same as table \ref{powpred} with equation (3) on
the pressure tank 6.}
\end{table}

\end{document}